\documentclass[journal,a4paper,12pt,onecolumn]{IEEEtran} 
\usepackage{booktabs} 
\usepackage[ruled]{algorithm2e}

\SetAlFnt{\small}
\SetAlCapFnt{\small}
\SetAlCapNameFnt{\small}
\SetAlCapHSkip{0pt}
\IncMargin{-\parindent}

\usepackage[cmex10]{amsmath}
\usepackage{amssymb}
\usepackage{mathtools}
\usepackage{mathrsfs}

\newtheorem{cons}{Constraint}

\usepackage[svgnames, x11names]{xcolor}

\usepackage[normalem]{ulem} 

\usepackage{blindtext}

\usepackage{listings}

\lstdefinelanguage{XML}
{
basicstyle=\ttfamily\footnotesize,
  morestring=[b]",
  moredelim=[s][\bfseries\color{Maroon}]{<}{\ },
  moredelim=[s][\bfseries\color{Maroon}]{</}{>},
  moredelim=[l][\bfseries\color{Maroon}]{/>},
  moredelim=[l][\bfseries\color{Maroon}]{>},
  morecomment=[s]{<?}{?>},
  morecomment=[s]{<!--}{-->},
  commentstyle=\color{gray},
  stringstyle=\color{blue},
  identifierstyle=\color{red}
}
\usepackage[pdftex]{graphicx}
\graphicspath{{./figures/}}
\DeclareGraphicsExtensions{.pdf}

\usepackage{subfig}

\usepackage{multirow}
\usepackage{rotating}
\usepackage{booktabs} 

\usepackage{array}

\usepackage{hyperref}

\usepackage{xspace}
\usepackage[nocompress]{cite}

\usepackage{enumitem}

\newcommand{\specialcell}[2][c]{%
  \begin{tabular}[#1]{@{}l@{}}#2\end{tabular}}

\usepackage{layout}

\usepackage{array}
\newcolumntype{L}[1]{>{\raggedright\let\newline\\\arraybackslash\hspace{0pt}}m{#1}}
\newcolumntype{C}[1]{>{\centering\let\newline\\\arraybackslash\hspace{0pt}}m{#1}}
\newcolumntype{R}[1]{>{\raggedleft\let\newline\\\arraybackslash\hspace{0pt}}m{#1}}

\begin{document}
%


\title{Distributed Scheduling of Event Analytics across Edge and Cloud}


\author{\IEEEauthorblockN{Rajrup Ghosh\IEEEauthorrefmark{1} and 
Yogesh Simmhan\IEEEauthorrefmark{1}}
\IEEEauthorblockA{\\\IEEEauthorrefmark{1}Department of Computational and Data Sciences, \\
Indian Institute of Science, Bangalore 560012, India\\
Email: rajrup@grads.cds.iisc.ac.in, simmhan@cds.iisc.ac.in}
}

\maketitle

\begin{abstract}
Internet of Things (IoT) domains generate large volumes of high velocity event streams from sensors, which need to be analyzed with low latency to drive decisions. Complex Event Processing (CEP) is a Big Data technique to enable such analytics, and is traditionally performed on Cloud Virtual Machines (VM). Leveraging captive IoT edge resources in combination with Cloud VMs can offer better performance, flexibility and monetary costs for CEP. Here, we formulate an optimization problem for \emph{energy-aware placement of CEP queries}, composed as an analytics dataflow, across a collection of edge and Cloud resources, with the goal of minimizing the end-to-end latency for the dataflow. We propose a Genetic Algorithm (GA) meta-heuristic to solve this problem, and compare it against a brute-force optimal algorithm (BF). We perform detailed real-world benchmarks on the compute, network and energy capacity of edge and Cloud resources. 
These results are used to define a realistic and comprehensive simulation study that validates the BF and GA solutions for $45$ diverse CEP dataflows, LAN and WAN setup, and different edge resource availability. We compare the GA and BF solutions against random and Cloud-only baselines for  
different 
configurations, for a total of $1764$ simulation runs. 
Our study shows that GA is within $97\%$ of the optimal BF solution that takes hours, maps dataflows with $4-50$ queries in $1-26~secs$, and only fails to offer a feasible solution $\le 20\%$ of the time.
\end{abstract}




%
%

\section{Introduction}\label{sec:intro}

Internet of Things (IoT) is a new computing paradigm where pervasive sensors and actuators deployed in the physical environment, with ubiquitous communication, allow us to observe, manage and enhance the efficiency of the system. The applications motivated by IoT spans cyber-physical city utilities such as smart water management~\cite{perera2014sensing}, health and lifestyle applications like smart watches~\cite{smart-health}, and even mobile platforms such as unmanned drones and self-driving cars. 
%
A key requirement for IoT applications is to apply analytics over the data collected from the distributed sensors to make intelligent decisions to control the system. Often, these decisions are performed on data that is \emph{continuously streaming} from the edge devices at high input rates. 
These analytics and decision making may also be time-sensitive, and require a \emph{low latency response}, such as in a smart power grid~\cite{simmhan:cise:2012}. 

Big Data platforms for stream and event processing enable \emph{continuous analytics} for IoT applications~\cite{siddhi11}. They are designed for low latency processing of data or event streams, such as from physical sensors or social network feeds~\cite{Ahmad:2004,storm-twitter}. In particular, \emph{Complex Event Processing (CEP) engines} allow users to define intuitive SQL-like queries over event streams that are executed on tuples as they arrive~\cite{cep-survey}. They are used to detect when thresholds are breached to trigger alerts, aggregate events over temporal windows, or identify events with a specific pattern of interest. The queries can be composed as a dataflow graph for online decision making in IoT applications~\cite{debs-challenge-soccer}. CEP queries are often implemented as deterministic/non-deterministic finite state automata by CEP engines~\cite{Hirzel12,Woods10}.

A common information processing architecture is to move data from thousands or millions of edge devices centrally into \emph{public Clouds}, where CEP or other analytics engines hosted on Virtual Machines (VM) can process the incoming streams, and data can be persisted for mining and visualization~\footnote{ https://aws.amazon.com/iot/how-it-works/}~\footnote{https://www.microsoft.com/en-in/server-cloud/internet-of-things/overview.aspx}. Here, the edge devices serve as ``dumb'' sensors that transmit the data to the Cloud for centralized analytics. 

However, moving event streams from the edge to the Cloud introduces data transfer latency, and the public network may become a bandwidth bottleneck for large deployments. There are associated monetary costs for the use of Cloud VMs and network. Edge devices are getting powerful, energy-efficient, and affordable. Their use is widespread as gateways in IoT deployments, and offer captive computing at low/no cost. Lastly, having a hub and spoke model where all data is ingested to the Cloud does not give us any fine-grained control over where data from the IoT deployment can go. 

In this article, we \emph{propose approaches for distributed event-based analytics across edge and Cloud resources to support IoT applications}. We consider a deployment with multiple event streams generated at the edge at high frequency, an analytics dataflow composed of CEP queries that should execute over these streams, and multiple edge devices and public Cloud VMs to perform the queries. Our goal is to find a distributed placement of these queries onto the edge and Cloud resources to \emph{minimize the end-to-end latency} for performing the event analytics. This placement must meet \emph{constraints} of throughput capacity on edge and Cloud machines, bandwidth and latency limits of the network, and energy capacity of the edge devices. The latter is particularly novel -- edge devices are often powered by batteries that are recharged, say through solar panels, and hence have a \emph{limited energy budget} between the recharge cycles.

\emph{Mobile Clouds}~\cite{clonecloud11}~\cite{mobile:fernando:2013} benefit from off-loading computing from smart phones to the Cloud, but they move parts of the application only to the Cloud and not other edge devices. Some support an edge-only solution using transient devices, but are not amenable to stream processing~\cite{Shi12}. \emph{Fog Computing} offers a low-latency ``data center'' close to the Edge with uplink to the Cloud, but lacks wide deployment, programming models and scheduling algorithms~\cite{varshney:icfec:2017}. Our work can extend to the Fog too. \emph{Peer-to-peer (P2P)} systems and query operator placement in Wireless Sensor Networks are relevant to IoT as well~\cite{p2p:record:2003,Srivastava05}, but do not consider recent evolutions like Cloud computing and large data rates like we do.
%
%
Our prior short paper tackles a simpler problem of bi-partitioning a CEP pipeline between a single edge device and the Cloud~\cite{Nithya14}. We address this more comprehensively here, and consider multiple edge devices, energy constraints and offer detailed experiments.

We make the following specific contributions in this article.
\begin{enumerate}
	\item We formulate the problem of query placement for a directed acyclic graph (DAG) with constant input rate onto distributed edge and Cloud resources, having computing, network and energy constraints, as a \emph{combinatorial optimization problem}, with the objective function being to minimize the end-to-end processing latency (\S~\ref{sec:problem}). 
	\item We propose a costly but optimal \emph{brute force approach} to solve this problem, and also a more practical solution based on the \emph{Genetic Algorithm meta-heuristic}  (\S~\ref{sec:approach}).
	\item We perform and present \emph{comprehensive, real-world micro-benchmarks} for a wide class of {21} CEP queries relevant to IoT, at different input event rates. These evaluate the throughput of a Raspberry Pi edge and Microsoft Azure Cloud VM, the energy capacity of the Pi, and network characteristics of edge and Cloud (\S~\ref{sec:bm}).
	\item Due to the lack of existing platforms for distributed event analytics across edge and Cloud, we instead conduct a \emph{detailed and realistic simulation study} using compute, energy and network property distributions measured by the micro-benchmarks. We evaluate the effectiveness of the query placement for 45 synthetic CEP dataflows, with varying numbers of queries, input rates, and resource availability. We offer a rigorous analysis of the results, using both quality metrics and timing analysis (\S~\ref{sec:sim}).
\end{enumerate}


\section{Background}\label{sec:background}
This problem is motivated by our prior work at the University of Southern California and the City of Los Angeles~\cite{simmhan:cise:2012}, and our current work at the Indian Institute of Science~\cite{smartx} on developing a Campus IoT fabric for emerging smart utility applications.
We use \emph{CEP engines} to perform continuous queries over one or more event streams, coming from smart meters and water level sensors, to detect patterns of interest~\cite{usc-smart-grid}. Each event is a tuple, 
and typically contains the timestamp, sensor ID and observed values. Such event patterns can predict energy surges, trigger water pumping operations, and notify consumers of surge pricing. 

\begin{table}[t]%
\centering
\caption{Example Siddhi CEP queries used in experiments\label{tbl:queries} 
} 
\begin{tabular}{ll}
	\hline
	{\bf Query Type} & {\bf Siddhi Query Definition}   \\ 
	\hline
	\hline
	{Filter} & {\specialcell[c]{\texttt{\textbf{define stream} inStream (height int);} \\\texttt{\textbf{from} inStream[height<150]~\textbf{select} height}\\\texttt{\textbf{insert into} outStream ;}}}\\ \hline
	
	\specialcell[c]{Sequence\\\emph{~Match 3 events}} & {\specialcell[c]{\texttt{\textbf{from every} e1 = inStream, e2 = inStream[e1.height==}\\ \texttt{e2.height], e3 = inStream[e3.height==e2.height]}\\\texttt{\textbf{select} e1.height \textbf{as} h1, e2.height \textbf{as} h2, e3.height \textbf{as} h3}\\\texttt{\textbf{insert into} outStream ;}}}\\ \hline
	
	\specialcell[c]{Pattern \\\emph{~Match 3 events}} & {\specialcell[c]{\texttt{\textbf{from every} e1=inStream -> e2=inStream[e1.height==}\\ \texttt{e2.height] -> e3=inStream[e2.height==e3.height]}\\\texttt{\textbf{select} e1.height \textbf{as} h1, e2.height \textbf{as} h2, e3.height \textbf{as} h3}\\\texttt{\textbf{insert into} outStream ;}}} \\ \hline
	
	\specialcell[c]{Aggregate (Batch) \\\emph{~Window Size = 60}}   & {\specialcell[c]{\texttt{\textbf{from} inStream \emph{\#window.lengthBatch(60)}} \\ \texttt{\textbf{select} avg(height) \textbf{as} AvgHeight~~\textbf{insert into} outStream ;}}}  \\ \hline 
	\specialcell[c]{Aggregate (Sliding) \\\emph{~Window Size = 60}} & {\specialcell[c]{\texttt{\textbf{from} inStream \emph{\#window.length(60)}} \\ \texttt{\textbf{select} avg(height) \textbf{as} AvgHeight~~\textbf{insert into} outStream ;}}}   \\ \hline
\end{tabular}
\end{table}%

CEP engines register queries and execute them continuously over event streams for days or weeks~\cite{siddhi11,Hirzel12}. CEP queries are of 4 four major types, as illustrated in Table~\ref{tbl:queries} using data from a water level sensor. \emph{Filter} queries match a property predicate against fields in the incoming event, and only those events that match the predicate are placed in the output stream. For example, `\texttt{height < 150}' in row 1 of Table~\ref{tbl:queries} is the predicate that is matched, detecting a situation where the water level may have dropped below a threshold. A \emph{Sequence} query matches predicates on consecutive events, and if all predicates match the sequence of events, those events are placed in the output stream. A \emph{Pattern} query is similar, except that the matching events do not have to be contiguous. For example, in row 3 of Table~\ref{tbl:queries}, we match events with the same heights for 3 \emph{successive} events \texttt{e1}, \texttt{e2} and \texttt{e3}, that is, there may be other non-matching events between \texttt{e1} and \texttt{e2}, and/or \texttt{e2} and \texttt{e3}. Lastly, \emph{Aggregate} queries apply an aggregation function, like average, on a window of events. The window gives the \emph{count} of consecutive events to be included, and may be formed by \emph{sliding} over the input stream incrementally, or by \emph{batching} events into non-overlapping windows.

The CEP queries can be composed into a \emph{directed acyclic graph (DAG)} or a dataflow, where vertices are queries and edges indicate events passed from the output of one query to the input of the downstream query. Multiple queries may run within the same CEP engine on a machine, or different queries may run on CEP engines in different machines and coordinate their execution using events streamed over the network. 

\section{Related Work}\label{sec:related}
There are three primary related research areas relevant to our article: mobile Clouds, Fog and Peer-to-Peer (P2P) computing, and query processing in sensor networks. 

\emph{Mobile Cloud Computing}~\cite{mobile:fernando:2013} has grown as a research area that lies at the intersection of mobile devices such as smart phones and the Cloud. The key idea is to use these personal devices as a thin client to access rich services hosted on Clouds, forming a variation of a client-server model. In addition, the concept of Cloudlets has been proposed as an additional layer that sits between the edge and the Cloud to help reduce latencies while offering superior computing power than the edge alone ~\cite{cloudlet:satyanarayanan:2009}. This is conceptually the computing equivalent of Content Distribution Networks that move data closer to the edge. Both these paradigms conceive of interactions between a single client and a remote Cloud/Cloudlet, which is in contrast to our approach of leveraging the collective capabilities of distributed edge devices and the Cloud.

Specific research papers have extended these broad concepts further.
CloneCloud is an application partitioning framework for mobile phones that off-loads a part of the application execution from the edge device to device ``clones'' in a public Cloud~\cite{clonecloud11}. Partitioning is done by migrating a thread from the mobile device at a chosen point in the application to the clone in the cloud. After execution on the cloud, it is re-integrated back onto the mobile device. It models energy as a function of CPU, display and network state, and this is considered in their partitioning strategy. 
A further extension tries to improve upon this by reducing state transfer costs for dynamic offloading~\cite{Yang14}. 
Although CloneCloud partitions an interactive application across mobile and Cloud, it does not address streaming analytics applications essential for the IoT domain, where factors like latency and throughput need to be addressed, nor does it use multiple edge devices.   
%

Others do deal with partitioning of data stream applications between mobile devices and Cloud to maximize the throughput of stream processing~\cite{yang12}. This framework considers sharing VM instances among multiple users in the Cloud to improve VM utilization, and solves the problem using a genetic partitioning algorithm like us. The empirical evaluation, however, uses a QR code recognition application, which is unlike the high rate event analytics that we support for IoT domains. Further, it does not consider distributing tasks to multiple edge resources either. 

Our own prior work~\cite{Nithya14} has considered a similar bi-partitioning of a CEP query pipeline between a single edge device and the cloud. Factors like compute time on each query, incoming event rate and latency between resources are used to find an edge-cut in the DAG such that 
latency is reduced. It also considers enforcing privacy constraints on event streams to determine if a stream is allowed on a resource. It does not, however, consider multiple edge devices and could be solved optimally using a dynamic programming solution. Energy constraints were not considered either, and there was no empirical evaluation. Besides the flexibility of multiple resources and energy constraints considered in our current article, we also present robust benchmarks and empirical evaluation. 

Distributed query processing on multiple edges and the Cloud have been considered for feed-following applications~\cite{Chandramouli13}. Here, database views of applications that follow social network feeds are distributed to edge devices, with query operators that are applied on the feeds by existing relational databases engines. The problem is modeled as a view placement problem with the goal of optimizing communication between sub-queries running on the edge and the Cloud. 
However, there is a difference between their view placement and our query placement problem on edge devices. In the former, the edge devices can communicate with each other only through the Cloud, causing a ``star'' network topology. This reduces the optimization problem to linear time. We instead allow the edges to communicate with each other, which is feasible as they are typically on the same private network. 

Serendipity~\cite{Shi12} is a more comparable work that uses remote computing collaboratively among closely connected mobile devices. It explores off-loading of computationally intensive tasks onto other intermittently connected mobile devices rather than the Cloud. In their model, jobs are distributed to nearby mobile devices with the aim of reducing the job completion time and conserving the device's energy. Unlike us, their approach does not work for streaming applications, as that would not be viable for transiently connected devices, and they do not consider Cloud resources either. 

More broadly, the concept of \emph{Fog Computing} is gaining traction in the IoT domain~\cite{fog:bonomi:2014}. Here, the Cloud with a massive centralized data center is supplanted by a fog of wireless edge devices that collaboratively offer computing resources. These have the benefit of low-latency communication and the ability to self-organize locally, but lack full (centralized) control and their availability is unreliable~\cite{fog:vaquero:2014}. Fog computing platforms are still in a nascent stage, but our approach to distributed analytics execution across edge devices (Fog) and the Cloud offers a model for coordinating their computation, and leveraging the best features of these two paradigms.

As such, there has historically been work on such \emph{Peer-to-Peer (P2P)} systems, where interconnected nodes can self-organize into a network topology to share files, CPU cycles, storage and bandwidth~\cite{p2p:record:2003}. Peers can offload execution of tasks to other peers to speed up their job completion, and significant work on lookup services such as distributed hash tables have taken place. 
This type of content search and retrieval requires guarantees on QoS parameters like timely results, utilization of resources, response time and correctness. For e.g., \cite{Lesser04}~\cite{Karrels13} give algorithms for distributing tasks to a set of peers using hierarchical coalition formation. Our work operates on a more deterministic set of edge devices (peers), but can benefit from the management services developed for \emph{ad hoc} P2P systems. Issues like energy and mobile devices were not relevant for those architectures but gain prominence in IoT ecosystems we target.  

Yet another related area from a decade back is on query processing in \emph{Wireless Sensor Networks (WSN)}. Here, distributed sensors (motes) deployed to measure environmental parameters assemble together to solve streaming analytics task~\cite{Gehrke03}~\cite{telegraphcq}. Query plan optimizing and placement techniques have been explored in the context of a large number of sensor nodes. A virtual tree topology is created with an elected leader node which receives query requests from users, and sends smaller tasks to worker nodes having the relevant streams~\cite{Srivastava05}.  Intermediate nodes in the tree can partially process the query or forward the results back to the leader to build the final result set. Like us, these strategies try to reduce the energy and communication costs on these embedded devices. Rather than consider individual queries that are partitioned into sub-queries, we consider a DAG of queries with placement at the granularity of query. CEP engines have a richer query model as well, and we also have access to Cloud VMs rather than execute exclusively on motes on the edge.  

\section{Problem Formalization}
\label{sec:problem}
In this section, we formally state the CEP query placement problem for a DAG, which we have motivated, and formulate it as a constrained optimization problem. The solution approaches to the problem are offered in the next section. 

\subsection{Preliminaries}
The streaming dataflow application 
is composed as a \textit{Directed Acyclic Graph (DAG)} of vertices and edges: \( \mathcal{G} = \langle \mathbb {V},\mathbb {E} \rangle \), where $\mathbb{V}=\{v_i\}$ is the set of \textit{CEP queries} that are the vertices of the DAG, and $\mathbb{E}$ is the set of \textit{event streams} that connect the output of query $v_i$ to the input of the next query $v_j$, and form the directed edges.  $\mathbb{E}$ is given by:
\[ \mathbb {E} =\{e_i ~|~ e_i= \langle v_i,v_j \rangle, v_i \in \mathbb{V}, v_j \in \mathbb{V} \cup \phi\} \]

The output event(s) of a query follow \emph{duplicate semantics} and are forwarded to all out-edges from a vertex. Multiple in-edges to a query follow \emph{interleave semantics}, meaning events from all in edges are appended to a single logical input stream for the vertex. The queries that receive the initial input event streams into the DAG are called \textit{source queries}, $\mathbb{V}^{SRC}$, and are characterized by having no in-edges and perform no computation. These are ``no-op'' tasks that generate and pass events downstream. Similarly, the queries that emit the output streams from the DAG are called \textit{sink queries}, $\mathbb{V}^{SNK}$, and have unbound out-edges that are incident on a dummy vertex $\phi$\footnote{The dummy vertex $\phi$ allows sink queries to have edges to other queries, but identify a separate unbounded edge for the final DAG outputs. Source queries are in fact equivalent to the dummy $\phi$ since they are a no-op.}. 
\begin{center}
	$\forall v_j \in \mathbb{V},\quad \mathbb{V}^{SRC} = \{v_j | ~\nexists e_i = \langle v_i,v_j \rangle \in \mathbb{E}\} \text{\quad and \quad} \mathbb{V}^{SNK} = \{ v_j | ~\exists e_j = \langle v_j, \phi \rangle \in \mathbb{E}\}$
\end{center}

A query executes on a specific \textit{computing resource} $r_k$ and the set of all computing resources available within the IoT infrastructure is given by $\mathbb{R} = \{r_k\}$. We consider two classes of computing resources -- \textit{edge devices} such as smart phones and Raspberry Pi's
, and \emph{Cloud resources} such as VMs, each having a specific computing capability. These two computing resources form two mutually exclusive sets, $\mathbb{R}_E$ for edge devices and $\mathbb{R}_C$ for Cloud VMs, respectively. Thus, $\mathbb{R}_E \cup \mathbb{R}_C=\mathbb{R}$ and $\mathbb{R}_E \cap \mathbb{R}_C=\varnothing$. A \textit{resource mapping function}, $\mathcal{M} :\mathbb{V} \rightarrow \mathbb{R}$, indicates the resource on which a query executes.

A \emph{query path} $p_i = \langle v_0,v_1 \rangle, ...., \langle v_k,v_{k+1} \rangle, ..., \langle v_n,v_{n+1} \rangle$ of length $n$ is a unique sequence of alternating and distinct vertices and edges, starting at a source vertex, $v_0 \in \mathbb{V}^{SRC}$ and ending at a sink vertex, $v_{n+1} \in \mathbb{V}^{SNK}$. $\mathbb{P}$ is the set of all paths in the DAG. 

A \textit{selectivity} function, $\sigma(v_i)$, for each query of the DAG gives a statistical measure of the average number of output events generated for every input event consumed by the query. Using duplicate semantics, the selectivity of each out-edge is same as the selectivity of the vertex writing to that edge. The \textit{stream rate} defines the number of events passing per unit time on a stream. The \textit{incoming rate} of the DAG $\Omega^{in}$, is the sum of the stream rates emitted by all source queries in the DAG. Similarly the \textit{outgoing rate} denoted by $\Omega^{out}$ is the sum of output rate of events emitted by the sink queries onto the dummy sink $\phi$. Then \textit{selectivity the whole DAG}, $\sigma(\mathcal{G})=\frac{\Omega^{out}}{\Omega^{in}}$.

The \textit{incoming rate}, $\omega^{in}_{i}$ for a vertex $v_i$ is the sum of stream rates on all in-edges, due to interleave semantics. Given duplicate semantics, the \textit{outgoing rate}, $\omega^{out}_{i}$ for a vertex $v_i$ is the product of its incoming stream rate  $\omega^{in}_{i}$ and its selectivity $\sigma(v_i)$. For simplicity, if the output rate for all source queries $v_k \in \mathbb{V}^{SRC}$ is uniform, we have $\omega^{out}_k = \frac{\Omega^{in}}{|\mathbb{V}^{SRC}|}$.

Using this, we recursively compute the input and output stream rates for downstream vertices $v_j$, and the outgoing stream rate $\Omega^{out}$ for the entire DAG $\mathcal{G}$ as:
\begin{align*}
\forall v_j \in \mathbb{V}, v_i \notin \mathbb{V}^{SRC},\qquad \omega^{out}_i = \omega^{in}_{i} \times \sigma(v_i) \qquad \omega^{in}_i = \sum \limits_{\langle v_j,v_i \rangle \in \mathbb{E}} \omega^{out}_j \qquad \Omega^{out} = \sum \limits_{v_i \in \mathbb{V}^{SNK}} \omega^{out}_{i}
\end{align*}

\textit{Compute latency}, $\lambda_{i}^{k}$, is the time taken to process one event by a query $v_i$ on an exclusive resource $r_k$.  If $n$ queries are placed on the same resource $r_k$, the latency for each query becomes $\sum_{i}^{n} \lambda_{i}^{k}, ~~\forall (v_i, r_k) \in \mathcal{M}$ due to round-robin scheduling~\footnote{For simplicity, we do not consider multi-thread execution of queries by a single CEP engine on a resource. Resources with multiple cores can instead be modelled as multiple resources with $\infty$ bandwidth.}. If $\lambda$ is the latency time in seconds taken by a query to process a single event on a resource, $\lambda^{-1}$ is the \emph{throughput} that can be processed by that query on that resource in $1$~second.

Let the \textit{size of an event} that is emitted by query $v_i$ on its out-edge(s) be denoted by $d_{i}$. The \emph{network latency} and \emph{network bandwidth} between two resources $r_m$ and $r_n$ is denoted by $l_{m,n}$ and $\beta_{m,n}$, respectively. 
Therefore, the \emph{end-to-end latency along a path} $p \in \mathbb{P}$ for a given resource mapping $\mathcal{M}$ can be defined as $L_p$, the sum of the compute latency and the network transfer time. The maximum of these latencies along all paths is $L_{\mathcal{G}}$, the \emph{end-to-end latency for the DAG}, also called the \textit{makespan}. The path which has this maximum time is called the \textit{critical path}.

\begin{center}
	$L_p =  \sum \limits_{\substack{ \langle v_i,v_j \rangle \in p_i\\(v_i, r_m) \in \mathcal{M}\\(v_j, r_n) \in \mathcal{M}}} \Big(\lambda_{i}^{m} + \big(l_{m,n} + \frac{d_{i}}{\beta_{m,n}}\big) \Big) \qquad\qquad L_{\mathcal{G}} = \max_{\forall p ~\in \mathbb{P}}(L_p)$
\end{center}

\subsection{Constraints}
\label{sec:problem:cons}
Based on the motivating scenario introduced in the background, we define several constraints that need to be satisfied when performing a mapping of queries to resources. 
\begin{cons}\label{cons:1}
	All source vertices should be mapped to an edge device, while the sink vertices should be on the Cloud.
	\begin{center}
		$v_i \in \mathbb{V}^{SRC} \implies r_k \in \mathbb{R}_E \qquad\text{and}\qquad v_i \in \mathbb{V}^{SNK} \implies r_k \in \mathbb{R}_C \quad \forall (v_i, r_k) \in \mathcal{M}$
	\end{center}
\end{cons}
This constraint ensures that source queries are co-located on the edge device that is generating the input event stream. Likewise, given that analytics performed after the CEP are hosted on the Cloud, the sink queries must be placed on the VM resource.

\begin{cons}\label{cons:2}
	Given an input rate $\omega^{in}_i$ on vertex $v_i$, if the vertex is \emph{exclusively} mapped to a resource $r_k$, it should not overwhelm the compute throughput capacity.
	\[\omega^{in}_{i} < \frac{1}{\lambda^k_i} ~~\forall v_i \in \mathbb{V}\]
	If \emph{multiple vertices} are running on the same resource $r_k$, then the input throughput $\omega^{in}_i$ on a vertex $v_i$ that the resource $r_k$ can handle is constrained by:
	\[\omega^{in}_{i} < \frac{1}{\sum \limits_{\substack{(v_j, r_k) \in \mathcal{M} \\v_j \notin \mathbb{V}^{SRC}}}\lambda^k_j} \Bigg(1 + \pi_m\Bigg)  \qquad \forall v_i \in \mathbb{V}, v_i \notin \mathbb{V}^{SRC},\quad m = \mid v_j\mid ~\forall (v_j, r_k) \in \mathcal{M}\]
\end{cons}

The maximum event rate that a resource $r_k$ can handle when exclusively running a query $v_i$ is given by the inverse of its latency $\frac{1}{\lambda^k_i}$, and for multiple queries it is the inverse of the sum of their latencies $\frac{1}{\sum\lambda^k_j}$. However, there is likely to be additional overheads in the latter case. Here, $\pi_m$ is a function of $m$ which denotes the \emph{parallelism overhead} of $m$ queries, and is obtained through empirical evaluations. Hence, we should ensure a mapping of a query to a resource such that it does not receive an input rate greater than the compute throughput supported by that query on that resource.

Edge resources are generally run on batteries with a fixed capacity. Let the \emph{power capacity} available for an edge device $r_k$ be $C_k$, given in mAh, when fully charged. Let the \emph{base load} (instantaneous current) drawn by an edge device $r_k$ when no queries are running be given by $\mu_B^k$, in mA. Let $\epsilon_{i}^{k}$ be the \emph{incremental power}, beyond $\mu_B^k$, drawn on the edge resource $r_k \in \mathbb{R}_E$ by a query $v_i$ to process a single input event, given in mAh. Let the \emph{time interval between charging} this edge device $r_k$, be denoted as $\tau_k$, given in seconds, be it through solar regeneration or by replacing the battery.  
\begin{cons}\label{cons:3}
	The queries running on a edge device $r_k$ should not fully drain out the battery capacity of that resource within the recharge time period $\tau_k$.
	
	\[\big( \mu_B^k \times \tau_k \big) \quad + ~ \sum \limits_{\substack{(v_i, r_k) \in \mathcal{M}\\v_i \notin \mathbb{V}^{SRC}\\r_k \notin \mathbb{R}_C}}(\omega^{in}_{i} \times \tau_k ) \times \epsilon_{i}^{k} \leq C_k\]
\end{cons}

We assume that DAGs once registered with the engine and placed on resources run for a much longer time (say days or weeks) than the recharge period (say $24$~hours). Thus our optimization plan should take this energy constraint into consideration~\footnote{For simplicity, we consider that discharging of a battery by an edge resource is linear with time and its full recharge is instantaneous at every time $\tau_k$. In practice, batteries have non-linear discharge cycles based on their present capacity, and batteries charged by solar panels may have charging/discharging constantly occurring during daytime.}.

\subsection{Optimization Problem}
\emph{Given a DAG $\mathcal{G} = \langle \mathbb{V}, \mathbb{E} \rangle$ and a set of edge and Cloud resources $\mathbb{R}$, find a resource mapping $\mathcal{M}$ for each query $v_i \in \mathbb{V}$ on to a resource $r_k \in \mathbb{R}$ such that the mapping meets the Constraints~\ref{cons:1}, \ref{cons:2} and \ref{cons:3} while \emph{minimizing} the end-to-end latency for the DAG.}

In other words, find the mapping that minimizes the end-to-end DAG latency,
\[ \widehat{L_\mathcal{G}}=\min_{\forall (\mathbb{V},\mathbb{R}) ~ \in \mathcal{M}}(L_\mathcal{G}) \]

Note that our optimization problem is orthogonal to the network topology, and the network characteristics between pairs of resources in $\mathbb{R}$ is captured by their latency and bandwidth distributions. Similarly, the resources themselves are abstracted based on their compute and energy capacity distributions for processing specific queries.


\section{Solutions to the Optimization Problem}
\label{sec:approach}
There have been a multitude of techniques that have been proposed to solve optimization problems, much like the ones we have used~\cite{Nocedal06}. Here, we present two approaches for solving the placement problem: one, a \emph{Brute Force (BF) approach} that gives the optimal solution while being computationally intractable for large problem sizes, and the other which translates the problem to a \emph{Genetic Algorithm (GA) meta-heuristic} and gives an approximate but fast solution.

\subsection{Brute Force Approach (BF)}
\label{sec:approach:bf}
Given the Constraint~\ref{cons:1} that source vertices $v_i \in \mathbb{V}^{SRC}$ are pinned to the edge devices and sink vertices $v_i \in \mathbb{V}^{SNK}$ are pinned to Cloud resources, our goal is to find a mapping for the $n$ intermediate vertices of the DAG to either edge or Cloud resources, where $n = \big(|\mathbb{V}| - (|\mathbb{V}^{SRC}| + |\mathbb{V}^{SNK}|)\big)$. In the process, we wish to minimize the end-to-end latency of the query DAG and also meet the other two constraints. 

The Brute Force (BF) approach is a na\"{i}ve technique which does a combinatorial sweep of the entire parameter space. Here, each of the $n$ vertices are placed in every possible $|\mathbb{R}|$ resources as a trial. For each trial, the constraints are evaluated and if all are satisfied, the end-to-end latency for the DAG $L_\mathcal{G}$ is calculated for this placement. If this latency improves the known minimum latency from earlier trials, then the current minimum latency is set to this smaller latency and the trial mapping stored as the current best mapping. Once all possible mappings are tried, the current minimum latency value is the best end-to-end latency, $\widehat{L_\mathcal{G}}$, and its respective mapping is returned. 

\subsubsection{Complexity Analysis}
\label{sec:approach:bf:cplx}
DAG scheduling in general is NP-complete except for some narrow conditions, which our problem does not meet~\cite{Kwok99}. We transform a previous DAG scheduling problem that is proved to be NP-complete into ours in polynomial time. Let $\mathscr{T} = \{T_1, T_2,..., T_n\}$ be a set of task that have a partial order to form a DAG, and $\mathscr{R} = \{R_1, R_2,..., R_r\}$ be resources with a bounded capacity $B_i$. Each task $T_i$ has a latency time $\tau_i$ and a capacity requirement of $R_j(T_i) \le B_j$ when run on resource $R_j$.~\cite{Garey75} have earlier shown that even \emph{checking the existence} of a valid schedule from tasks $\mathscr{T}$ to resources $\mathscr{R}$ while meeting a deadline $D$, the resource capacity bounds, and the partial task ordering is NP-Complete, for more than two resources. We can get the fastest schedule by testing different integer values of $D$ to find the smallest with a valid schedule.
%
%

We transform this known NP-complete problem to our optimization problem, which is more complex, in polynomial time, as follows. We map each task $T_i \in \mathscr{T}$ to a query $v_i \in \mathbb{V}$ with running time $\tau_i$ replacing the sum of the compute and network latencies, $\lambda_{i}^{m} + \big(l_{m,n} + \frac{d_{i}}{\beta_{m,n}}\big)$. We also transform the resource bounds $B_j$ into the compute and energy bounds in Constraints~\ref{cons:2} and ~\ref{cons:3}. Given a mapping of queries to resources $\mathcal{M}: \mathbb{V} \rightarrow \mathbb{R}$, we can test its validity as a solution by checking constraints at each query and finding the critical path in the DAG in $O((|\mathbb{R}| \times |\mathbb{V}|)+\mathbb{|E|})$ time, that is polynomial. Hence, this shows that our optimization problem is NP-complete.

The BF algorithm gives a provably optimal solution since it considers all possible solutions. However, its computational cost is high. Specifically, the asymptotic time-complexity of the Brute Force algorithm is exponential, at $\mathcal{O}\big((|\mathbb{V}|+|\mathbb{E}|)\times|\mathbb{R}|^{n}\big)$.

\subsection{Genetic Algorithm based Optimization Problem Solver}
\label{sec:approach:ga}





Finding an optimal placement of the query to resources is a non-linear optimization problem without a real-valued solution, which makes it difficult to use heuristics like integer linear programming~\cite{clonecloud11}. Many NP-complete problems have been practically solved using evolutionary meta-heuristic algorithms like Genetic Algorithm (GA) and Particle Swarm Optimization (PSO)~\cite{Michalewicz96}. 
Since GA has solved hard graph-based problems like Job Scheduling and Travelling Salesman Problem (TSP) with considerable accuracy in practice, we chose this technique to solve this optimization problem. Our 
problem poses an extra challenge of satisfying the constraints too -- converged solutions from such meta-heuristics may cause compute, network, and/or energy violations. GA offers the flexibility of being modified to produce solutions which satisfy multiple constraints, which we use. 


There are four integral components to a GA approach~\cite{Michalewicz96}. \emph{Chromosomes} contain solutions to the problem being solved. A \emph{Population} is the set of all chromosomes whose solutions are being considered. A \emph{Generation} is the number of evolutions (iterations) that the chromosomes in the population have undergone. The \emph{Objective Function} gives the measure of fitness of a chromosome. Defining the GA solution requires us to map our placement problem to each of these stages.

Let a chromosome $Q = \{q_0, q_1, ... q_{n-1}\}$ give the placement of a set of $n = \big(|\mathbb{V}| - (|\mathbb{V}^{SRC}| + |\mathbb{V}^{SNK})|\big)$ queries onto a set of resources $\mathbb{R}$, where $n$ is the number of variables in the GA. The chromosome's values $q_i$ are encoded with an integer value in the range $[0, |\mathbb{R}|-1]$ 
such that it represents the resource number to which the $i^{th}$ query gets mapped to. A set of chromosomes form a population and the population size, $p$, is a fixed value across the generations. 
The $0^{th}$ generation of the population is initialized randomly with $p$ chromosomes. In every generation, an optimization function $F$ gives the \emph{fitness value} $F(j)$ for the $j^{th}$ chromosome $c_j$ in the population. Since we want to minimize the end-to-end latency of the DAG and GA attempts to maximize the fitness value, we define the fitness value for a chromosome solution by subtracting the DAG's end-to-end latency for this placement solution from a large positive constant. 

Apart from the population which offers the current candidate solutions, we also maintain a \emph{best-fit chromosome} which is the solution with the best fitness value seen so far across all generations. After each generation, the current population's new chromosomes are compared with the best-fit chromosome to see if an improved solution has been discovered, and if so, the best-fit chromosome is updated to this. 

We use a ``roulette wheel'' algorithm to select the chromosomes from the current population to use for evolution into the next generation's population. Alternatively, linear rank-based selection and binary tournament selection may also be used~\cite{Hogenboom:2009,Shukla:2015}. For roulette wheel, we first calculate the \emph{total fitness value} for the current population, $\overline{f}$. Then, we calculate the \emph{probability mass function (PMF)}, $\rho_j$, which gives the probability of selecting a chromosome $c_j$ from the population $(c_0, c_1, ..., c_{p-1})$. Next, we compute the \emph{cumulative distribution function (CDF)}, $\delta_j$ of this PMF. These are given by:
\begin{center}
	$\overline{f} =  \sum\limits_{j=0}^{p-1} F(j) \qquad\qquad \rho_j = \mathcal{P}(J=j) = \frac{F(j)}{\overline{f}} \qquad\qquad \delta_j = \sum\limits_{k=0}^{j} \rho_k$
\end{center}

A random real number $x$ in the range $[0..1]$ is then generated. If $x \leq \delta_0$, we select $c_0$ into the population; otherwise if $x$ falls in the range $(\delta_{j-1}, \delta_{j}]$, we select $c_j$ into the population. This selection step is repeated $p$ times to generate the next population. The selections are independent, and some chromosomes that have a greater PMF may get selected multiple times. Thus, the chance of choosing a chromosome is proportional to its fitness value, and hence inversely proportional to the end-to-end latency of that solution which we want to minimize~\cite{Steinbrunn1997}. 

After a new population has been generated, we apply two recombination operators to further its evolution: \emph{crossover} and \emph{mutation}. Crossover picks each chromosome into the crossover set with a probability $\chi$, thus giving a crossover set size of $p \times \chi$. Chromosomes in this set are randomly paired to form ``couples'' for crossover; if the crossover set has an odd number of chromosomes, the last added chromosome is dropped. During crossover between a pair of chromosomes $c_i = (q_0, q_1, ...,q_{m}, q_{m+1}, ..., q_{n-1})$ and $c_j = (q'_0, q'_1, ...,q'_{m}, q'_{m+1}, ..., q'_{n-1})$, a random crossover point $m$ is selected in the range $[0..n-1]$. Then, the crossover results in the new chromosomes $\widehat{c_i} = (q'_0, q'_1, ...,q'_{m}, q_{m+1}, ..., q_{n-1})$ and $\widehat{c_j} = (q_0, q_1, ...,q_{m}, q'_{m+1}, ..., q'_{n-1})$. 
The mutation operation helps jump (out of local minimas) to regions of the solution space which may not have been searched before. The probability for mutation $\mu$ decides whether a query $q_i$ in a chromosome will change its resource placement value or not, and if it mutates, the new value for the query becomes a random integer in the range $[0..(|\mathbb{R}|-1)]$. We expect $\mu \times (n \times p)$ number of queries to change their resource mapping values in each population. 

For every generation, we repeat the steps: roulette wheel population selection from the previous generation's population; crossover to generate new chromosomes; mutation of these chromosomes; and potential update of the best-fit chromosome based on the fitness values for chromosomes in this population. These operations are repeated for $g$ generations, which may be a constant or based on a convergence function.

Our optimization problem requires us to enforce constraints. There are three approaches to doing this -- remove solutions which violate constraints in each generation, give a penalty to the fitness value of the violating chromosome, or have an encoder-decoder scheme so that invalid solutions do not occur in the first place~\cite{Michalewicz96}. The first approach can constrain the search space of GA and cause the population to die out, while the third approach is difficult, sometimes impossible, to formulate cleanly and also increases the time complexity. Instead, we use a \emph{high-valued penalty function} on invalid solutions to reduce the chance that they will be selected and will cause them to eventually be pushed out. It also holds the possibility that the generation having such invalid solutions can still evolve to a valid one. 

We add a penalty value of $log_2(1+\gamma\times F(j))$ to each chromosome $c_j$, for each constraint that is violated, where $\gamma=1.5$. As the fitness value is large (since the end-to-end latency is subtracted from a large positive constant), we use $log_2$ to subdue the effect of the constant and also make the penalty function non-linear. The penalty function is cumulative -- when a chromosome violates both Constraints~\ref{cons:2} and~\ref{cons:3}, the penalty doubles. We ensure that the constraint penalty is large enough to even accommodate viable placements that exhibit a \emph{ping-pong-ping} effect between edge and Cloud, i.e., 4 sequential tasks are placed in Edge-Cloud-Edge-Cloud, causing 3 network trips. 

\subsubsection{Complexity Analysis for GA}
\label{sec:approach:ga:cplx}

In each generation of the GA, we need to find the objective value for all the chromosomes present in the population. This means evaluating the critical path for the DAG based on each mapping solution (chromosome) present in the population. Since we have $g$ generations, a population size of $p$ chromosomes in each generation, and the time to find the longest path in the DAG is $\mathcal{O}(|\mathbb{V}|+|\mathbb{E}|)$ for each candidate solution, the asymptotic time complexity of the GA approach is $\mathcal{O}(g \times p \times (|\mathbb{V}|+|\mathbb{E}|))$.

\section{Micro-benchmarks on Resource Usage for Event Analytics}
\label{sec:bm}

We perform a series of micro-benchmark experiments to measure and build a distribution of the \emph{latency}, $\lambda_i^k$, for a query $v_i$ running on resource $r_k$ to process an event, and its \emph{incremental energy consumption}, $\epsilon_{i}^{k}$. 
We also measure and construct distributions for \emph{network latency} $l_{m,n}$ and \emph{bandwidth} $\beta_{m,n}$  between pairs of resources $r_m, r_n$. These offer real-world characteristics of the compute capacity used on the edge and Cloud resources, their local area network characteristics, and the energy usage of edge devices for event analytics. These empirical distributions are useful in themselves for IoT deployment studies, and also inform the design and execution of our simulation study to evaluate the proposed query placement solutions, presented in the next section. 

\subsection{Experimental Setup}
We run experiments with different CEP query configurations, on edge and Cloud resources. We use the popular Raspberry Pi 2 Model B v1.1 as our edge device, and a Standard D2 VM in Microsoft Azure's Southeast Asia data center as our Infrastructure-as-a-Service public Cloud. The Pi has a 900MHz quad-core ARM Cortex-A7 CPU and 1GB RAM, while the Azure D2 VM has a 2.2Ghz dual-core (4 hyper-threads) Intel Xeon E5-2660 CPU and 7GB RAM. The Azure VM has as many hyper-threads as the Pi's cores, and is rated at about twice the clockspeed. It also has seven times the physical memory. Both run Linux OS distributions.


We use \emph{WSO$_{2}$'s Siddhi} as our CEP engine~\cite{siddhi11} on both Pi and Azure. Siddhi is written in Java, open sourced, and used for IoT applications~\footnote{http://wso2.com/library/articles/2014/12/article-geo-fencing-for-iot-with-wso2-cep/}. Queries are written in Siddhi's APIs and compiled to executable JARs that are run on the resources. Pi runs it in Oracle JDK SE 1.8 for ARM and Azure uses OpenJDK SE 1.7. 

We generate input event streams with synthetic integer values to represent a sensor's observation stream. Events are pre-fetched into memory, and replayed as 4-byte integers to Siddhi. 
The event values are generated to meet the selectivity needed for a query, as discussed later in Table~\ref{tbl:queries}. 
Output patterns matched for a query are returned through a callback. Counters maintained at the input and output streams measure the event rate per second, and are used to find the latency and throughput for each query.

\begin{table}[t]
\centering
\caption{Summary of query configurations used in micro-benchmarks\label{tbl:query-config}
}
\resizebox{!}{0.27\textwidth}
{
\begin{tabular}{l r l r r r r}
	\hline
	{\bf \specialcell[t]{Query ID}} & {\bf \specialcell[t]{Selectivity\\($\sigma$)}} & {\bf \specialcell[t]{Input Event generation\\for required selectivity}} & {\bf \specialcell[t]{Pattern/\\Window\\ length}} & {\bf \specialcell[t]{Peak Rate\\(Pi)\\{[}e/sec{]}}} & {\bf \specialcell[t]{Peak Rate\\(Azure)\\{[}e/sec{]}}} & {\bf \specialcell[t]{Energy\\used by Pi\\{[}mA{]}}}
	\\ \hline
	\hline
	\textbf{Fil 1.0} & {1.0} & {Random integer $<$ 150} & {-} & {114,334} & {337,357} & {337.04}
	\\ \hline
	\textbf{Fil 0.5} & {0.5} & {Random integer [0-299]} & {-} & {152,026} & {401,454} & {336.91}
	\\ \hline
	\textbf{Fil 0.0} & {0.0} & {Random integer $>=$ 150} & {-} & {253,766} & {514,599} & {337.41}
	\\ \hline
	\textbf{Seq3 1.0} & {1.0} & {Equal integers (10,10,10,...)} & {3} & {37,790} & {248,153} & {340.91}
	\\ \hline
	\textbf{Seq3 0.5} & {0.5} & {\specialcell[c]{5 equal integers followed by a \\different integer (3,3,3,3,3,10,...)}} & {3} & {47,042} & {297,712} & {342.45}
	\\ \hline
	\textbf{Seq3 0.0} & {0.0} & {Unequal integers (3,7,9,...)} & {3} & {67,101} & {375,508} & {342.62}
	\\ \hline
	\textbf{Seq5 1.0} & {1.0} & {Equal integers (10,10,10,10,10,...)} & {5} & {27,499} & {210,138} & {341.27}
	\\ \hline
	\textbf{Seq5 0.5} & {0.5} & {\specialcell[c]{9 equal integers followed by a \\different integer (3,3,3,3,3,3,3,3,3,12,...)}} & {5} & {34,247} & {250,769} & {342.42}
	\\ \hline
	\textbf{Seq5 0.0} & {0.0} & {Unequal integers (3,4,7,8,9,...)} & {5} & {53,475} & {331,334} & {344.44}
	\\ \hline
	\textbf{Pat3 1.0} & {1.0} & {Equal integers (10,10,10,...)} & {3} & {37,816} & {245,899} & {340.88}
	\\ \hline
	\textbf{Pat3 0.5} & {0.5} & {\specialcell[c]{Sequence of 3 equal and \\ 3 random integers (3,4,3,5,3,100,...)}} & {3} & {151} & {634} & {351.32}
	\\ \hline 
	\textbf{Pat3 0.0} & {0.0} & {Random integers} & {3} & {103} & {462} & {343.75}
	\\ \hline
	\textbf{Pat5 1.0} & {1.0} & {Equal integers (10,10,10,10,10,...)} & {5} & {27,692} & {210,960} & {352.33}
	\\ \hline
	\textbf{Pat5 0.5} & {0.5} & {\specialcell[c]{Sequence of 5 equal and  5 random \\integers (3,4,3,5,3,6,3,10,3,11,...)}} & {5} & {151} & {631} & {351.59}
	\\ \hline
	\textbf{Pat5 0.0} & {0.0} & {Random integers} & {5} & {104} & {459} & {352.45}
	\\ \hline
	\textbf{Agg B 60} & {$1/60$} & {Random integers} & {60} & {128,053} & {331,670} & {393.68}
	\\ \hline
	\textbf{Agg B 600} & {$1/600$} & {Random integers} & {600} & {129,529} & {333,295} & {396.55}
	\\ \hline
	\textbf{Agg B 6000} & {$1/6000$} & {Random integers} & {6,000} & {122,558} & {327,424} & {387.84}
	\\ \hline
	\textbf{Agg S 60} & {1.0} & {Random integers} & {60} & {63,221} & {241,126} & {393.92}
	\\ \hline
	\textbf{Agg S 600} & {1.0} & {Random integers} & {600} & {62,096} & {239,917} & {393.72}
	\\ \hline
	\textbf{Agg S 6000} & {1.0} & {Random integers} & {6,000} & {59,175} & {238,720} & {393.41}
	\\ \hline
\end{tabular}}
\end{table}


We design the \emph{query benchmark} by configuring the selectivity and length of patterns matched or aggregated for the four major query types, to give 21 different queries as summarized in Table~\ref{tbl:query-config}. 
We consider 3 configurations for \emph{filter} queries with different selectivities: $\sigma=0.0$ which does not match any input events, $\sigma=0.5$ which matches about half the input events, and $\sigma=1.0$ which matches all input events. 
For \emph{sequence} queries, we consider two queries with sequence lengths of 3 and 5, and within each have selectivities of $0.0, 0.5$ and $1.0$. \emph{Pattern} queries of lengths of 3 and 5 are considered, with three selectivities each.
\emph{Aggregate} queries are designed with window widths of $60, 600$ and $6000$, emulating different temporal sampling frequencies for sensors. We include both \emph{sliding} and \emph{batching} window variants. 

These 21 queries are used to benchmark several common performance parameters used to solve the optimization problem.  
We measure the \textbf{peak throughput rate} of a query on the Pi and VM by replaying input events through Siddhi queries without pause. The compute capability of the resource decides the maximum input rate that is sustained. Since Siddhi is used in a single thread, the inverse of the peak throughput for a query $v_i$ on a resource $r_k$ gives the expected latency per event, $\lambda_i^k$. 

We measure the \textbf{energy usage} of the Pi as the current drawn (milli-Ampere, mA), measured using a high precision multimeter which samples 4~values per second. The energy usage is measured under a \emph{base-load}, with no queries running, and a \emph{load} condition for each query. We also measure the energy use for \emph{input event rates} of $100~e/sec$, $1000~e/sec$, and $10,000~e/sec$ per query, besides the peak rate, since in the real world, the input rate to a query may be lower than the peak supported rate. 
Each query runs for $5~mins$ thus giving $1,200$ samples of current drawn for our distribution. The energy usage for Azure is not relevant here as it is not a constraint. 


We measure \textbf{network latency} using the \texttt{nping} command that sends a 40~Bytes TCP packet from a source to a destination machine, and the destination responds with a 44~Bytes TCP packet. This gives the round trip time (rtt) for the packet, and the network latency in a single direction as $\frac{1}{2}\times rtt$. 
\textbf{Network bandwidth} is measured using an \texttt{iperf} server on a destination machine, to which the source machine connects and downloads data for $1~min$ ($\approx 150-600~MB$). 
We deploy 4 Raspberry Pis at different network locations in our campus along with 4 Azure VMs. Each pair measures their inter-device latency and bandwidth using \texttt{nping} ($\approx 1/min$) and \texttt{iperf} ($\approx 10/mins$) for a 24~hour duration
, while ensuring that no two pairs overlap. This gives $10,681$ latency and $894$ bandwidth samples between the edge pairs within campus, and similarly $5,240$ and $775$ samples from campus to Cloud.


\subsection{Observations and Analysis}
\label{sec:bm:results}

\begin{figure*}[t]
	\centering
	\subfloat[Peak query rate on Pi \& Azure]{
		\includegraphics[width=0.24\textwidth]{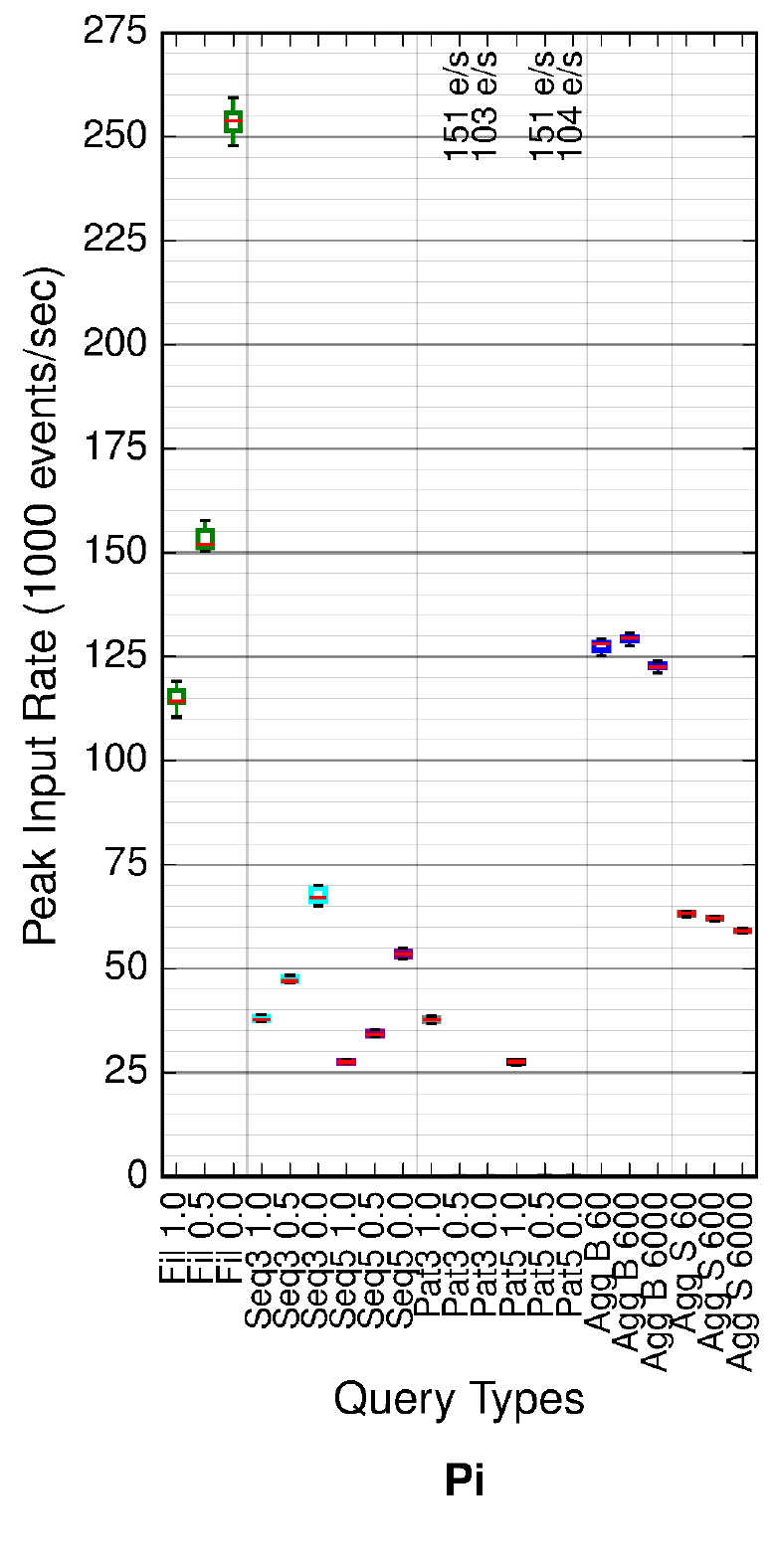}\enskip
		\includegraphics[width=0.24\textwidth]{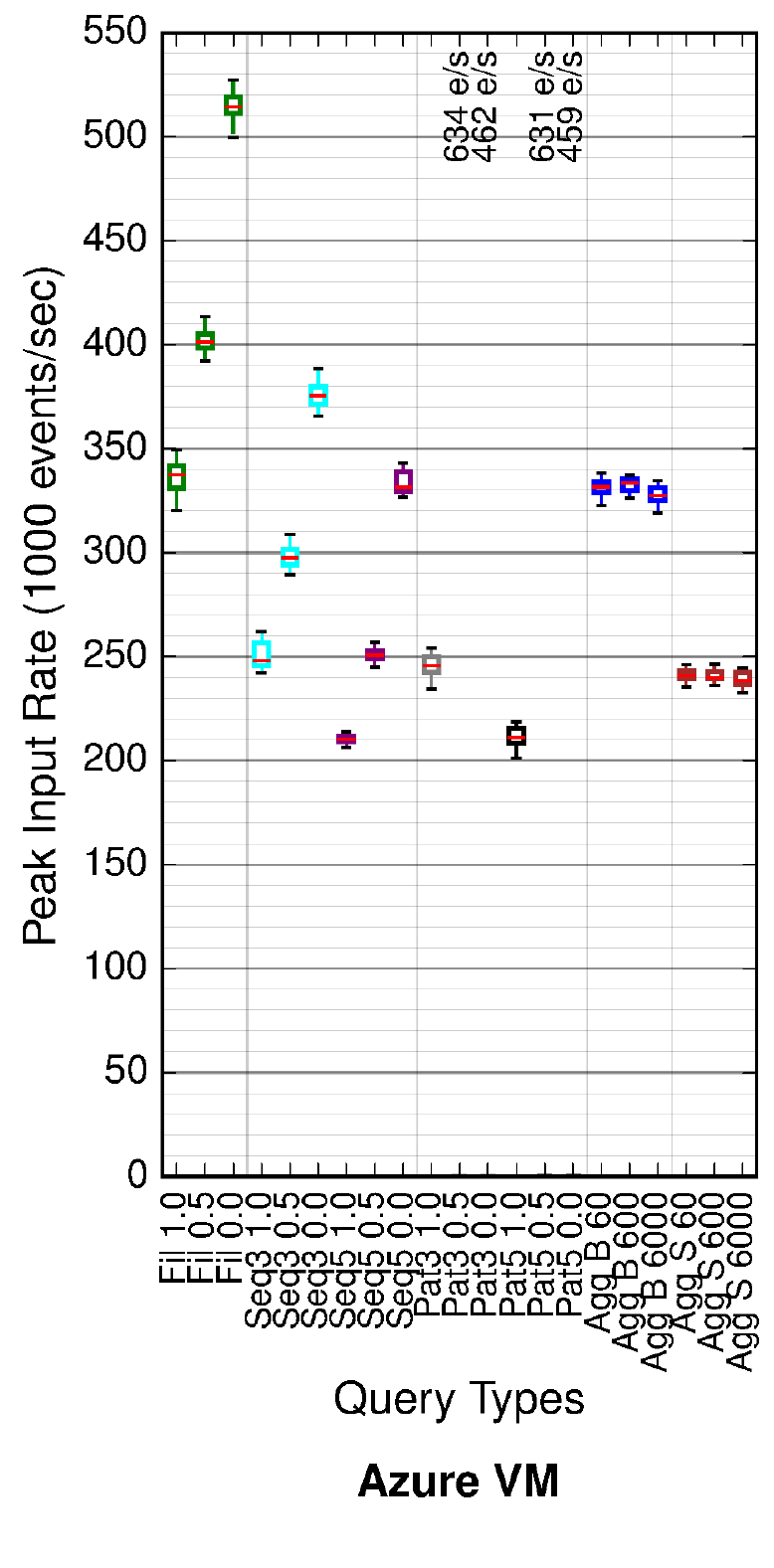}
		\label{fig:throughput}
	}\qquad
	\subfloat[Energy at peak \& variable rate]{
		\includegraphics[width=0.43\textwidth]{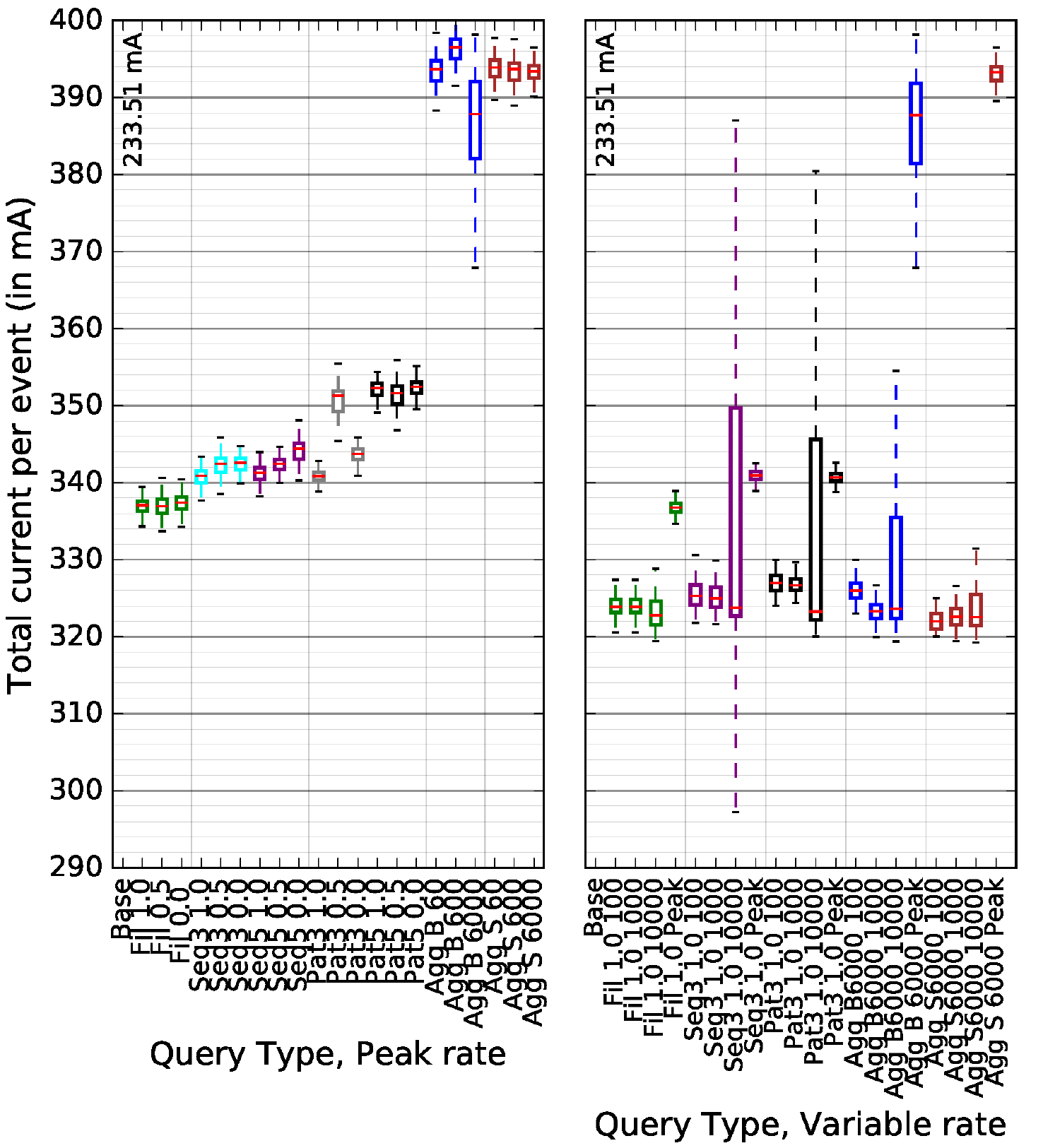}
		\label{fig:energy:peak-diff}
	}\\
	\subfloat[Latency]{
		\includegraphics[width=0.18\textwidth]{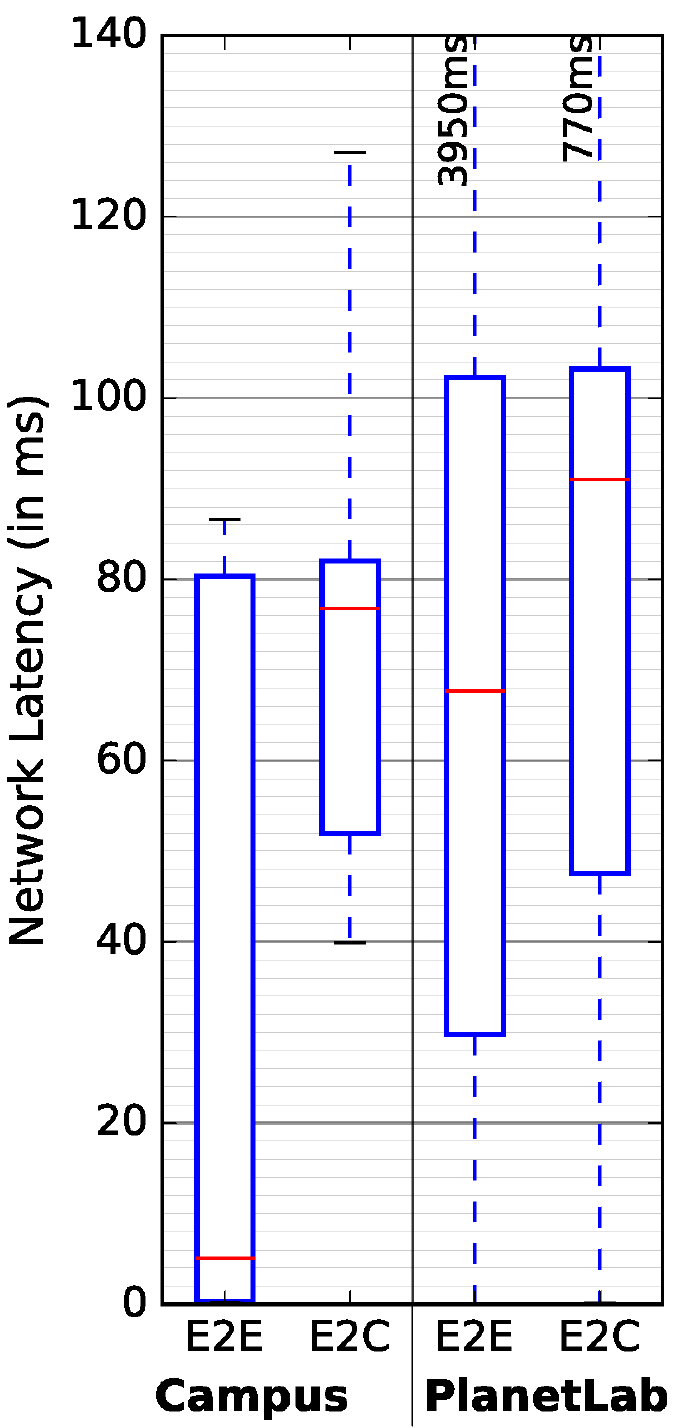}
		\label{fig:network:new:latency}
	}\enskip
	\subfloat[Bandwidth]{
		\includegraphics[width=0.18\textwidth]{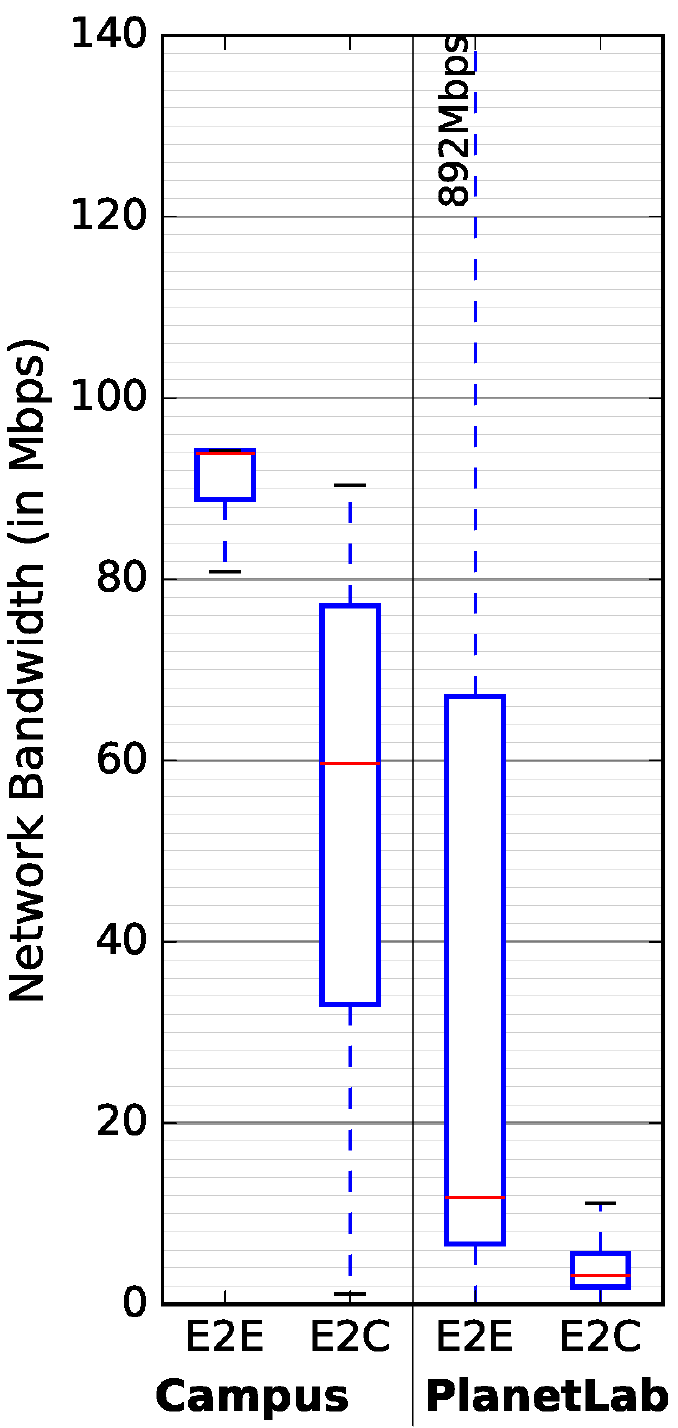}
		\label{fig:network:new:bandwidth}
	}
	\caption{\emph{Peak rate (a)}, \emph{Pi Energy usage (b)} and \emph{Network Profile (c,d)} from benchmarks}
	\label{fig:energy}
\end{figure*}


The peak input rates that can be sustained for different CEP queries on Pi and Azure are shown in Fig.~\ref{fig:energy} as Box and Whiskers plots of the distributions. In general, the filter and batch aggregate queries support a higher peak rate than the other query types, staying above $110,000~e/sec$ and $320,000~e/sec$ for Pi and Azure, respectively. The Pi is about $3\times$ slower than Azure in processing equivalent queries, which is understandable given their different CPU architectures and clock speeds ($900$~MHz vs. $2.2$~GHz). 

The peak rates of filter and sequence queries are inversely correlated with their selectivity. For example, \texttt{Seq5 1.0} sequence query of length $5$ with $\sigma=1.0$ supports a median rate of $27,499~e/sec$ on Pi (Fig.~\ref{fig:throughput}) while \texttt{Seq5 0.0}, which has the same length but lower selectivity at $\sigma=0.0$, supports a higher rate of $53,475~e/sec$. A similar trend is seen for the Azure VM too in Fig.~\ref{fig:throughput}. As the selectivity reduces, fewer output event objects have to be generated, which allows more input events to be processed. 

However, this does not hold for pattern queries and the peak rate actually decreases sharply as the selectivity decreases, from $27,692~e/sec$ for \texttt{Pat5 1.0} on the Pi to barely $104~e/sec$ for \texttt{Pat5 0.0}. Pattern queries allow other events to occur between successive matching events. So as fewer events match, Siddhi's state machine maintains more partially matched states, and every future event checks with each state to see if the full pattern matches. This increases the resource usage and lowers the throughput. 

We also observe that for sequence and pattern queries, the peak rate decreases with an increase in the pattern length, for example, on the Pi reducing from a median $37,790~e/sec$ for \texttt{Seq3 1.0}, which matches 3 events, to $27,499~e/sec$ for \texttt{Seq5 1.0}, which matches 5 events. This is understandable -- as the match size increases, more in-memory states have to be maintained to match subsequent events against. 

In case of aggregate queries, batch windows support a higher peak rate than sliding windows since the latter process many more windows than the former, and their selectivities are also very different. For example, the batch aggregate query rate is double that of the sliding aggregate query on the Pi, though it is less pronounced on Azure. However, the peak rate supported does not change as we increase the window width, for both batch and sliding windows. Even though the window size grows, the number of aggregation operations performed remain almost the same.
%
%
All these distributions are tight, ensuring reproducibility, and their relative trends are consistent for both Pi and Azure assuring us that these are characteristics of the query and not a device artifact. 


We report the \emph{total energy used by the Pi} for each input event for different queries at the \emph{peak rate}, given as current drawn (in mA) in Figure~\ref{fig:energy:peak-diff}. Subtracting the \emph{base load} current drawn by a freshly booted Pi, $\approx233$~mA, gives the incremental current for the query per event. Multiplying this value by the duration for which the query runs helps match the Constraint~\ref{cons:3} of battery capacity for the Pi, given in mA$\cdot$h. 
In general, we do not see a significant difference in the energy used by filter, sequence and pattern queries, largely falling between $336-353$~mA. Different aggregate queries have comparable energy levels as well, with their boxes between $382-398$~mA, though they are higher than filter due to the floating-point operations required for aggregation. 

The distribution of the total energy use for query types with $\sigma=1.0$ but for \emph{different input event rates} of $100~e/sec$, $1000~e/sec$, $10,000~e/sec$ and their peak rate is shown in Figure~\ref{fig:energy:peak-diff}. We see that the median current drawn by the Pi is relatively the same for all rates, other than the peak, which has a higher consumption. For example, \texttt{Fil 1.0 100}, \texttt{Fil 1.0 1000} and \texttt{Fil 1.0 10000}, which are the filter query with $\sigma=1.0$ and at rates of $100~e/sec$, $1000~e/sec$, and $10,000~e/sec$, all draw between {$322-324$~mA} while at the peak rate, the same query \texttt{Fil 1.0 Peak} uses a higher {$338$~mA}. However, we do see that for input rates of $10,000~e/sec$, several queries have a wider distribution, displaying the transient energy usage behaviour for the Pi under certain cases.



The \emph{network latency} between the Pi's on campus is variable (Fig.~\ref{fig:network:new:latency}, left), in the range $0.31-80~ms$, though the median is lower at $5.08~ms$. This is because the Pi's are at different segments in the campus topology, with $4$ hops within the private network. The Pi to Azure latency is in general higher at $51.93-82.06~ms$, with a median of $76.77~ms$. 
This reflects the latency from Bangalore, India to the Microsoft data center in Singapore where the VMs are placed, across the public Internet. 
The \emph{network bandwidth} between the Pi's on campus is $88-94~Mbps$ (Fig.~\ref{fig:network:new:bandwidth}, left) -- close to the network interface limit of $100~Mbps$ for the Pi's. These values do not exhibit the variability seen in the latency. In comparison, the bandwidth between the Pi's and the Azure VMs is lower, at a median $60~Mbps$, and varies widely by $\pm 20~Mbps$ due to higher congestion in the public Internet relative to the private campus network.

Constraint~\ref{cons:2} requires the empirical evaluation of the \emph{parallelism overhead} $\pi_m$. We run $m=2$ to $m=10$ queries concurrently in Siddhi on the Pi and the Azure VM, and find a linear fit of the parallelism overhead for $m$ queries relative to a single query performance, given as a \% penalty on the input rate supported. We have:
\begin{center}
	$\pi_m = -1.12 \times (m-1) - 5.68,~~\forall r_k = Pi \quad\text{and}\quad \pi_m = -0.35 \times (m-1) - 3.80,~\forall r_k = Azure$
\end{center}

\section{Simulation Study of Distributed Query Placement}
\label{sec:sim}
Our problem formalization and solution are generalizable to any edge or Cloud resource, network, and DAG. However, since it is not possible to formally bound the quality of GA solutions~\cite{ga-quality1,ga-quality2}, we instead perform detailed simulations to evaluate the placement of CEP queries across the edge and Cloud resources as provided by the proposed algorithms. These comprehensive simulations use realistic query, DAG, resource and network distributions, sourced from our micro-benchmarks and from public datasets. We ensure that every variable present in the optimization problem is varied. We evaluate the relative quality and speed of solutions given by the brute-force and our GA meta-heuristic approaches.

\subsection{Experimental Setup}
\subsubsection{DAG Generation and Static Characteristics}
Our evaluation considers a broad collection of synthetically generated DAGs composed out of the CEP queries introduced and benchmarked above. We use the \emph{Random Task and Resource Graph (RTRG) tool}~\cite{RTRG}, developed for embedded systems research, to generate dataflows with different numbers of CEP queries (vertices). We generate DAGs with a maximum vertex out-degree of $1$--$5$ edges. We then map the benchmarked CEP queries onto each vertex in the DAG. We first randomly set a vertex to one of the query types -- filter, sequence, pattern, batch aggregate or sliding aggregate, with equal probability. Next, we uniformly select a variant of this query type from Table~\ref{tbl:query-config}, to give coverage to query type and selectivity in each DAG~\footnote{For these experiments, we consider all queries in Table~\ref{tbl:query-config} except \texttt{Pat3 0.5, Pat3 0.0, Pat5 0.5} and \texttt{Pat5 0.0}. These four have a sharply lower peak throughput rate compared to the other queries, and their inclusion makes it difficult to automatically generate synthetic DAGs having a feasible solution. This gives us $17$ queries in all as candidates to map onto vertices in the DAG.}. We generate DAGs with $4-50$ vertices, of which
$1$ or $4$ are source queries ($\mathbb{V}^{SRC}$) -- the latter only for larger DAGs. DAGs may consume multiple streams, and it impacts the downstream rate and the selectivity, and also varies the number of unconstrained queries available for placement.  

We avoid local effects of the random DAG generator tool by generating 3 DAGs for each configuration to give a total of $45$ DAGs (Table~\ref{tbl:dag}). We see a fair coverage of the different query types in each DAG. The selectivity $\sigma(v_i)$ of each query $v_i$ is used to generate the \emph{overall selectivity for the DAG} recursively. The DAGs' selectivities also have a wide range, from $\sigma=0.04 - 458.28$. This, when combined with the input rate to the DAG, determines the \emph{output rate of the DAG}~\footnote{In case of multiple sinks in the DAG, the output rate is determined as the sum of the output rates from all of them, and the selectivity of the DAG reflects this as well.}.
For a sample input rate of $1000~e/sec$, the DAGs' expected output rates range from $20-114,000~e/sec$~\footnote{DAGs with multiple source vertices divide the input rate evenly between the sources.}.
We identify as \emph{max query}, the one with the highest relative input rate in the DAG that may be a bottleneck, and tabulate its \emph{input selectivity} and \emph{peak input rate}. In some cases, like query \texttt{50\_4\_3}, the rate for the max query is much higher than the DAG's output rate.

\begin{table}
\centering
\caption{Configuration of DAGs used in simulation study. Counts of source and sink vertices, and different query types present in each DAG is listed. Selectivity and rates for the DAG and the query in the DAG with the maximum input rate are also shown.}  

\resizebox{!}{0.47\textwidth}{
\begin{tabular}{r | r r | r r r r r | r r | r r}
	\hline
	{\bf \specialcell[t]{DAG\\ID$^1$}}     & {\bf \specialcell[t]{Sources}} & {\bf \specialcell[t]{Sinks}} & {\bf Filter} & {\bf Seq.} & {\bf Pattern} & {\bf Agg B} & {\bf Agg S} &  {\bf \specialcell[t]{Max Qry\\I/P $\sigma$}}  &  {\bf \specialcell[t]{DAG $\sigma$}}  &  {\bf \specialcell[t]{Max Qry\\I/P Rate$^2$}}  &  {\bf \specialcell[t]{DAG \\O/P Rate$^2$}}  \\
	\hline
	\hline
	4\_1\_1  & 1                 & 1                  & 0      & 1        & 2       & 0     & 0     &   1.50         &   1.50               &   1,500    &   1,500    \\ \hline
	4\_1\_2  & 1                 & 2                  & 1      & 0        & 1       & 1     & 0     &   1.00         &   1.00               &   1,000    &   1,000    \\ \hline
	4\_1\_3  & 1                 & 1                  & 1      & 1        & 0       & 0     & 1     &   2.00         &   2.00               &   2,000    &   2,000    \\ \hline \hline
	6\_1\_1  & 1                 & 3                  & 1      & 2        & 1       & 1     & 0     &   6.00         &   6.00               &   6,000    &   6,000    \\ \hline
	6\_1\_2  & 1                 & 3                  & 1      & 2        & 1       & 1     & 0     &   2.50         &   0.06               &   2,500    &   60       \\ \hline
	6\_1\_3  & 1                 & 3                  & 1      & 1        & 3       & 0     & 0     &   9.50        &   6.00               &   9,500    &   6,000    \\ \hline  \hline
	8\_1\_1  & 1                 & 2                  & 1      & 3        & 2       & 0     & 1     &   2.00         &   2.00               &   2,000    &   2,000    \\ \hline
	8\_1\_2  & 1                 & 2                  & 0      & 3        & 2       & 1     & 1     &   12.0        &   12.0              &   12,010   &   12,010   \\ \hline
	8\_1\_3  & 1                 & 1                  & 0      & 1        & 2       & 3     & 1     &   1.00         &   1.00               &   1,000    &   1,000    \\ \hline \hline
	10\_1\_1 & 1                 & 2                  & 4      & 0        & 4       & 0     & 1     &   40.5        &   40.5              &   40,500   &   40,500   \\ \hline
	10\_1\_2 & 1                 & 1                  & 0      & 3        & 3       & 1     & 2     &   4.13         &   2.10               &   4,130    &   2,100    \\ \hline
	10\_1\_3 & 1                 & 2                  & 0      & 4        & 3       & 0     & 2     &   56.0        &   56.0              &   56,000   &   56,000   \\ \hline
	10\_4\_1 & 4                 & 3                  & 1      & 1        & 1       & 2     & 1     &   18.9       &   18.9             &   4,720    &   4,720    \\ \hline
	10\_4\_2 & 4                 & 2                  & 1      & 4        & 0       & 0     & 1     &   21.7        &   12.0              &   5,430    &   3,000    \\ \hline
	10\_4\_3 & 4                 & 1                  & 0      & 1        & 3       & 1     & 1     &   116       &   116             &   29,000   &   29,000   \\ \hline \hline
	12\_1\_1 & 1                 & 2                  & 4      & 3        & 3       & 0     & 1     &   41.6       &   41.6             &   41,600   &   41,600   \\ \hline	
	12\_1\_2 & 1                 & 2                  & 2      & 2        & 3       & 3     & 1     &   1.55       &   0.01             &   1,550   &   10   \\ \hline	
	12\_1\_3 & 1                 & 3                  & 2      & 0        & 5       & 2     & 2     &   29.0       &   0.01             &   29,000   &   10   \\ \hline	
	12\_4\_1 & 4                 & 1                  & 2      & 1        & 3       & 1     & 1     &   42.3       &   21.1             &   10,570   &   5,280   \\ \hline	
	12\_4\_2 & 4                 & 2                  & 1      & 5        & 0       & 0     & 2     &   49.0       &   14.5             &   12,250   &   3,630   \\ \hline	
	12\_4\_3 & 4                 & 2                  & 1      & 3        & 3       & 1     & 0     &   56.0       &   56.0             &   14,000   &   14,000   \\ \hline \hline	
	20\_1\_1 & 1                 & 2                  & 1      & 6        & 6       & 2     & 4     &   4.11         &   2.95               &   4,110    &   2,950    \\ \hline
	20\_1\_2 & 1                 & 3                  & 2      & 5        & 5       & 5     & 2     &   14.8        &   14.8              &   14,780   &   14,780   \\ \hline
	20\_1\_3 & 1                 & 2                  & 3      & 3        & 7       & 5     & 1     &   6.13         &   6.13               &   6,130    &   6,130    \\ \hline
	20\_4\_1 & 4                 & 2                  & 1      & 2        & 7       & 2     & 4     &   458       &   458             &   114,570  &   114,570  \\ \hline
	20\_4\_2 & 4                 & 2                  & 0      & 5        & 7       & 2     & 2     &   78.0        &   15.7              &   19,500   &   3,930    \\ \hline
	20\_4\_3 & 4                 & 1                  & 3      & 4        & 2       & 3     & 4     &   186       &   62.8              &   46,510   &   15,700   \\ \hline \hline
	30\_1\_1 & 1                 & 1                  & 1      & 6        & 10      & 7     & 5     &   84.3        &   1.68               &   84,260   &   1,680    \\ \hline
	30\_1\_2 & 1                 & 1                  & 3      & 11       & 10      & 3     & 2     &   7.55         &   7.55               &   7,550    &   7,550    \\ \hline
	30\_1\_3 & 1                 & 2                  & 4      & 10       & 8       & 2     & 5     &   2.00         &   0.72               &   2,000    &   720      \\ \hline
	30\_4\_1 & 4                 & 1                  & 3      & 5        & 8       & 6     & 4     &   40.7        &   0.20               &   10,180   &   50       \\ \hline
	30\_4\_2 & 4                 & 2                  & 2      & 11       & 4       & 5     & 4     &   16.0        &   0.80               &   4,000    &   20       \\ \hline
	30\_4\_3 & 4                 & 1                  & 4      & 8        & 8       & 4     & 2     &   155       &   16.0              &   38,760   &   4,160    \\ \hline \hline
	40\_1\_1 & 1                 & 2                  & 2      & 12       & 13      & 8     & 4     &   9.95        &   0.04               &   9,950    &   40       \\ \hline
	40\_1\_2 & 1                 & 1                  & 5      & 11       & 9       & 6     & 8     &   151       &   76.6              &   150,690  &   76,600   \\ \hline
	40\_1\_3 & 1                 & 1                  & 5      & 8        & 8       & 5     & 13    &   24.4        &   0.32               &   24,410   &   320      \\ \hline
	40\_4\_1 & 4                 & 2                  & 4      & 15       & 9       & 4     & 4     &   104       &   4.44               &   26,000   &   1,110    \\ \hline
	40\_4\_2 & 4                 & 2                  & 1      & 7        & 14      & 9     & 5     &   16.0        &   6.44               &   4,000    &   1,610    \\ \hline
	40\_4\_3 & 4                 & 1                  & 3      & 8        & 11      & 8     & 6     &   875       &   2.44               &   218,840  &   610      \\ \hline \hline
	50\_1\_1 & 1                 & 1                  & 10     & 14       & 10      & 4     & 11    &   122       &   72.9              &   121,580  &   72,950   \\ \hline
	50\_1\_2 & 1                 & 1                  & 6      & 15       & 15      & 4     & 9     &   64.7        &   64.7              &   64,670   &   64,670   \\ \hline
	50\_1\_3 & 1                 & 2                  & 7      & 17       & 12      & 4     & 9     &   6.00         &   1.03               &   6,000    &   1,030    \\ \hline
	50\_4\_1 & 4                 & 2                  & 3      & 14       & 19      & 5     & 5     &   48.5        &   0.12               &   12,120   &   30       \\ \hline
	50\_4\_2 & 4                 & 3                  & 3      & 11       & 15      & 12    & 5     &   305       &   0.44               &   76,320   &   110      \\ \hline
	50\_4\_3 & 4                 & 2                  & 9      & 13       & 10      & 5     & 9     &   1,003     &   102             &   250,780  &   25,600   \\ \hline
	\multicolumn{12}{l}{\em $^1$~The $1^{st}$ number in the DAG ID is the number of vertices, $2^{nd}$ is the number of source vertices, and $3^{rd}$ is a count for the 3 versions.}\\
	\multicolumn{12}{l}{\em $^2$~Based on a DAG input rate of $1000~e/sec$}
\end{tabular}}
\label{tbl:dag}
\end{table}



\subsubsection{Dynamic Characteristics of DAGs and Network}\label{sec:sampling}
Besides the above static characteristics of the DAG, dynamic runtime characteristics of the edge, Cloud and network can vary in the real-world. For each DAG, we sample from real-world distributions the values of the \emph{latency} ($\lambda_i^k$) of each query $v_i$ running on each resource $r_k$; the \emph{network latency and bandwidth} ($l,\beta$) for each out-edge from a vertex (edge-edge, edge-Cloud); and the \emph{energy usage} ($\epsilon_i^k$) on the edge device for each query, at the input rate it is processing~\footnote{For simplicity, we assume that all edge devices have the same computing capacity, and similarly for the Cloud VMs. However, the analytical model does consider edge and Cloud resources of different capabilities.}.

Besides the network benchmarks from our campus Local Area Network (LAN), our simulations also consider public measurements from the widely-used \emph{PlanetLab} project~\cite{PlanetLab}, to capture the performance of edge devices in a Wide Area Network (WAN). Specifically, for \emph{edge to edge latency}, we use the dataset from~\cite{e2e1} that gives the pair-wise \emph{rtt} between $490$ PlanetLab nodes at global universities, over 
a $9$ day period. 
We extract $4,312,980$ valid measurements into a 
distribution. We source \emph{edge to edge bandwidths} from~\cite{e2e2}, which provides 
$2,448$ pair-wise measurements 
from $50$ random PlanetLab nodes. 
%
Lastly, ~\cite{wisc} offer detailed \emph{latency and bandwidth measurements between edge and Cloud}, for $80$ geographically distributed PlanetLab nodes and $40$ Amazon EC2 Cloud instances in $8$ regions.
This gives $867,489$ valid samples of the edge to Cloud latency, and $771,433$ data-points for their bandwidth distribution. 
Figs.~\ref{fig:network:new:latency}(right) and~\ref{fig:network:new:bandwidth}(right) show the latency and bandwidth distributions for these PlanetLab datasets that have been used in prior research, and are representative of the heterogeneity of network traffic in wide-area IoT deployments. As we see, their latency distributions are wider and higher, while their bandwidths are lower, relative to our campus private network.

We simulate the runtime variability of these parameters by \emph{sampling} from the box plot distributions 
in Figs.~\ref{fig:energy}. 
We first pick one of the two quartile ranges, $(Q1-Q2)$ or $(Q2-Q3)$, with equal chance. Then, with a uniform probability, we select a value from that inter-quartile range.
This technique is simple, reproducible from the box plots, and also captures the variability in the measured parameter values. 

For each vertex  $v_i$ in each DAG, we apply the sampling technique on these box plots to determine its runtime parameters: $\langle \lambda_i^{pi}, \lambda_i^{azure}, \epsilon_i^{azure} \rangle$, indicating the latencies for running this query on the Pi and Azure VM, and the energy consumed when running it on the Pi, respectively. Similarly, for each edge between vertices $v_i,v_j$ in each DAG, we sample and determine its network characteristics: $\langle l_{i,j}^{pi-pi}, l_{i,j}^{pi-azure}, \beta_{i,j}^{pi-pi}, \beta_{i,j}^{pi-azure} \rangle$, which are the latencies and bandwidths from edge to edge and edge to Cloud, respectively. Separate simulations are done for \emph{campus LAN} and \emph{PlanetLab WAN} setups.

\subsubsection{Input Rates to Generated DAGs}
Given the diversity in the structure and selectivities of the generated DAGs, we need to carefully determine meaningful input rates $\Omega^{in}$ to them such that a feasible solution is highly likely, and realistic.
We select two different static input stream rates for our study, $100~e/sec$ and $1000~e/sec$. These ensure that sufficient queries can run on the edge  without forcing all of them to the Cloud. At rates of $100~e/sec$ and $1000~e/sec$, $90\%$ and $95\%$ of all queries, respectively, present in the DAGs will receive an input rate smaller than the $Q1$ input rate supported for that query on the Pi, thus forcing no more than $5-10\%$ of queries to run in the Cloud.   

We also check if each DAG meets these two tests and otherwise regenerate them: 1)~A DAG should have no query whose effective input rate for a DAG input of $1000~e/sec$ is greater than the $Q3$ peak input rate for it on Azure, ensuring the VM can handle this query's throughput, 2)~We eliminate trivial DAGs whose effective selectivity is zero.
%
%





\subsubsection{Edge Resources in Deployment}
As the queries in a DAG increase, its resource needs will increase too. The sink queries need to run in the Cloud, so we are assured of having one Cloud VM available. However, each additional VM will have a monetary cost. So, we limit this study to use a single Azure VM, i.e., $|\mathbb{R}_C| =1$. On the other hand, an IoT deployment may have hundreds of gateway edge devices. So we consider 3 scenarios for the edge devices available for the query placement. In a \emph{liberal} setup, the number of edge devices plus the single Azure VM equals the number of queries ($|\mathbb{R}_E| = |\mathbb{V}|-1$). In a \emph{centrist} setup, the number of edge devices is one-half the number of queries in the DAG ($|\mathbb{R}_E| = \frac{|\mathbb{V}|}{2}$), while in a \emph{conservative} setup, the number of edge devices is one-quarter the number of queries in the DAG ($|\mathbb{R}_E| = \frac{|\mathbb{V}|}{4}$). 

We assume a $24~hour$ battery recharge cycle for the edge device, i.e., {$\tau_k=86,400~sec$} for all edges $r_k$. From Fig.~\ref{fig:energy:peak-diff}, we see that the mean of the median current draw by the Pi at the peak rate across all queries is $358~mA$. So, in $24~hours$, an active Pi would, on average, consume $\sim8,600~mAh$. We use this as the battery capacity parameter {$C_k$}.

\subsubsection{Brute Force and Genetic Algorithm Configuration}
Both BF and GA algorithms are implemented using \texttt{C++}. All experiments to solve the optimization problem are run on a server with an AMD Opteron 6376 CPU with 32~cores rated at 2.3GHz, having 128GB of RAM and running CentOS 7. We configure the GA with a population size  $p = 50$, crossover probability $\chi = 0.50$, and mutation probability $\mu = 0.15$. Rather than fix a static number of generations, $g$, we test if the solution has converged as follows. After running the GA for a minimum of $15,000$ generations to avoid local convergence effects, we check after each generation if the best fitness value has not changed in the last $50\%$ of generations. We set an upper bound of $1,000,000$ generations. 

\subsection{Observations and Analysis}
\label{sec:sim:results}

\subsubsection{End-to-end Latencies of the Solutions}
Here, we evaluate the effectiveness of the GA algorithm in offering a low-latency for the DAG and a feasible solution, and compare its qualitative performance with the BF and a baseline algorithm.

\newcommand{\myleftsize}{0.31}
\newcommand{\mymidsize}{0.275}
\newcommand{\myrightsize}{0.312}

\begin{figure*}[t!]
	\centering
	\subfloat{
		\includegraphics[width=\myleftsize\textwidth]{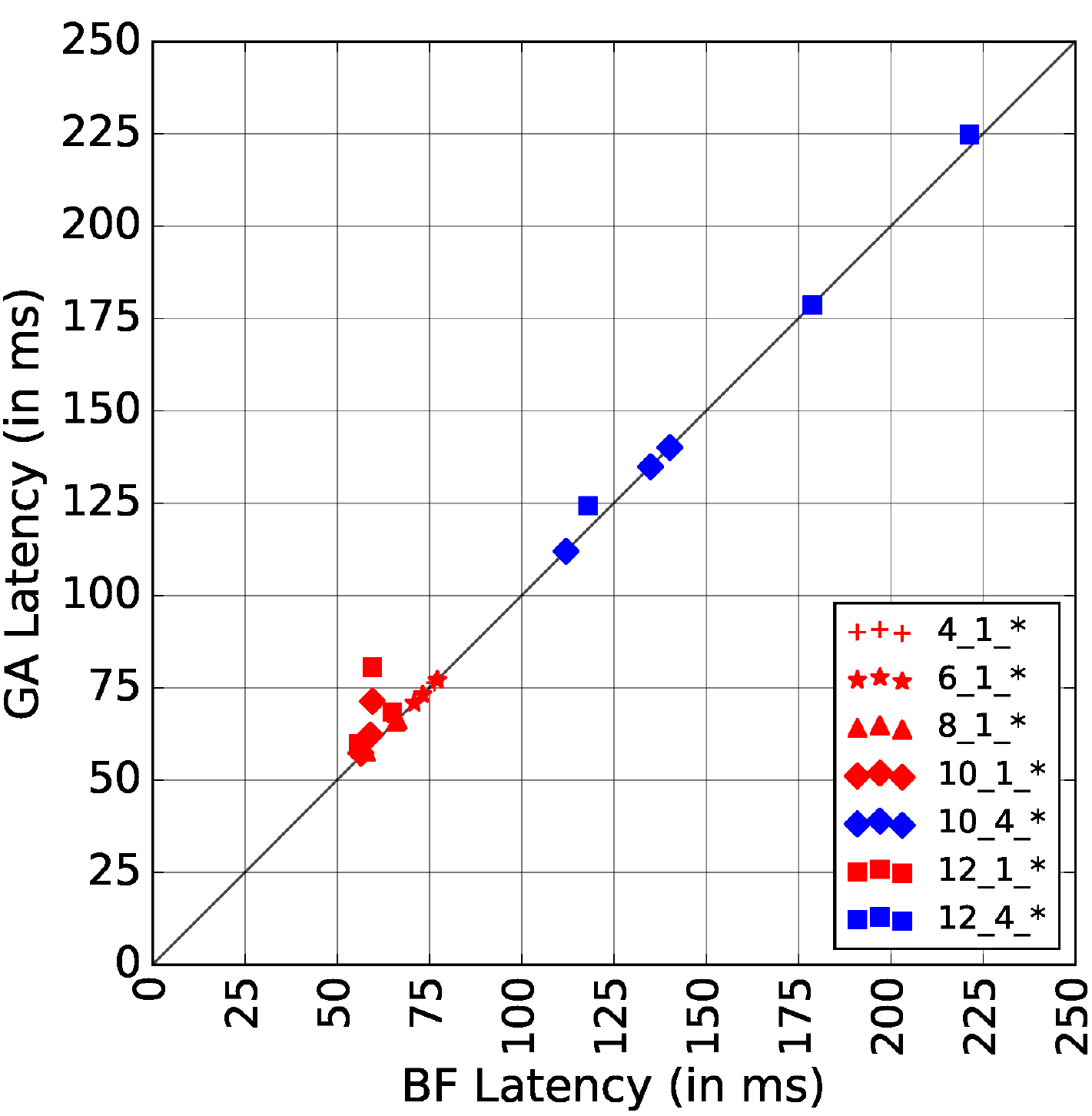}%
		\label{fig:GAvsBF_ObjVal:campus:100:N}
	}~
	\subfloat{
		\includegraphics[width=\mymidsize\textwidth]{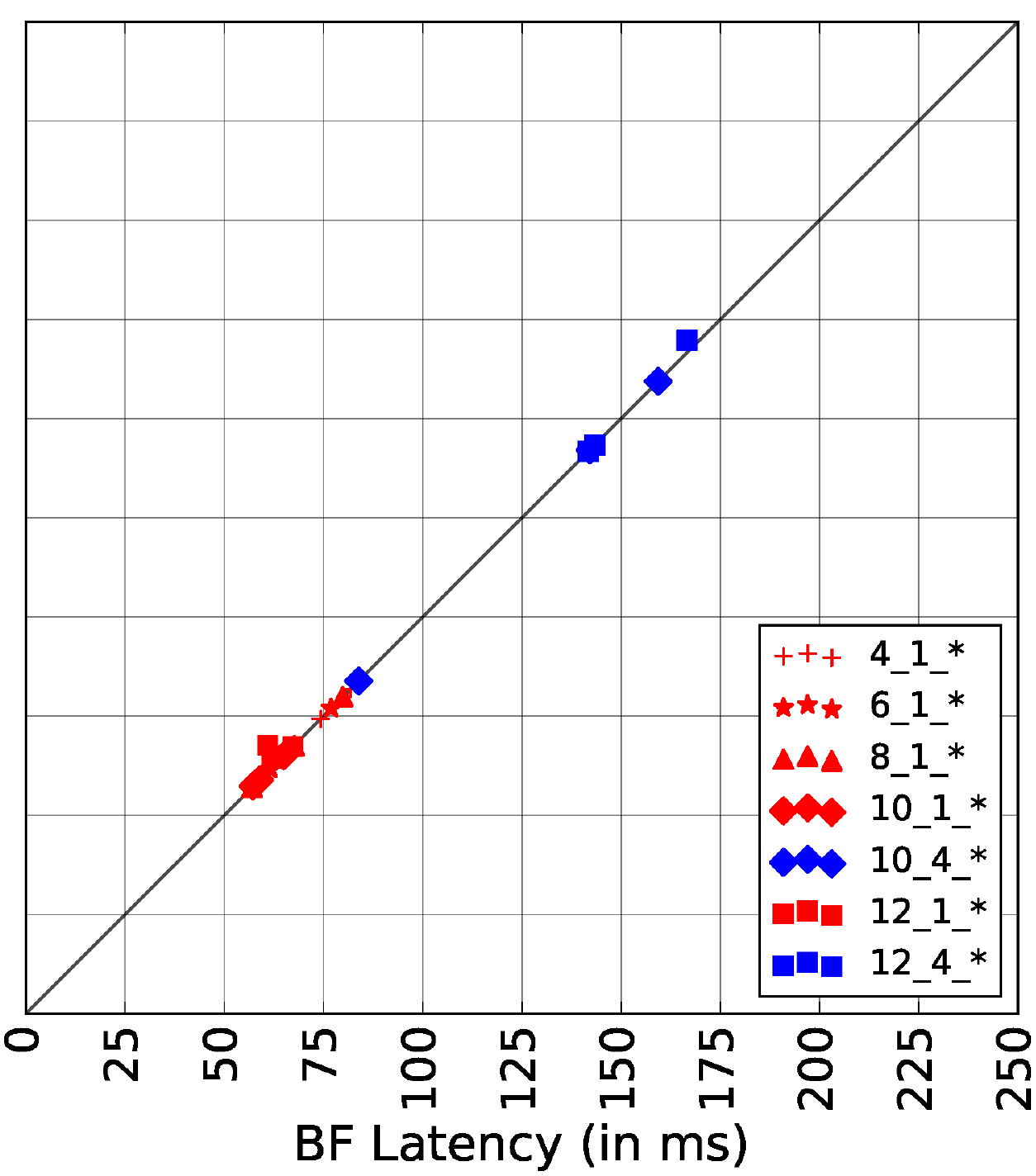}
		\label{fig:GAvsBF_ObjVal:campus:100:Nhalf}
	}~
	\subfloat{
		\includegraphics[width=\myrightsize\textwidth]{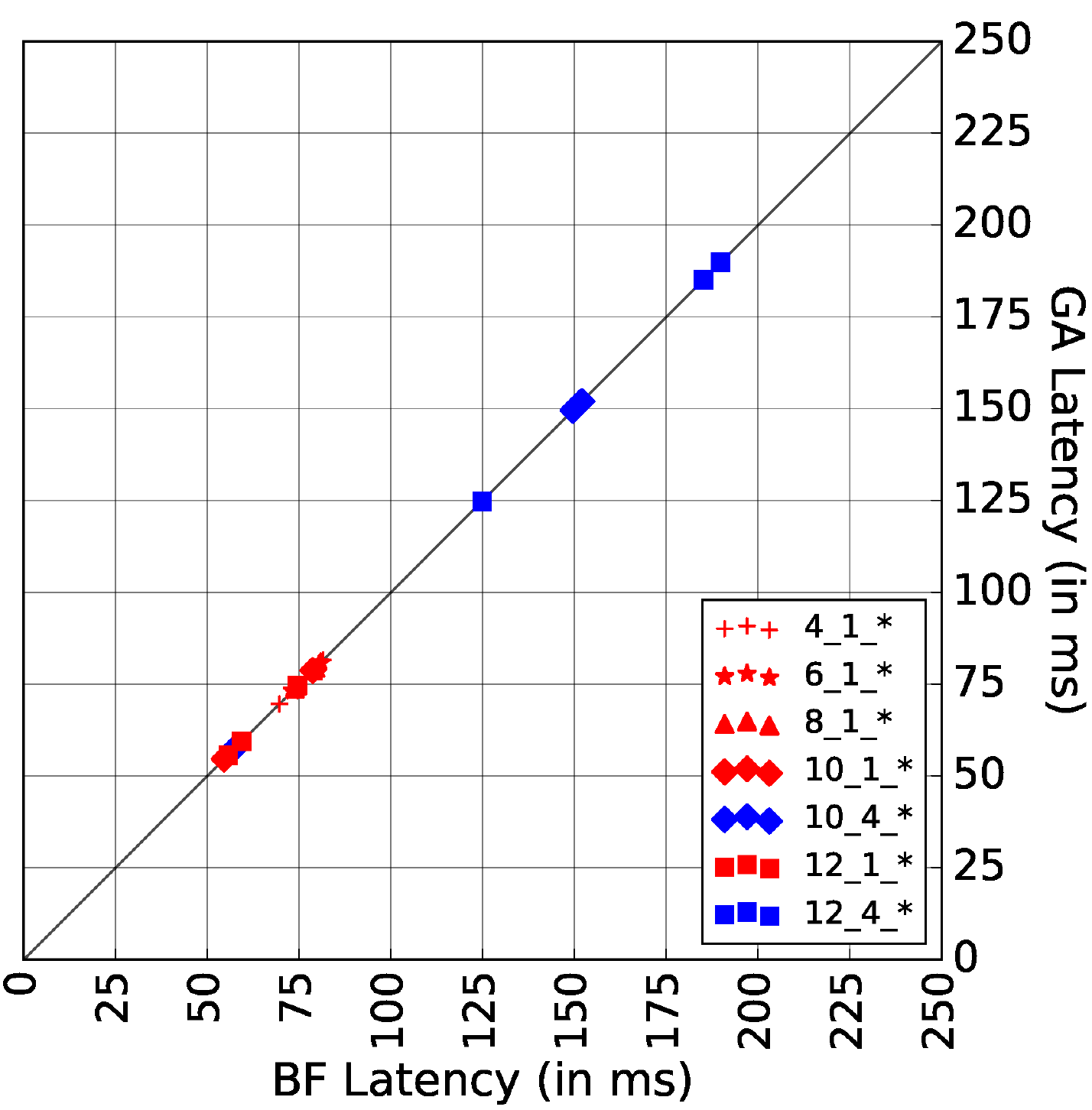}
		\label{fig:GAvsBF_ObjVal:campus:100:Nfourth}
	}\quad
	\subfloat{
		\footnotesize
		\begin{rotate}{270}\hspace{-1.45in}(i)~~$\Omega^{in} = 100~e/s$\end{rotate}
	}
	\\
	\addtocounter{subfigure}{-4}
	\subfloat[\emph{Liberal}]{
		\includegraphics[width=\myleftsize\textwidth]{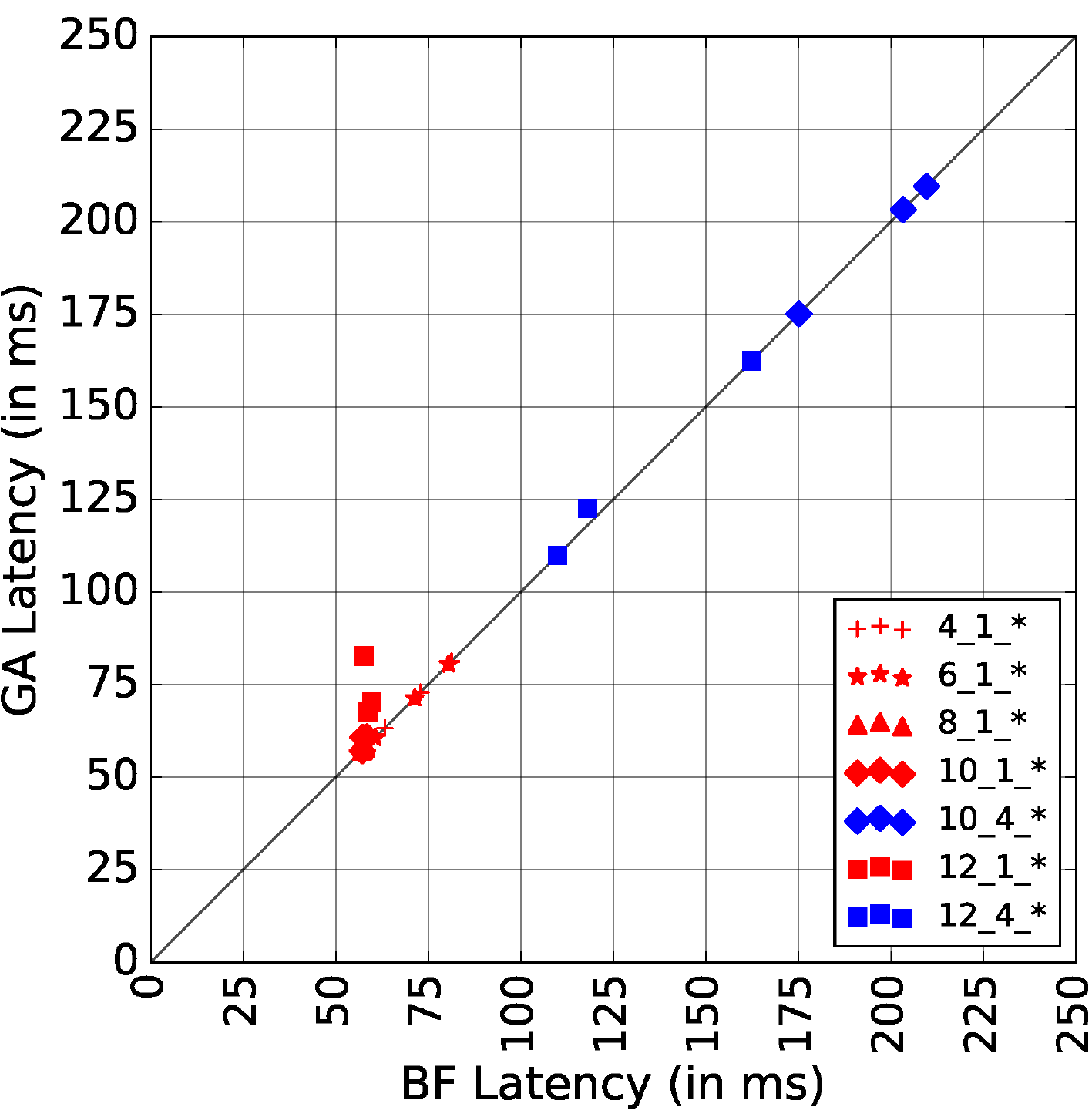}
		\label{fig:GAvsBF_ObjVal:campus:1000:N}
	}
	\subfloat[\emph{Centrist}]{
		\includegraphics[width=\mymidsize\textwidth]{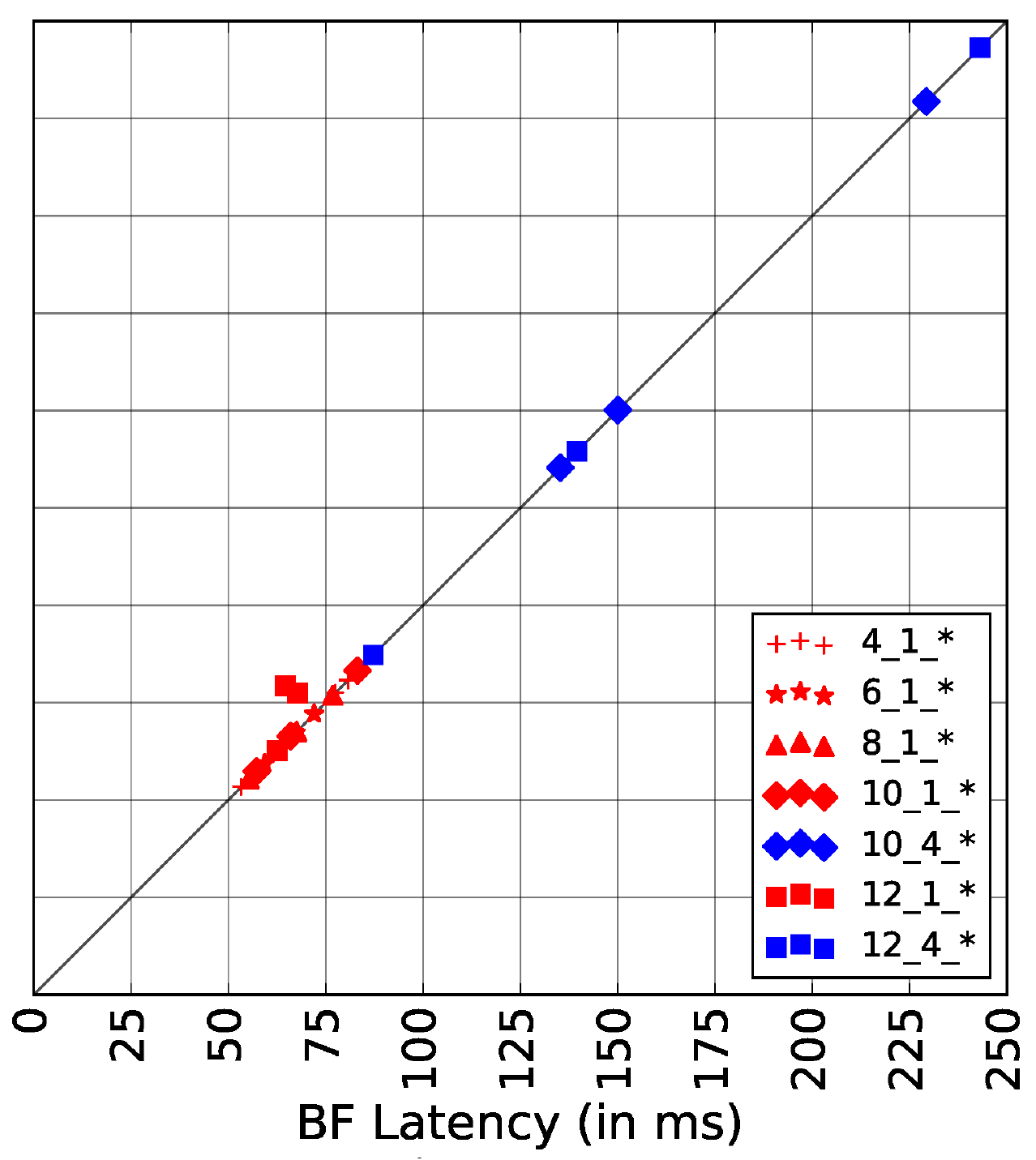}
		\label{fig:GAvsBF_ObjVal:campus:1000:Nhalf}
	}
	\subfloat[\emph{Conservative}]{
		\includegraphics[width=\myrightsize\textwidth]{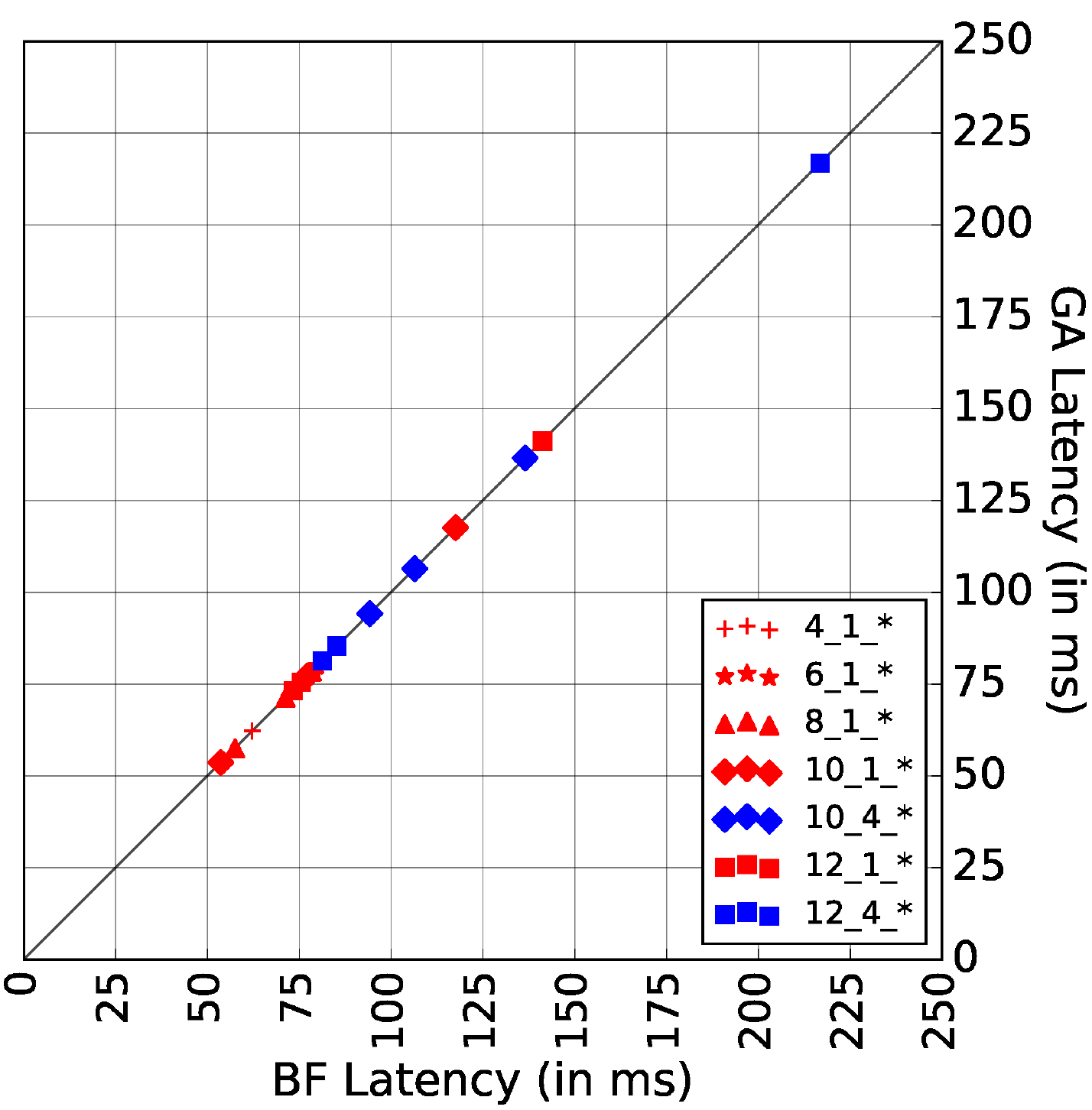}
		\label{fig:GAvsBF_ObjVal:campus:1000:Nfourth}
	}\quad
	\subfloat{
		\footnotesize
		\begin{rotate}{270}\hspace{-1.45in}(ii)~~$\Omega^{in} = 1000~e/s$\end{rotate}
	}
	\caption{Comparing \emph{end-to-end latency} of GA and BF using \emph{Campus LAN} for different resource setups and input rates. Each plot has $21$ DAGs with $4$--$12$ queries that BF could solve within $12~hrs$.}
	\label{fig:latency:ga-bf-new:campus}
\end{figure*}

\begin{figure*}[t!]
	\centering
	\subfloat{
		\includegraphics[width=\myleftsize\textwidth]{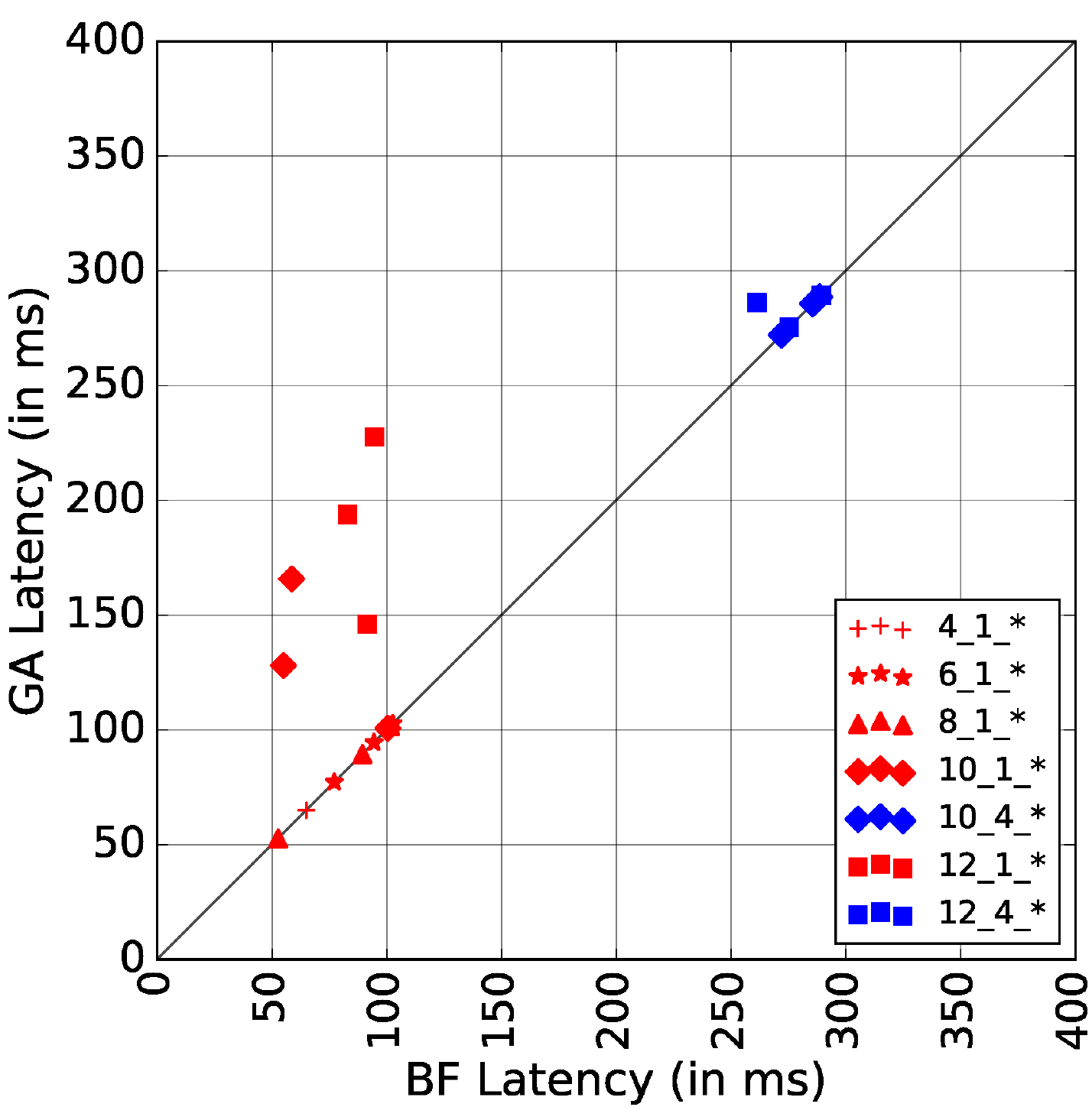}
		\label{fig:GAvsBF_ObjVal:planetlab:100:N}
	}
	\subfloat{
		\includegraphics[width=\mymidsize\textwidth]{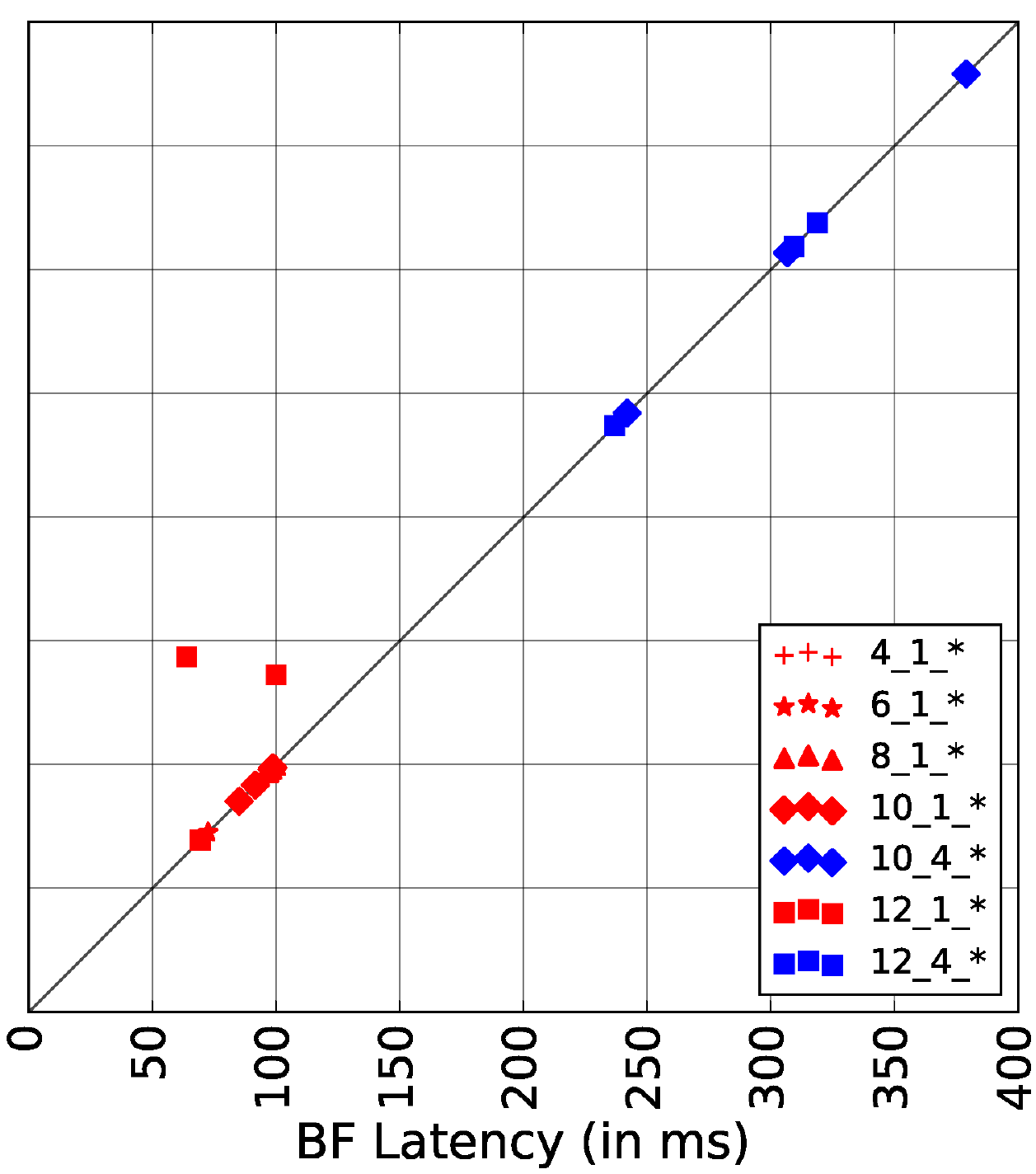}
		\label{fig:GAvsBF_ObjVal:planetlab:100:Nhalf}
	}~
	\subfloat{
		\includegraphics[width=\myrightsize\textwidth]{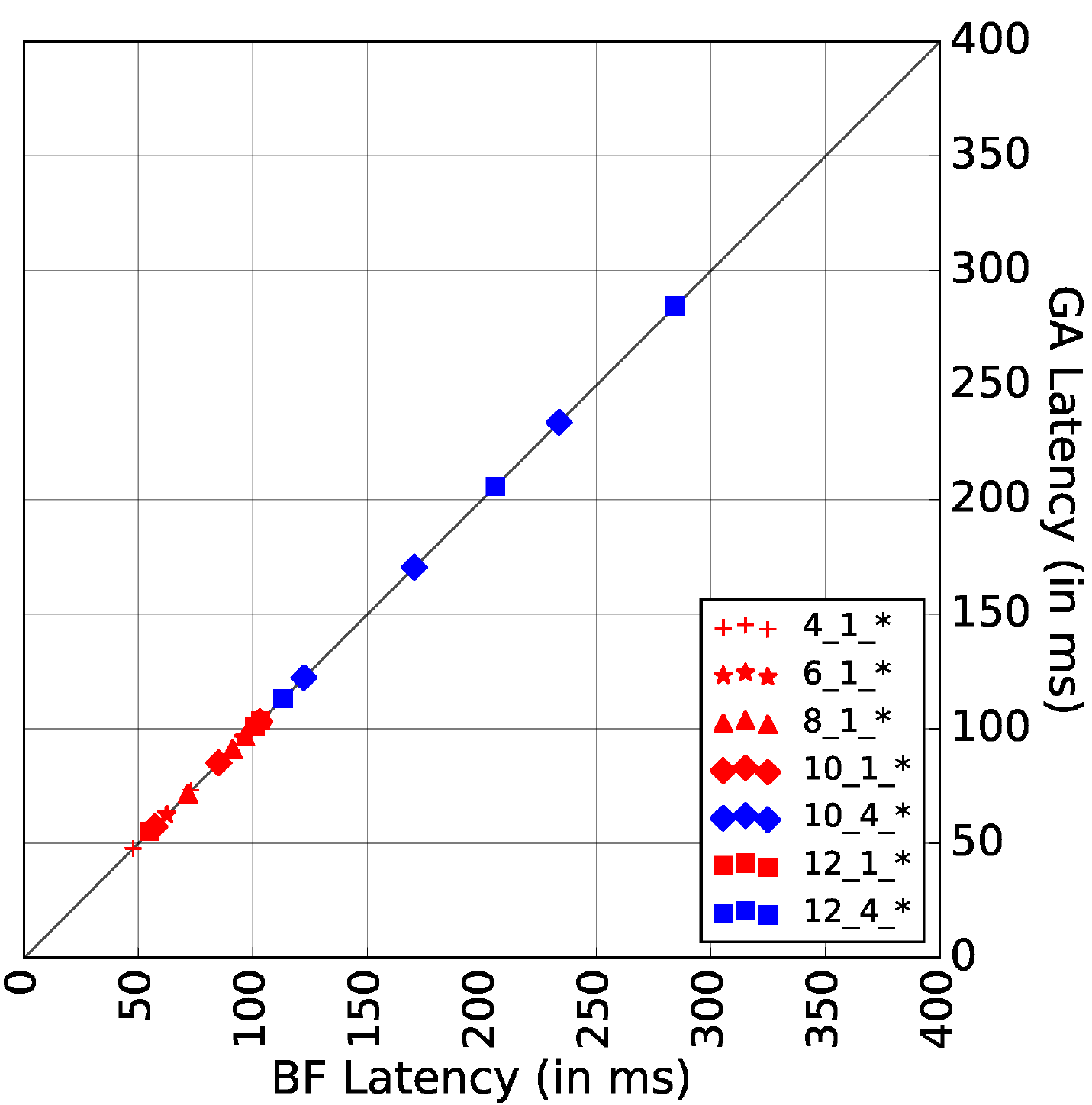}
		\label{fig:GAvsBF_ObjVal:planetlab:100:Nfourth}
	}\quad
	\subfloat{
		\footnotesize
		\begin{rotate}{270}\hspace{-1.45in}(i)~~$\Omega^{in} = 100~e/s$\end{rotate}
	}
	\\
	\addtocounter{subfigure}{-4}
	\subfloat[\emph{Liberal}]{
		\includegraphics[width=\myleftsize\textwidth]{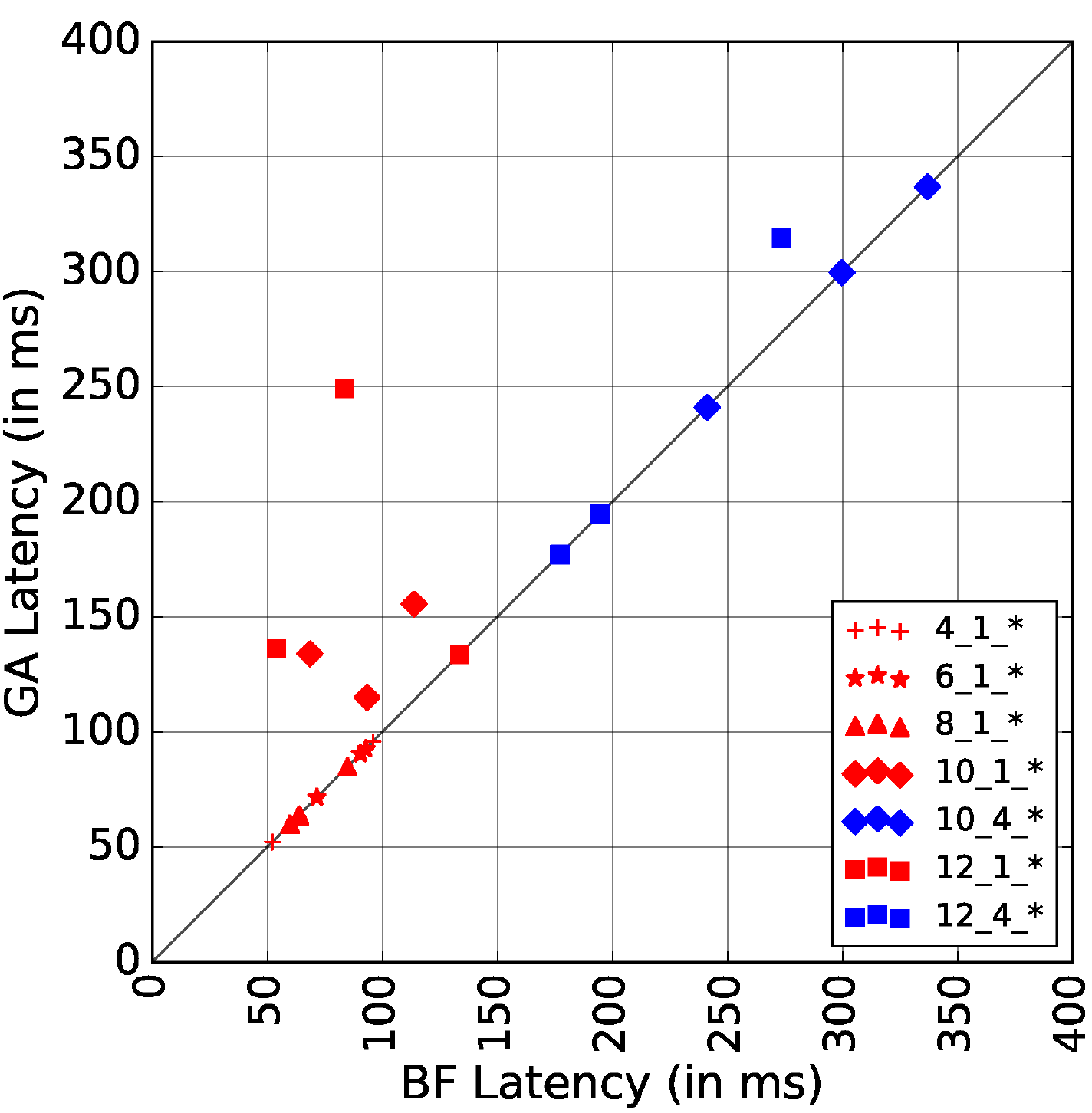}
		\label{fig:GAvsBF_ObjVal:planetlab:1000:N}
	}
	\subfloat[\emph{Centrist}]{
		\includegraphics[width=\mymidsize\textwidth]{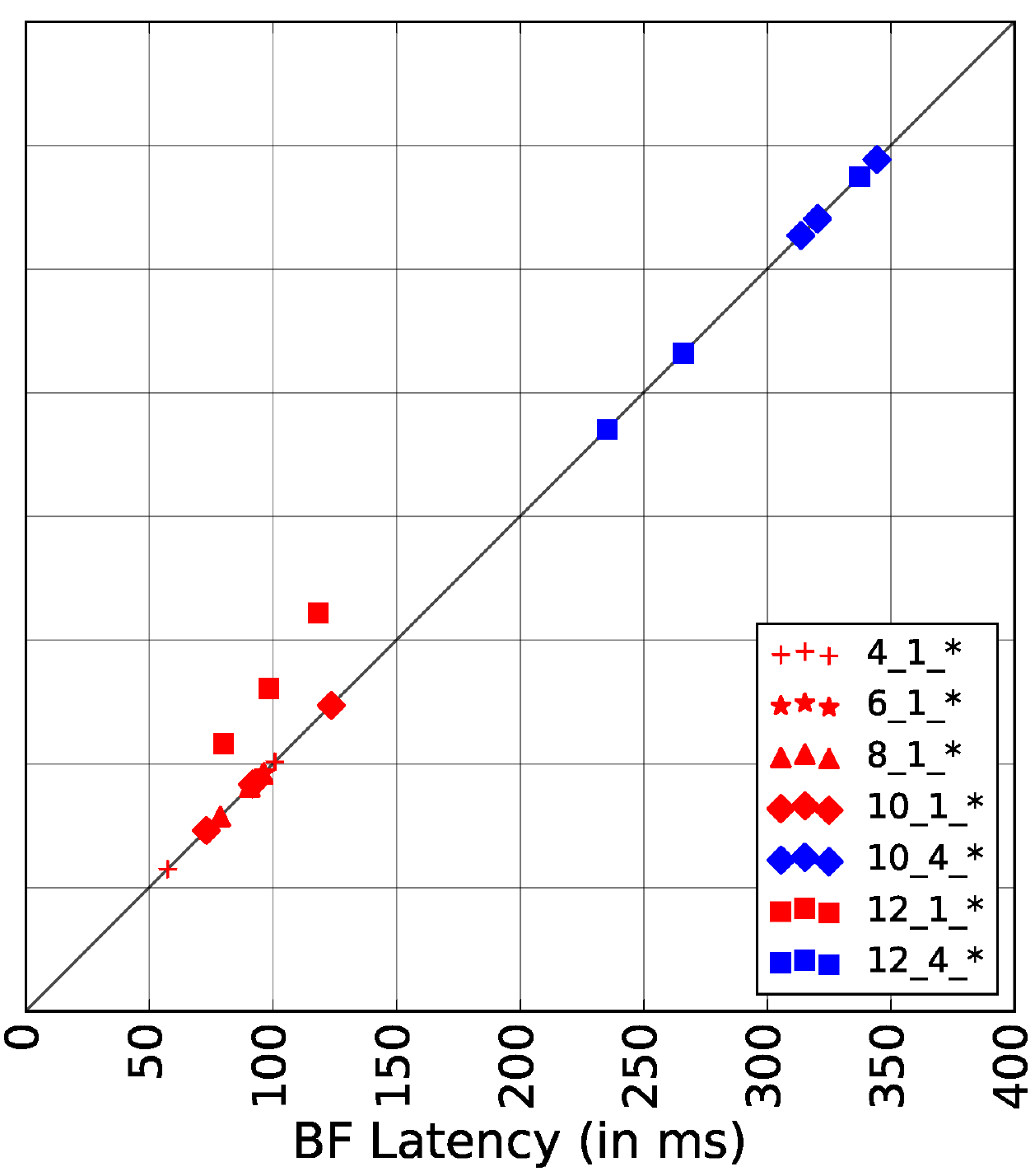}
		\label{fig:GAvsBF_ObjVal:planetlab:1000:Nhalf}
	}
	\subfloat[\emph{Conservative}]{
		\includegraphics[width=\myrightsize\textwidth]{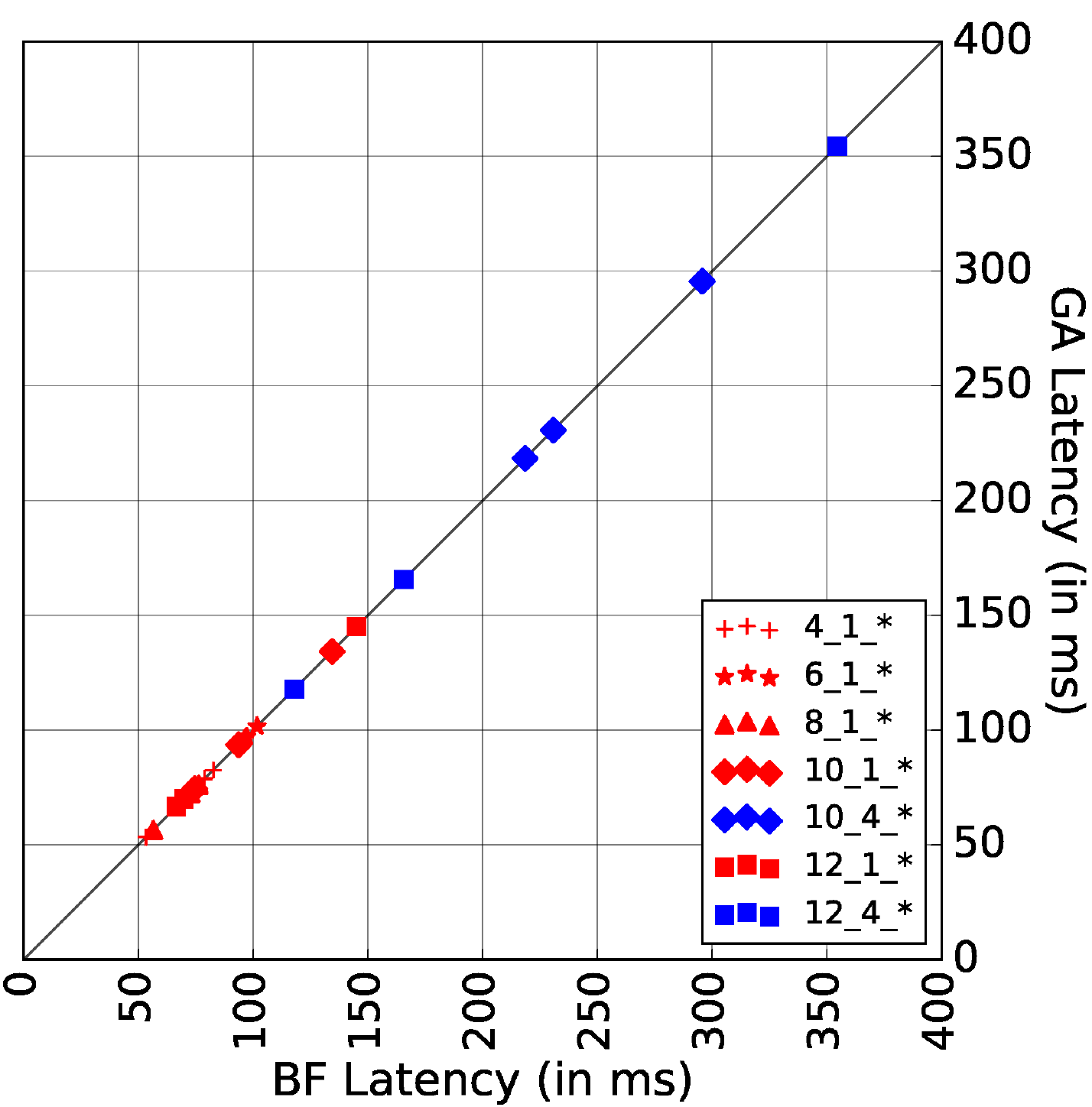}
		\label{fig:GAvsBF_ObjVal:planetlab:1000:Nfourth}
	}\quad
	\subfloat{
		\footnotesize
		\begin{rotate}{270}\hspace{-1.45in}(ii)~~$\Omega^{in} = 1000~e/s$\end{rotate}
	}
	
	\caption{Comparing \emph{end-to-end latency} of GA and BF using \emph{PlanetLab WAN} network for different resource setups and input rates. Each plot has $21$ DAGs with $4$--$12$ queries that BF could solve within $12~hrs$.}
	\label{fig:latency:ga-bf-new:planet}
\end{figure*}



%
\noindent \textbf{Comparing GA with BF.}
Figs.~\ref{fig:latency:ga-bf-new:campus} and~\ref{fig:latency:ga-bf-new:planet} show scatter plots comparing latencies for the GA solution with the optimal BF solution, for \emph{Campus LAN} and \emph{PlanetLab WAN} networks respectively. Here, we show results for DAG with $\le12$ queries since BF takes $>12~hours$ to run for larger DAGs. The plots are for \emph{(a) liberal}, \emph{(b) centrist} and \emph{(c) conservative} edge resource availability, with input rates of \emph{(i)~$\Omega^{in}=100~e/sec$} and \emph{(ii)~$1000~e/sec$}. The shape indicates the number of queries in the DAG, while the color gives the number of source queries.

We see that GA (Y axis) performs very well, falling close to the 1:1 line indicating that its solutions are close to BF's optimal (X axis). When the input rate is $100~e/sec$ for DAGs with liberal setup on campus, GA converges to near-optimal values for DAG sizes up to $8$ queries, but for $10$ queries, the solution is marginally higher ($+4.5\%$) than optimal. With fewer edge resources in the centrist setup, 
GA gives the exact optimal solution for all but 8 DAGs, and near-optimal for these 8 ($+6.8\%$). Reducing the resources further in the conservative setup gives a perfect GA solution in all cases. Intuitively, by limiting the search space for GA -- with fewer resources or fewer queries in the DAG to place -- we improve its chances of converging to the optimal result.

\newcommand{\mylsize}{0.335}
\newcommand{\mymsize}{0.28}
\newcommand{\myrsize}{0.322}

\begin{figure*}[t!]
	\centering
	\subfloat{
		\includegraphics[width=\mylsize\textwidth]{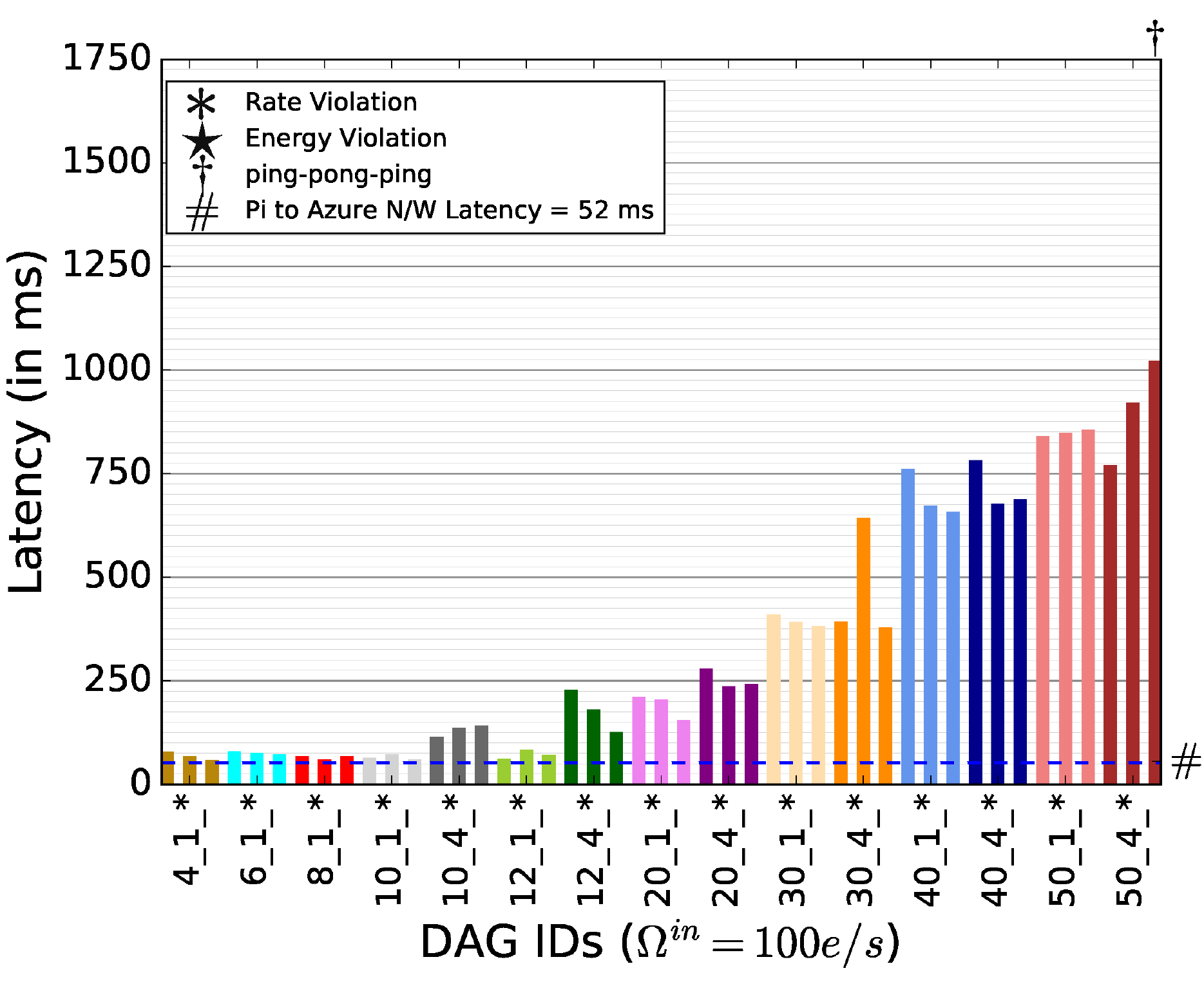}~
		\label{fig:latency:ga:campus:100:N}
	}%
	\subfloat{
		\includegraphics[width=\mymsize\textwidth]{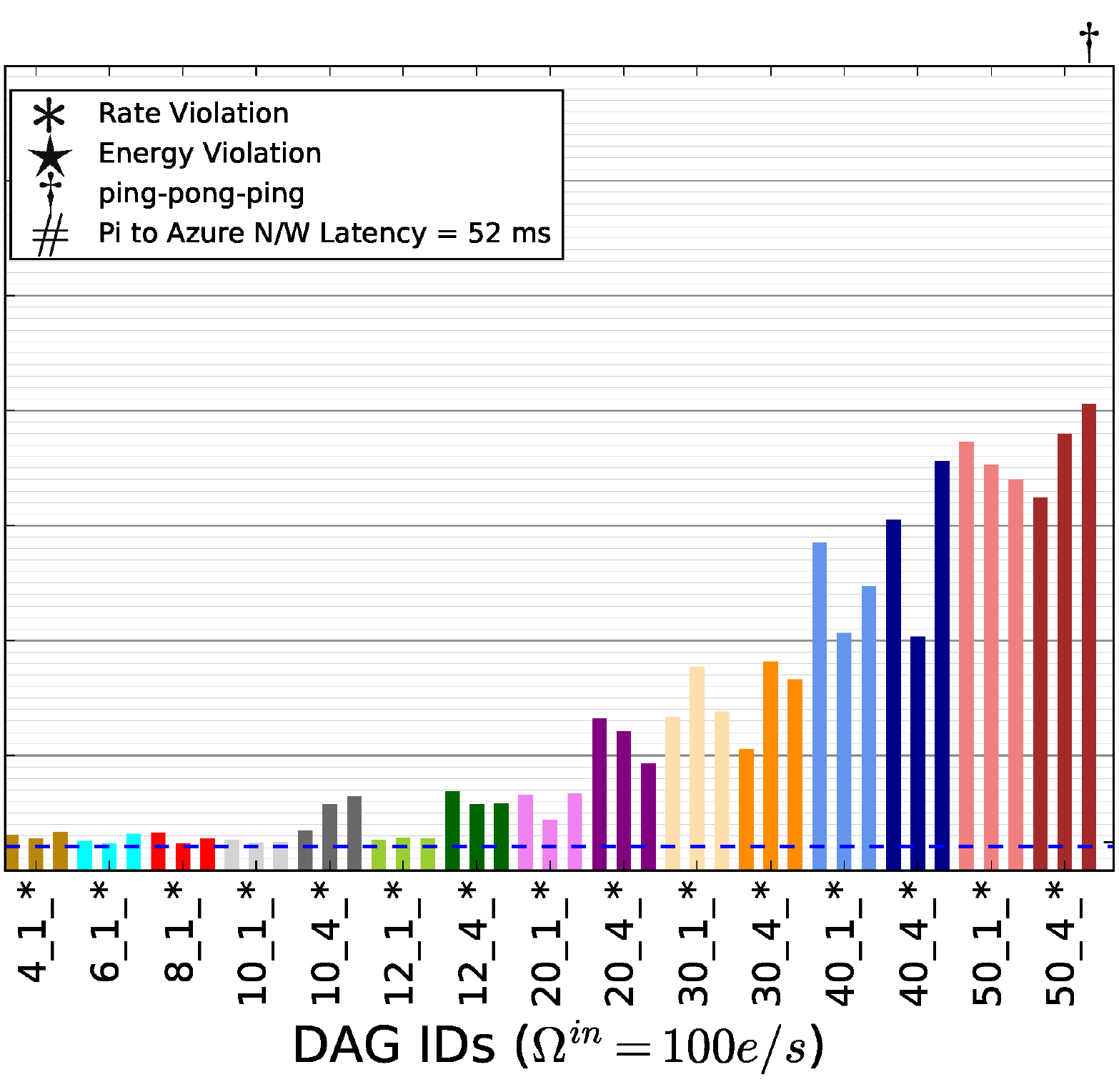}~
		\label{fig:latency:ga:campus:100:Nhalf}
	}%
	\subfloat{
		\includegraphics[width=\myrsize\textwidth]{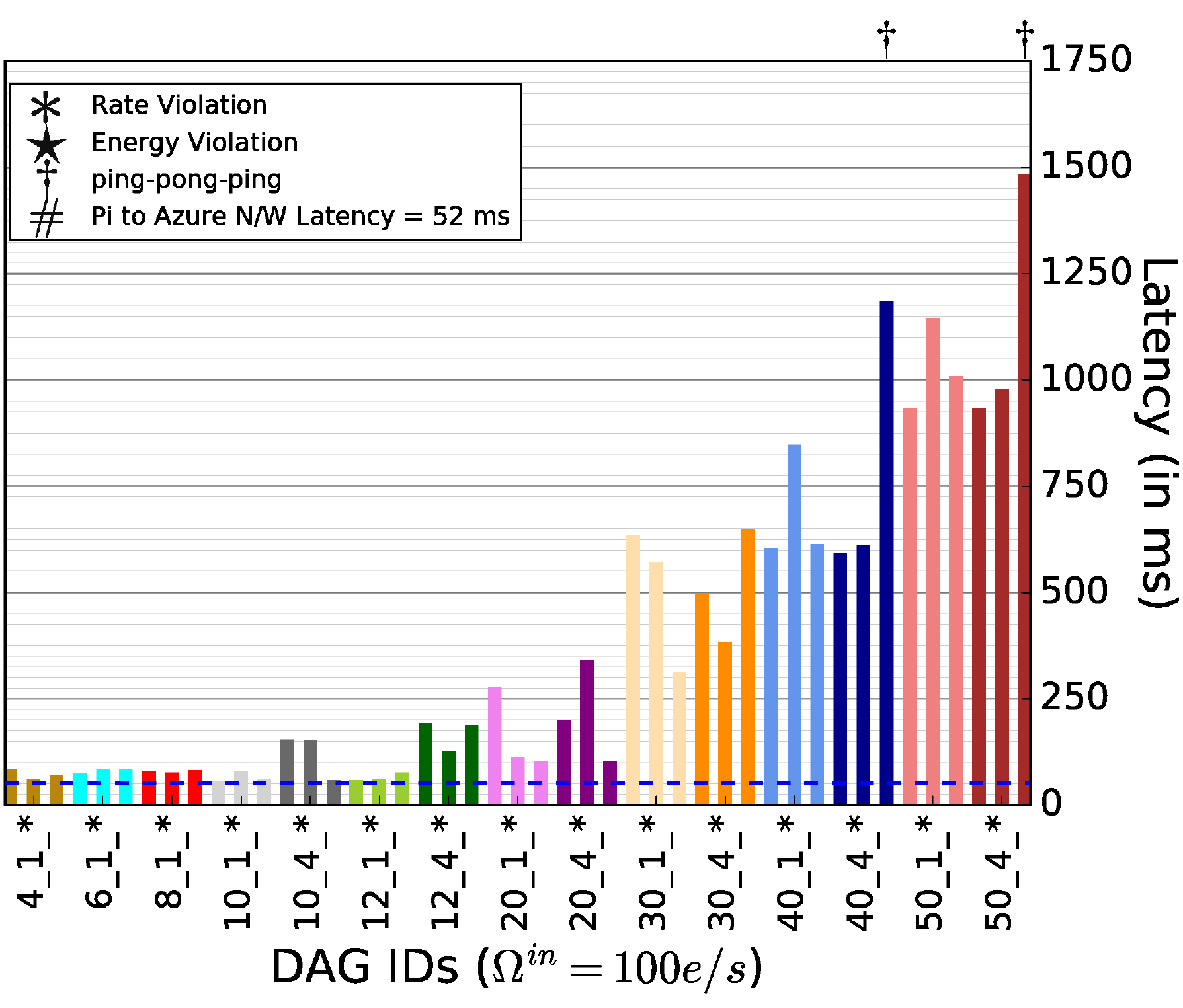}
		\label{fig:latency:ga:campus:100:Nfourth}
	}
	\subfloat{
		\footnotesize
		\begin{rotate}{270}\hspace{-1.35in}(i)~~$\Omega^{in} = 100~e/s$\end{rotate}
	}\\
	\addtocounter{subfigure}{-4}
	\subfloat[\emph{Liberal}]{
		\includegraphics[width=\mylsize\textwidth]{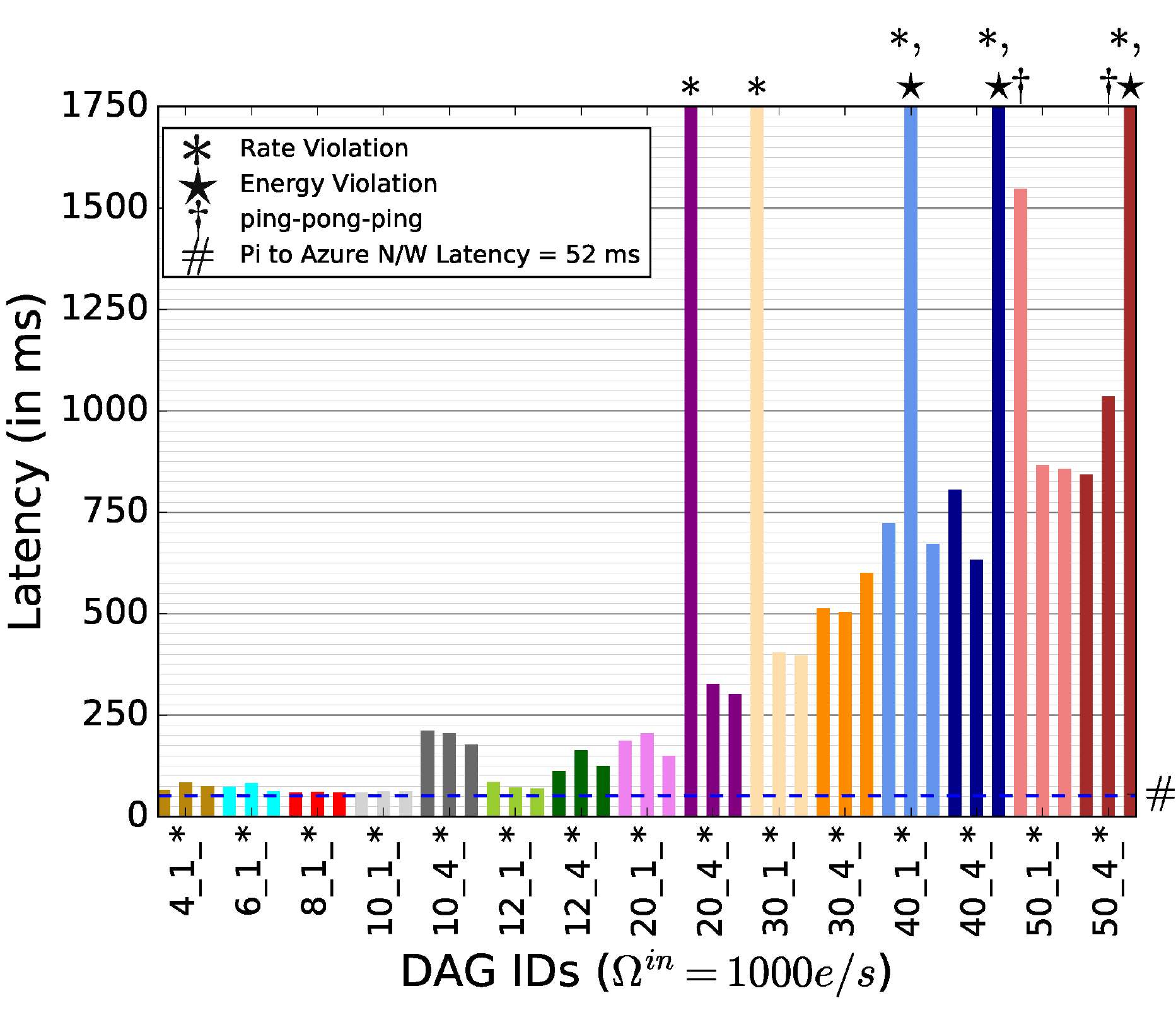}%
		\label{fig:latency:ga:campus:1000:N}
	}%
	\subfloat[\emph{Centrist}]{
		\includegraphics[width=\mymsize\textwidth]{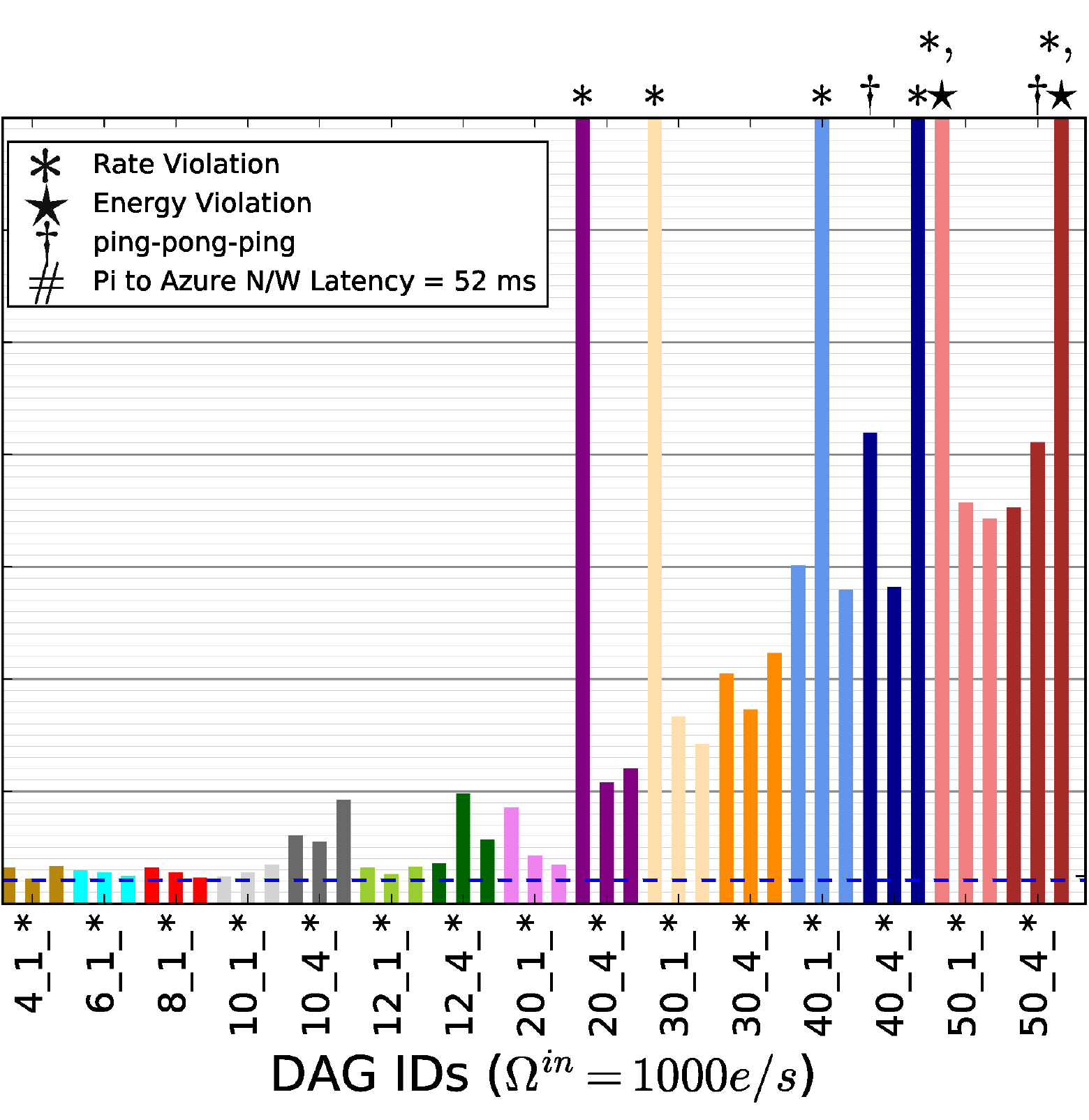}%
		\label{fig:latency:ga:campus:1000:Nhalf}
	}%
	\subfloat[\emph{Conservative}]{
		\includegraphics[width=\myrsize\textwidth]{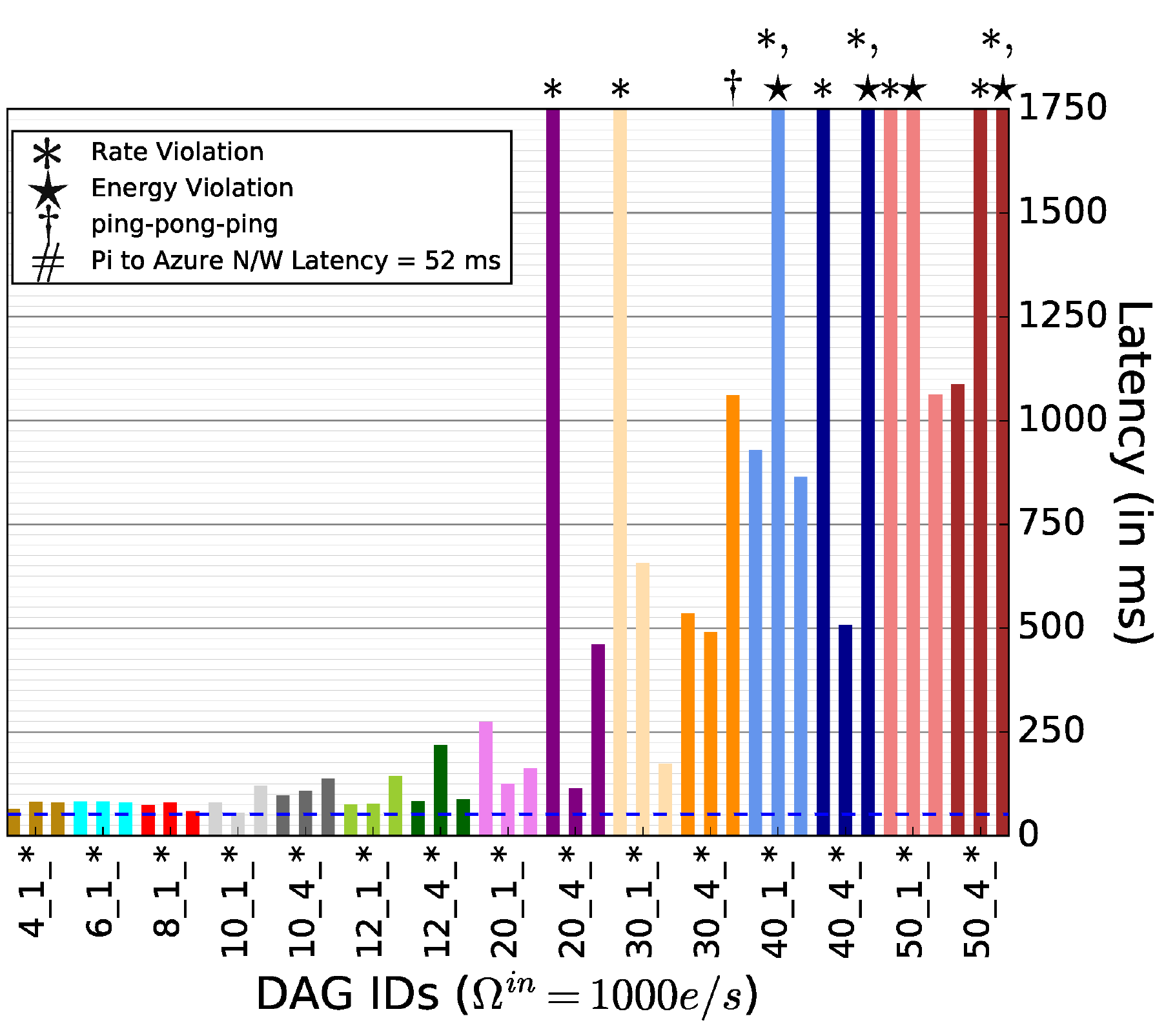}
		\label{fig:latency:ga:campus:1000:Nfourth}
	}
	\subfloat{
		\footnotesize
		\begin{rotate}{270}\hspace{-1.35in}(ii)~~$\Omega^{in} = 1000~e/s$\end{rotate}
	}
	
	\caption{\emph{End-to-end latency from GA solution} using \emph{Campus LAN} for different resource setups and input rates. Each plot shows all $45$ DAGs, with $4$--$50$ queries each.}
	\label{fig:latency:ga-new:campus}
\end{figure*}

\renewcommand{\mylsize}{0.335}
\renewcommand{\mymsize}{0.28}
\renewcommand{\myrsize}{0.322}

\begin{figure*}[t!]
	\centering
	\subfloat{
		\includegraphics[width=\mylsize\textwidth]{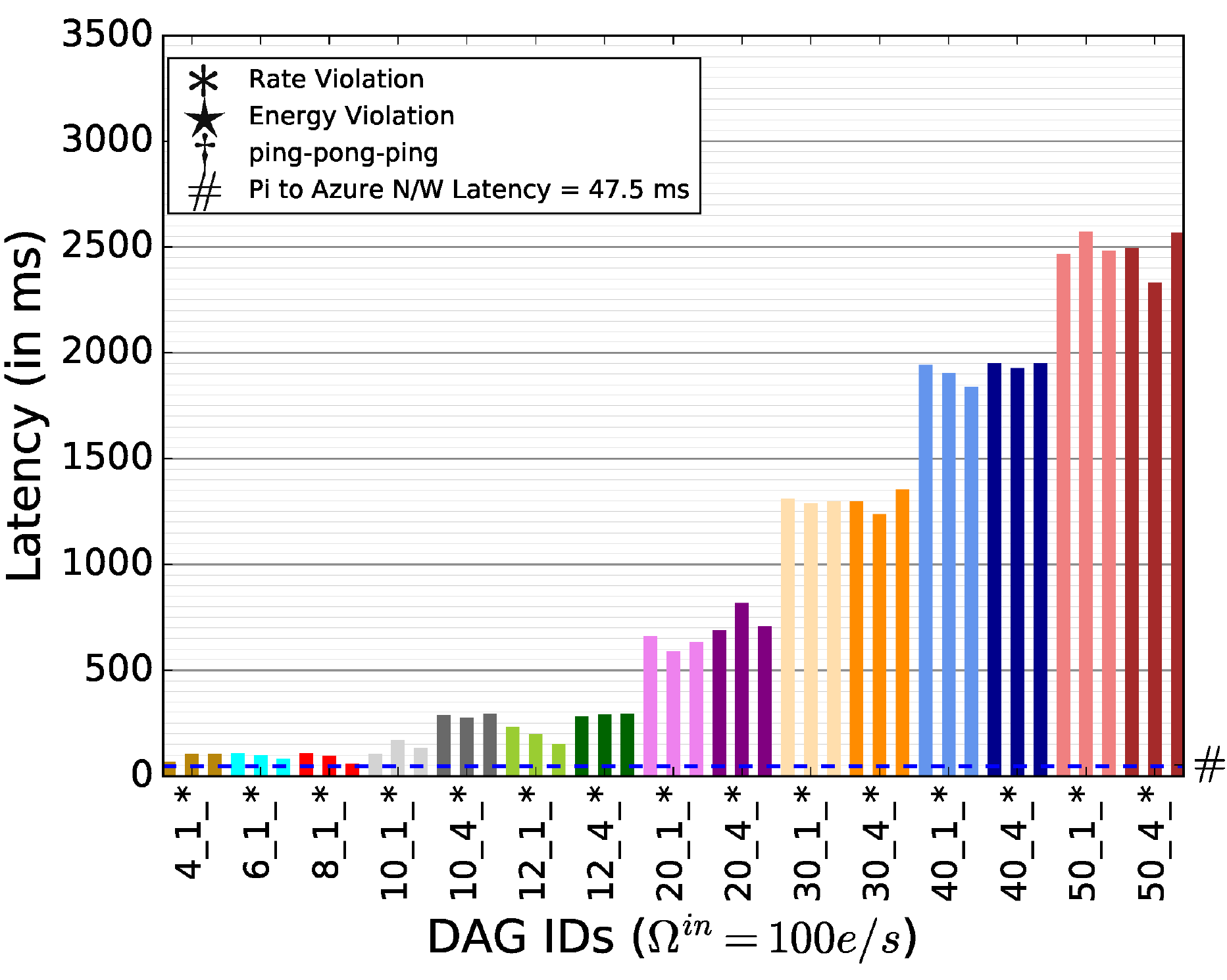}
		\label{fig:latency:ga:planetlab:100:N}
	}\enskip
	\subfloat{
		\includegraphics[width=\mymsize\textwidth]{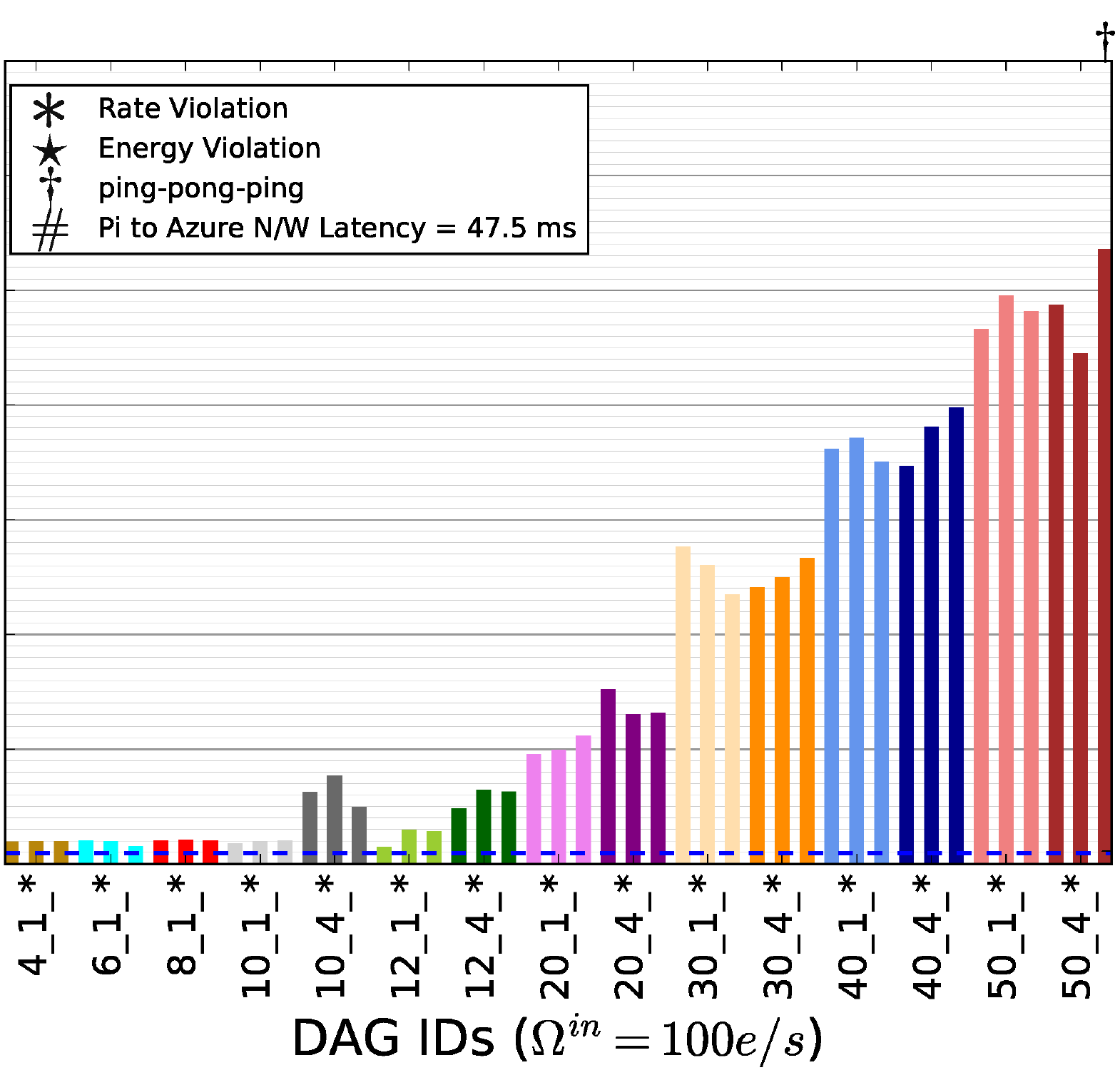}
		\label{fig:latency:ga:planetlab:100:Nhalf}
	}\enskip
	\subfloat{
		\includegraphics[width=\myrsize\textwidth]{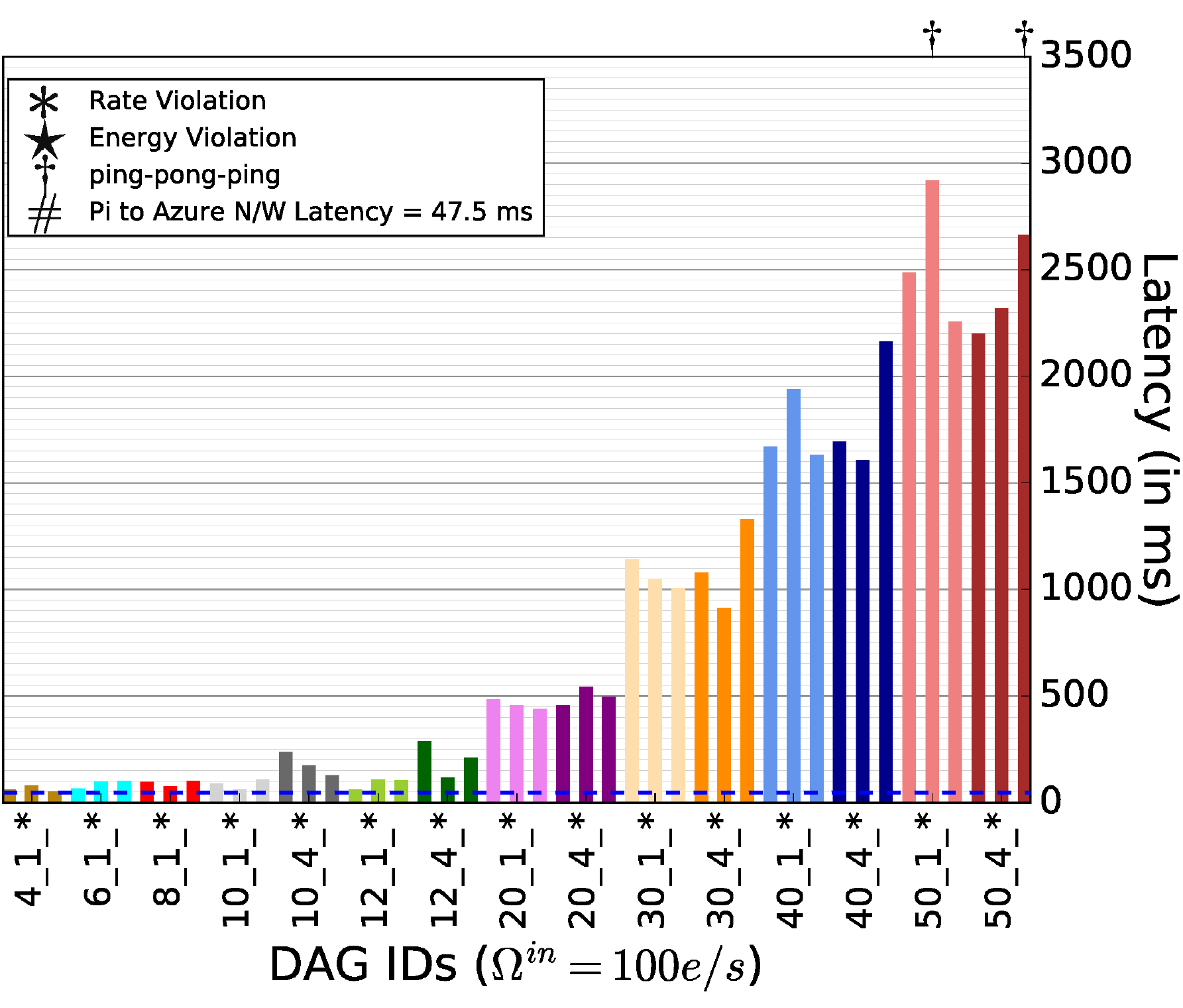}
		\label{fig:latency:ga:planetlab:100:Nfourth}
	}
	\subfloat{
		\footnotesize
		\begin{rotate}{270}\hspace{-1.35in}(i)~~$\Omega^{in} = 100~e/s$ \end{rotate}
	}\\
	\addtocounter{subfigure}{-4}
	\subfloat[\emph{Liberal}]{
		\includegraphics[width=\mylsize\textwidth]{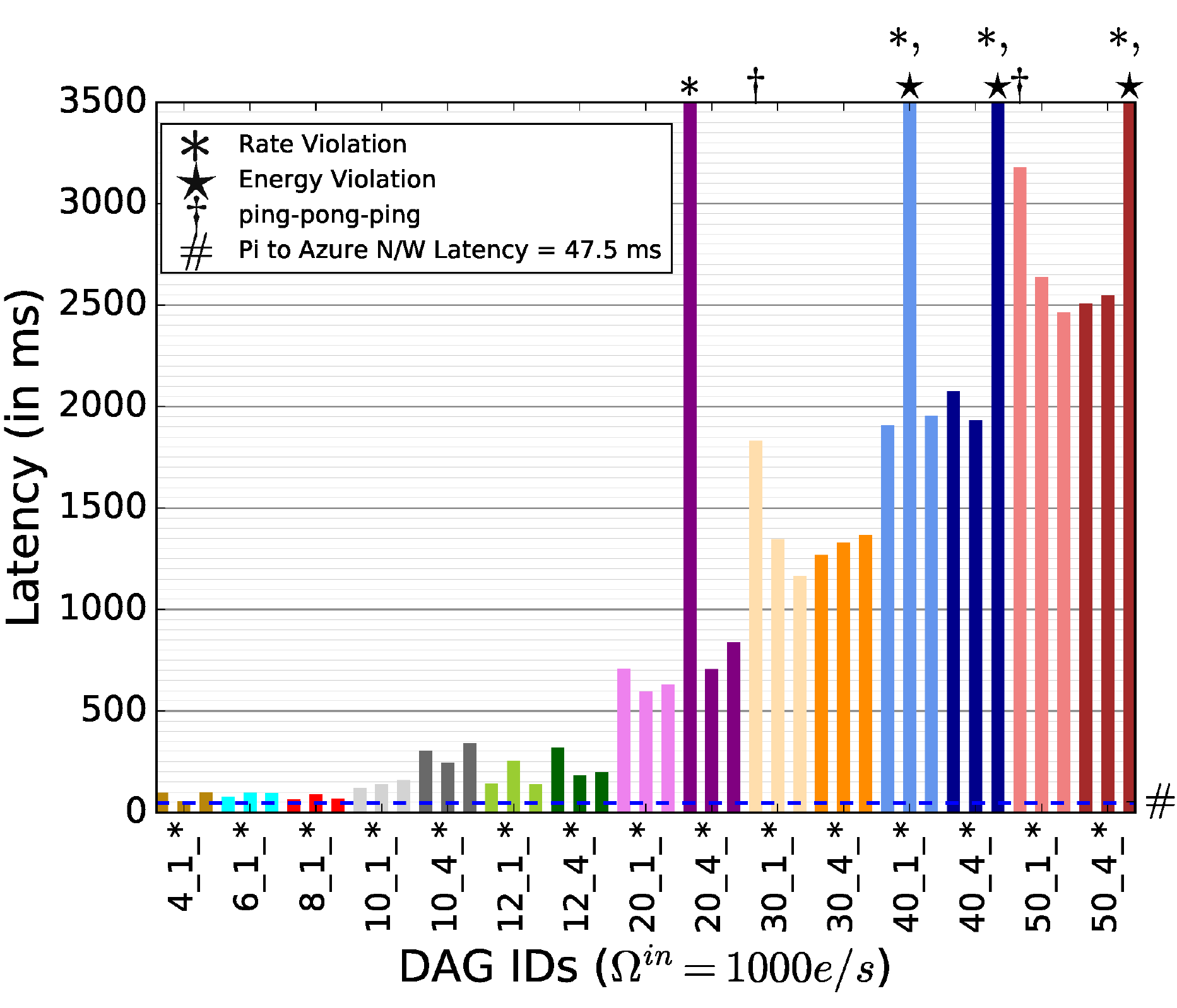}
		\label{fig:latency:ga:planetlab:1000:N}
	}
	\subfloat[\emph{Centrist}]{
		\includegraphics[width=\mymsize\textwidth]{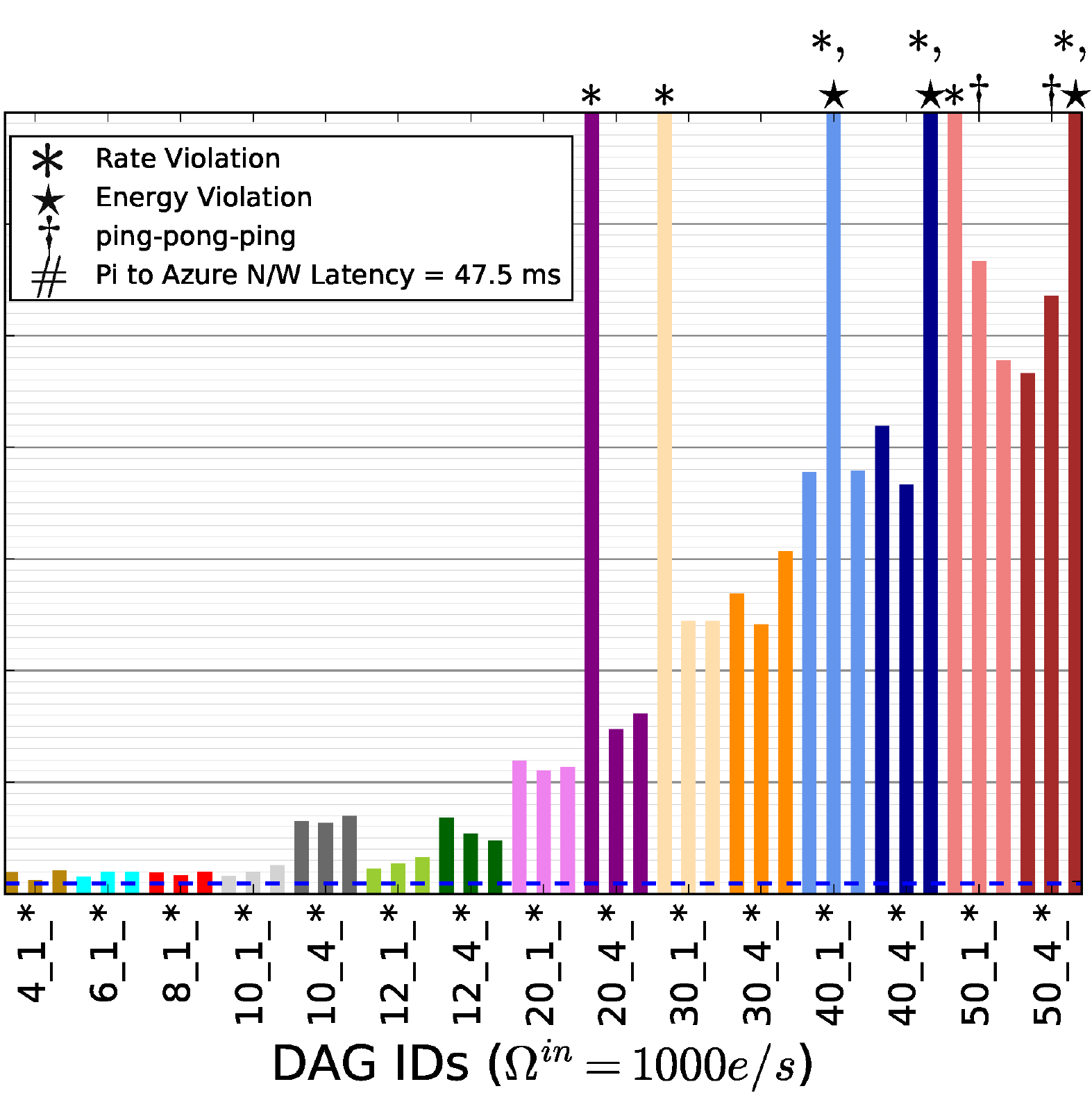}
		\label{fig:latency:ga:planetlab:1000:Nhalf}
	}
	\subfloat[\emph{Conservative}]{
		\includegraphics[width=\myrsize\textwidth]{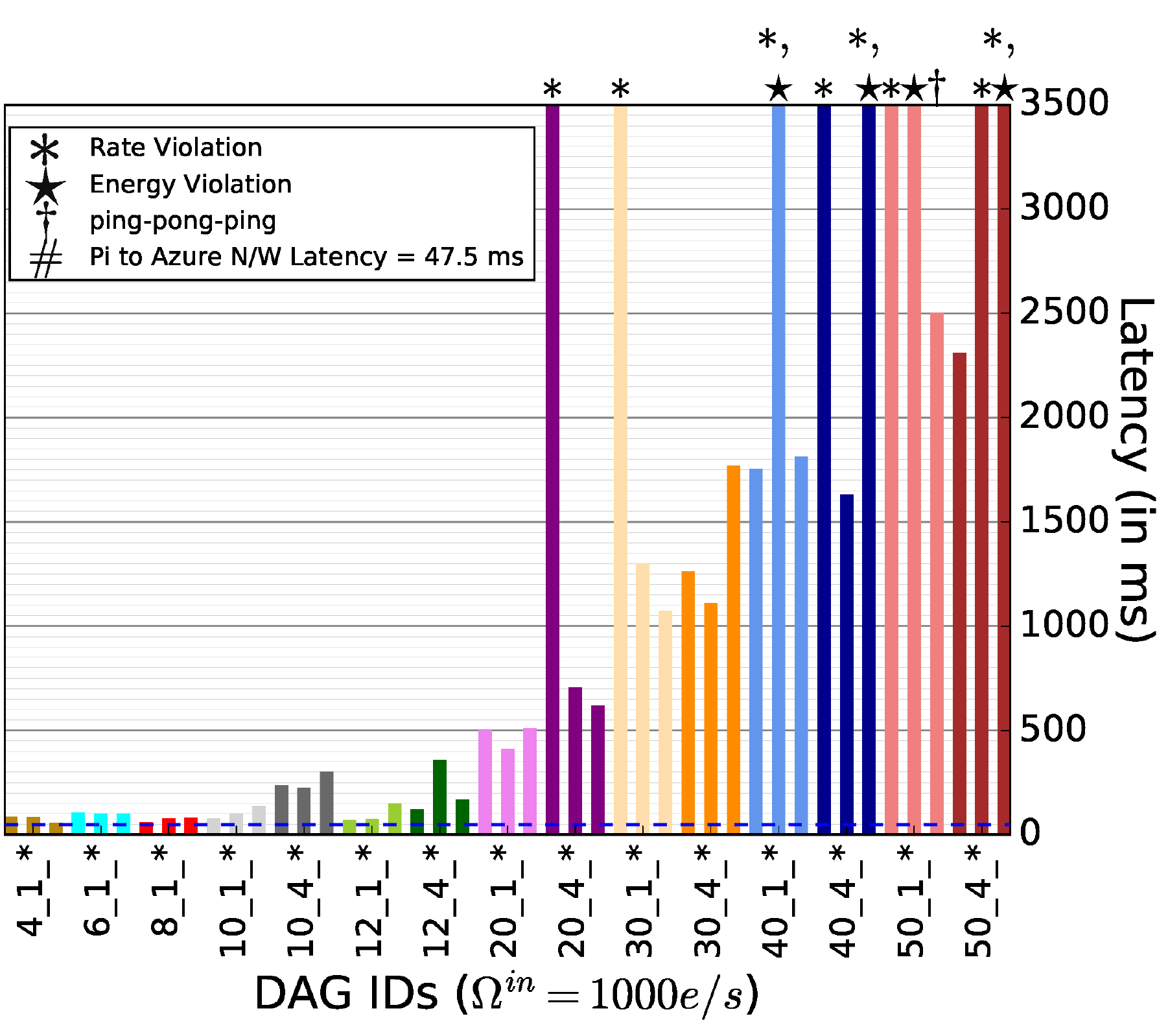}
		\label{fig:latency:ga:planetlab:1000:Nfourth}
	}
	\subfloat{
		\footnotesize
		\begin{rotate}{270}\hspace{-1.35in}(ii)~~$\Omega^{in} = 1000~e/s$\end{rotate}
	}
	\caption{\emph{End-to-end latency from GA solution} using \emph{PlanetLab WAN} for different resource setups and input rates. Each plot shows all $45$ DAGs, with $4$--$50$ queries each.}
	\label{fig:latency:ga-new:planet}
\end{figure*}

A dominating factor in the end-to-end latency is the network latency. Our micro-benchmarks show that the $Q1$ latency between a campus Pi and Azure is $52~ms$. Since we limit the source queries to be on the edge and the sink to be on the Cloud, this $Q1$ latency is the minimum latency for any placement solution. But based on the data center used, this network latency can be much lower, and the benefits of a near-optimal solution will be more significant. For example, a recent Azure South India Data Center shows a $40~ms$ latency, and others report values of $25-80~ms$~\cite{gaming2}.

\noindent\textbf{GA Solution on Large DAGs. }
While BF is intractable for DAGs with $>12$ queries, we can find GA solutions for all $45$ DAGs that are simulated. Figs.~\ref{fig:latency:ga-new:campus} and~\ref{fig:latency:ga-new:planet} show the latencies on campus and PlanetLab for all these DAGs, with different input rates, and resource availability.
We see that the latencies increase as the DAG size grows due to the longer critical path through more numbers of queries between the source and sink. There are also minor variations in the latencies for DAGs with similar configurations (shaded similarly) due to the random DAG generation. In liberal and centrist setups at $100~e/s$, the end-to-end latency is mostly $\le 1000~ms$ on campus (Fig.~\ref{fig:latency:ga-new:campus}(i)) and $\le 2500~ms$ on PlanetLab (Fig.~\ref{fig:latency:ga-new:planet}(i)), reflecting the LAN and WAN network latencies that accumulate between queries. It is occasionally higher for the conservative setup with fewer edge devices. In general, at $\Omega^{in}=100~e/sec$ the edges can retain most of the queries without overwhelming their compute or energy capacities. However, for $1000~e/sec$ input rate shown in Figs.~\ref{fig:latency:ga-new:campus}(ii) and ~\ref{fig:latency:ga-new:planet}(ii), we see two types of outliers.

\emph{One,} the GA solution converges to a \emph{valid} solution but with a higher latency for DAGs like \texttt{50\_4\_2} on liberal and centrist on campus ($\dagger$ at the top of the figures).  
This happens when the solution places successive queries on the edge followed by Cloud, and back to the edge and then Cloud in a \emph{ping-pong-ping} manner. This causes the edge-to-Cloud network latency to be paid $3\times$, causing an increase of $\approx 230~ms$ in Pi to Cloud latency for DAGs like \texttt{50\_4\_2} and \texttt{40\_4\_3}. DAGs on conservative resources like \texttt{50\_4\_3} at $100~e/sec$ and \texttt{30\_4\_3} at $1000~e/s$ are worse with $5\times$ and $7\times$ network penalties on campus, respectively.
This arises when the GA solution causes constraint violations on the edge for some query in the DAG and is forced to move it to the Cloud, but its subsequent query is moved back to an edge to avoid a constraint violation on the VM.

\emph{Two,} queries of some DAGs have high input rates of $\ge 100,000~e/sec$ (Table~\ref{tbl:dag}, \emph{Max Rate} column) due to high selectivities of previous queries, compounded by the DAG input rate of $1000~e/sec$. For such DAGs like \texttt{20\_4\_1, 30\_1\_1, 40\_1\_2,} and \texttt{50\_4\_3}, the GA often converges to an invalid solution for all resource configurations, with energy and/or throughput rate violations ($\star/*$ at the top of the figures). There is no tangible impact of the network setup on the number of violations. The latency value of the GA solution is in the tens of seconds (truncated in the plots), which reflects the penalty applied by the GA for solutions that violate constraints. In the absence of a BF solution for these DAGs, we cannot state if the GA is unable to find a valid solution, or if a valid solution does not even exist. Our experiments do show that occasionally, rerunning the GA several times helps identify a better or feasible solution, and this approach can be used, given the low GA computational cost. Otherwise, we inform the user that the given resources -- compute capacity, energy capacity or resource count, are insufficient to meet the requirements of the DAG. This can help users with capacity planning.


\begin{figure*}[htb!]
	\centering
	\subfloat{
		\includegraphics[width=0.225\textwidth]{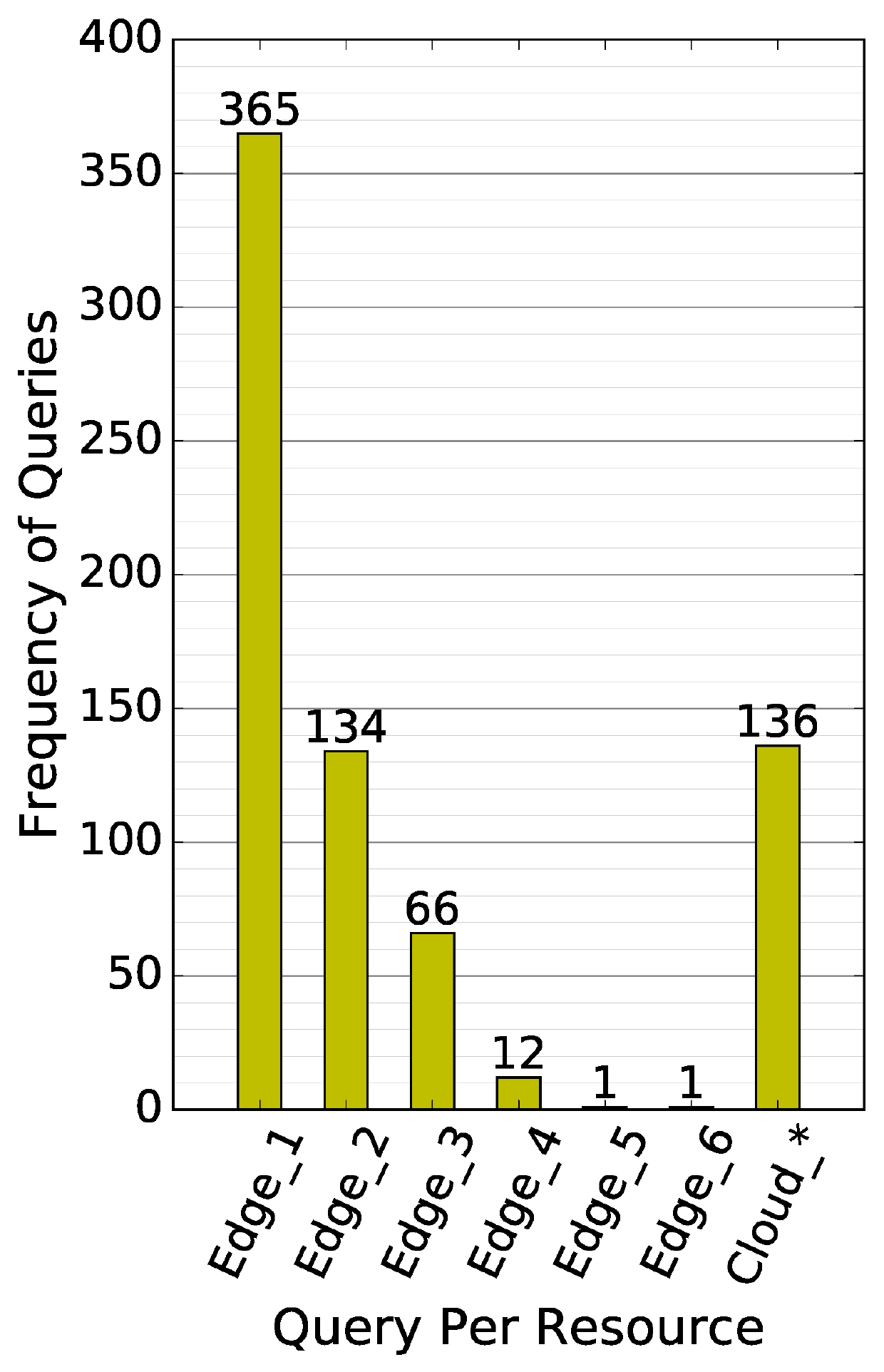}
		\label{fig:queries-edge:100:campus:lib}
	}
	\subfloat{%
		\includegraphics[width=0.225\textwidth]{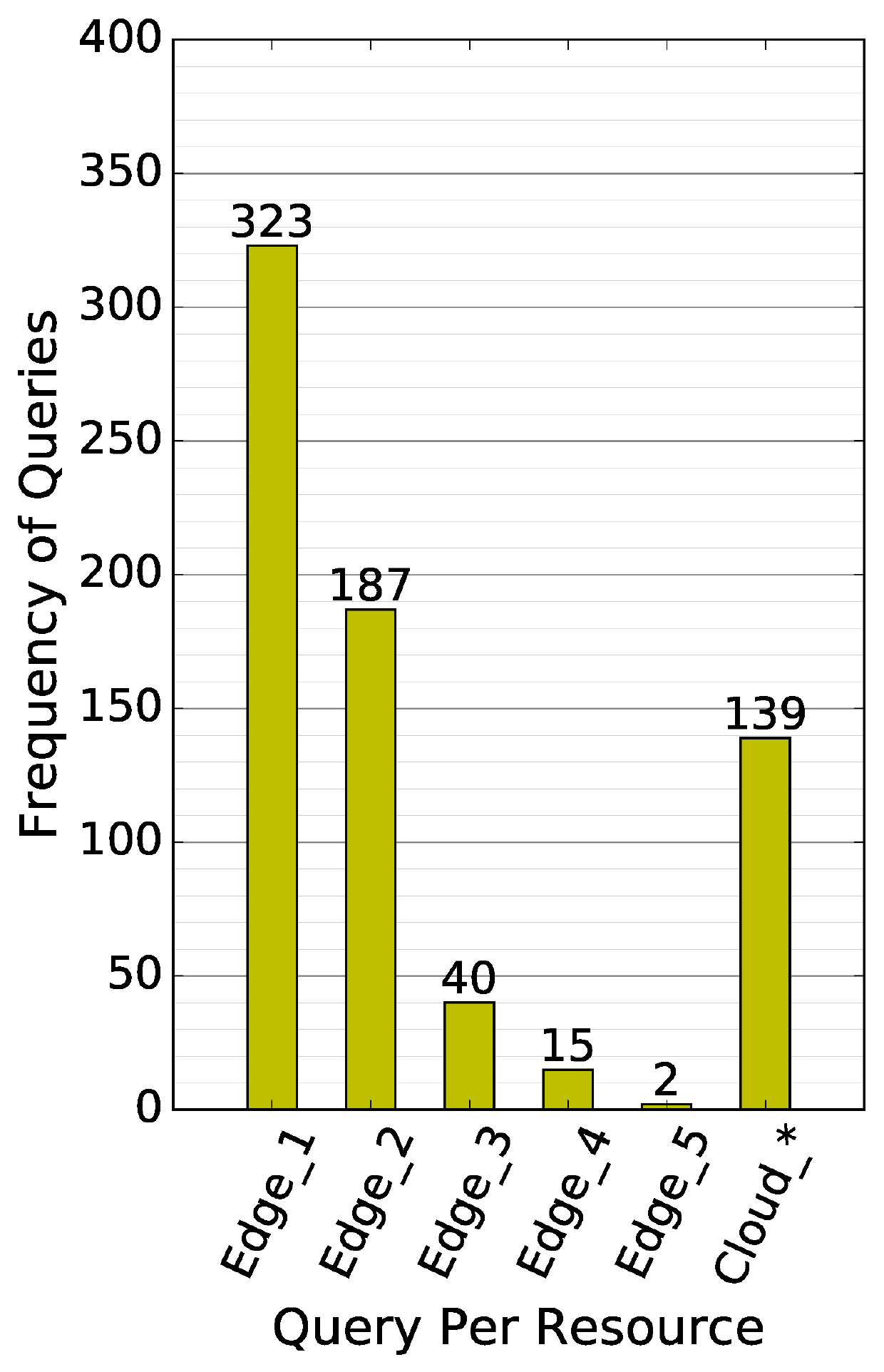}
		\label{fig:queries-edge:1000:campus:lib}
	}~
	\subfloat{
		\includegraphics[width=0.225\textwidth]{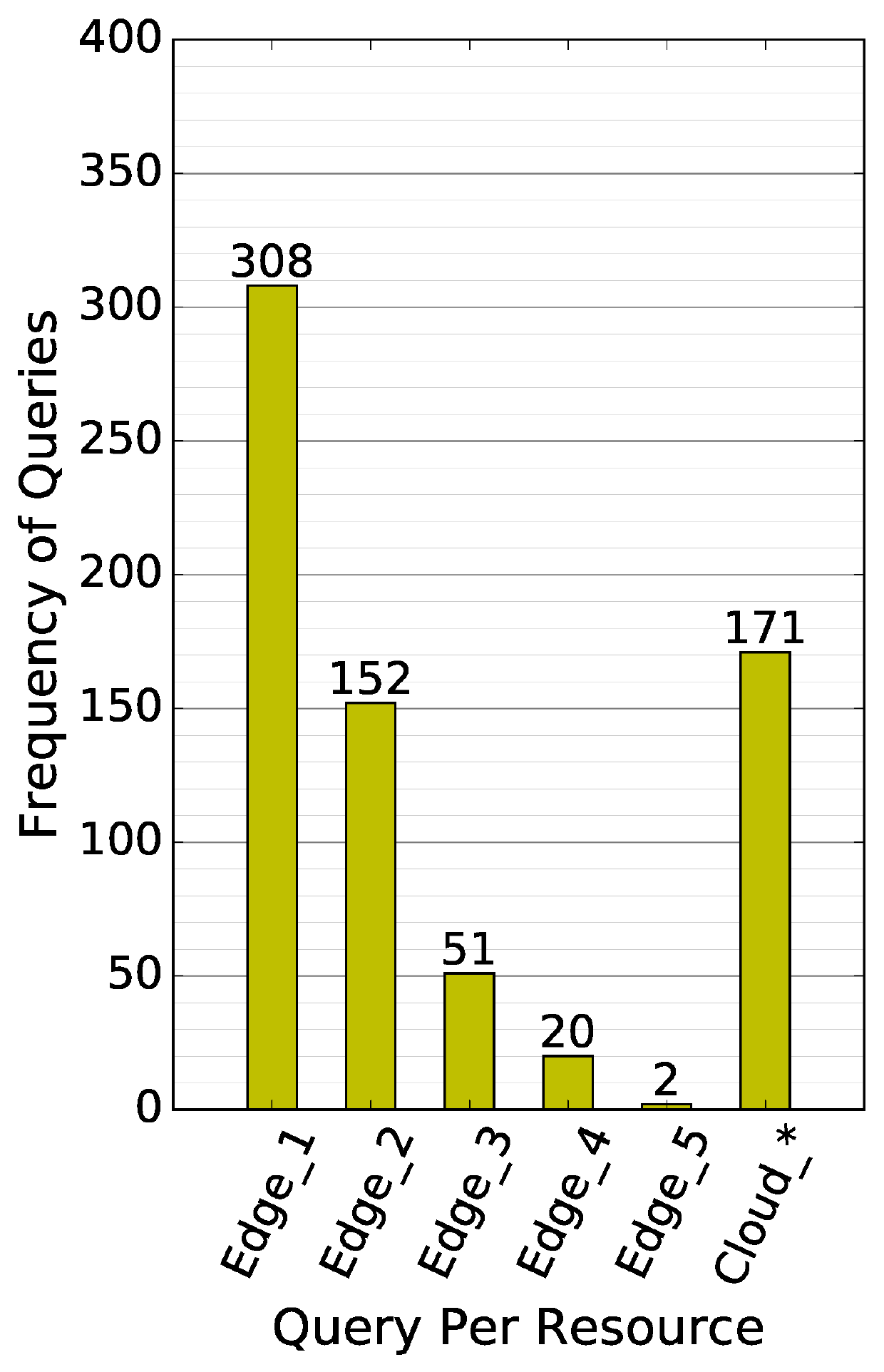}
		\label{fig:queries-edge:100:planetlab:lib}
	}
	\subfloat{%
		\includegraphics[width=0.225\textwidth]{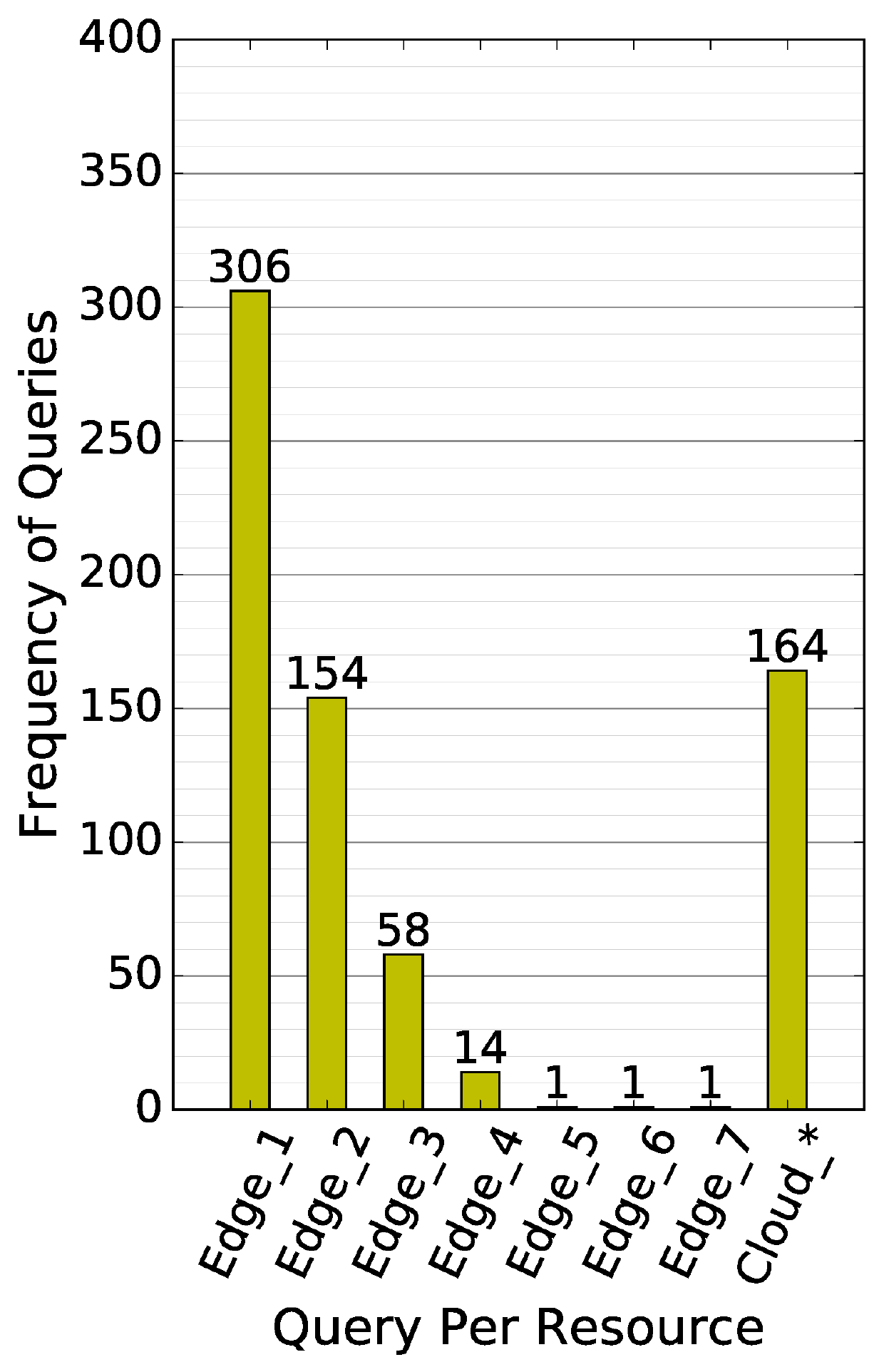}
		\label{fig:queries-edge:1000:planetlab:lib}
	}
	\subfloat{
		\footnotesize
		\begin{rotate}{270}\hspace{-1.4in}(i)~\emph{Liberal}\end{rotate}
	}\\
	\subfloat{
		\includegraphics[width=0.225\textwidth]{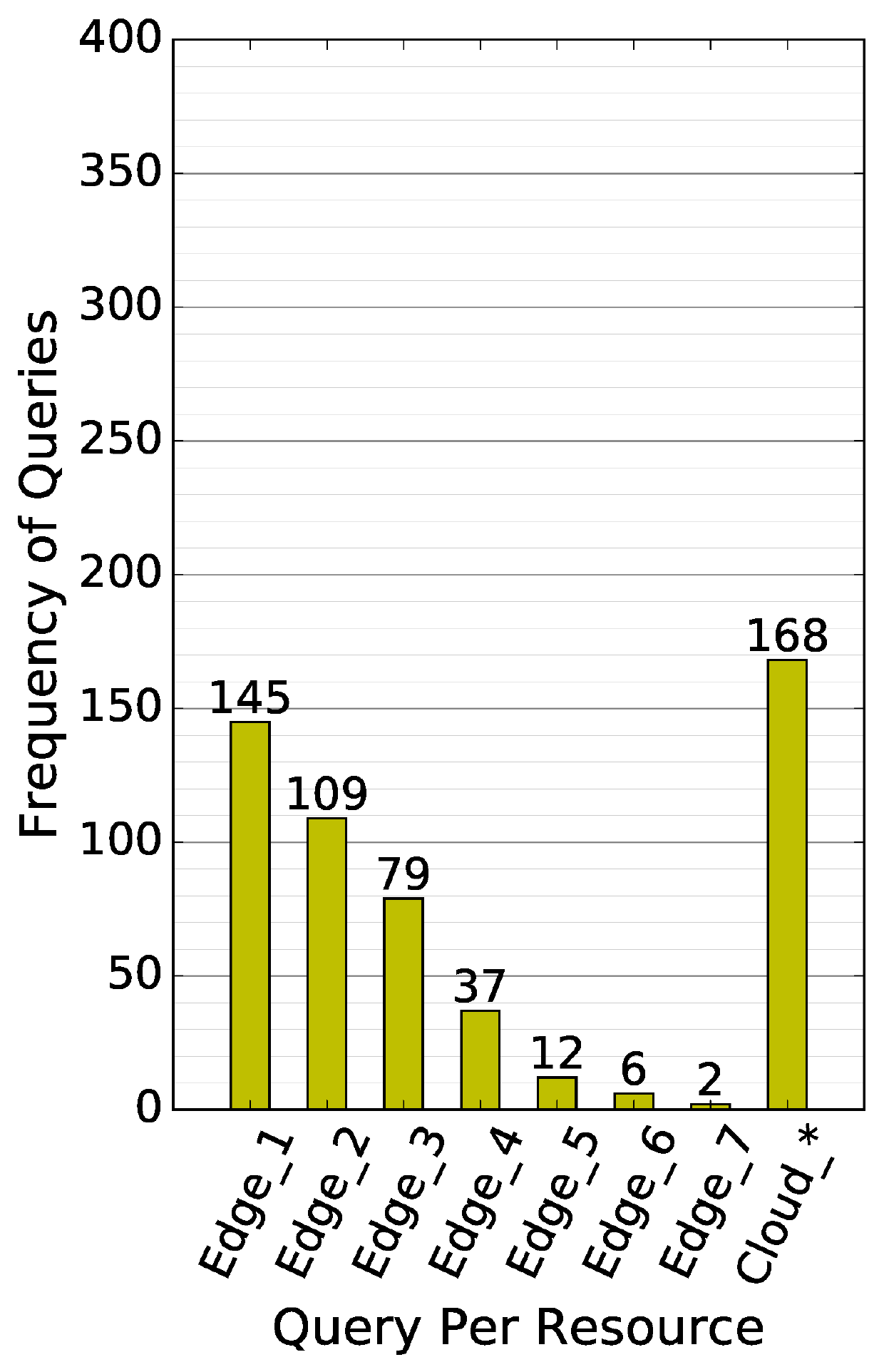}
		\label{fig:queries-edge:100:campus:cent}
	}
	\subfloat{%
		\includegraphics[width=0.225\textwidth]{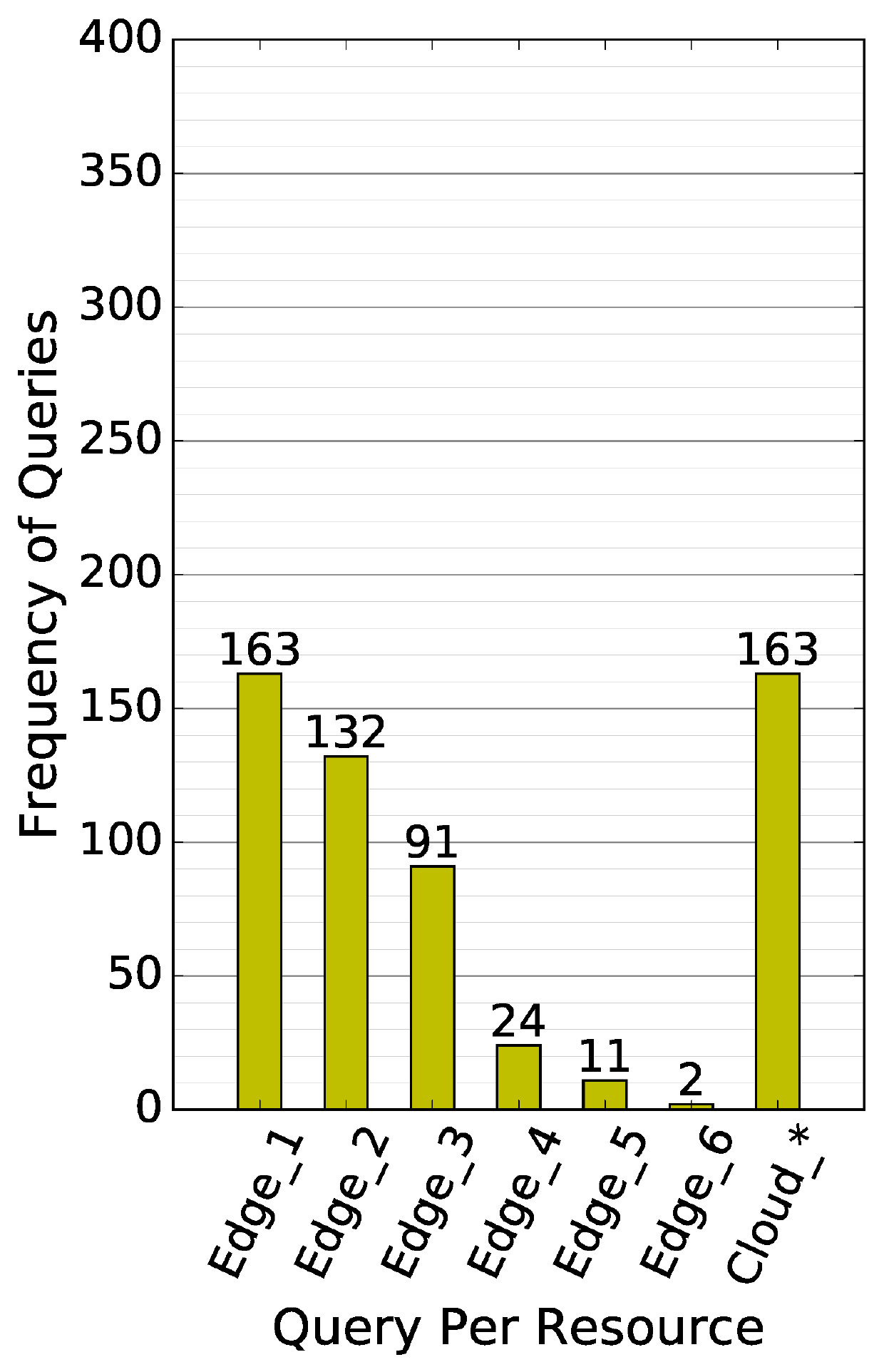}
		\label{fig:queries-edge:1000:campus:cent}
	}~
	\subfloat{
		\includegraphics[width=0.225\textwidth]{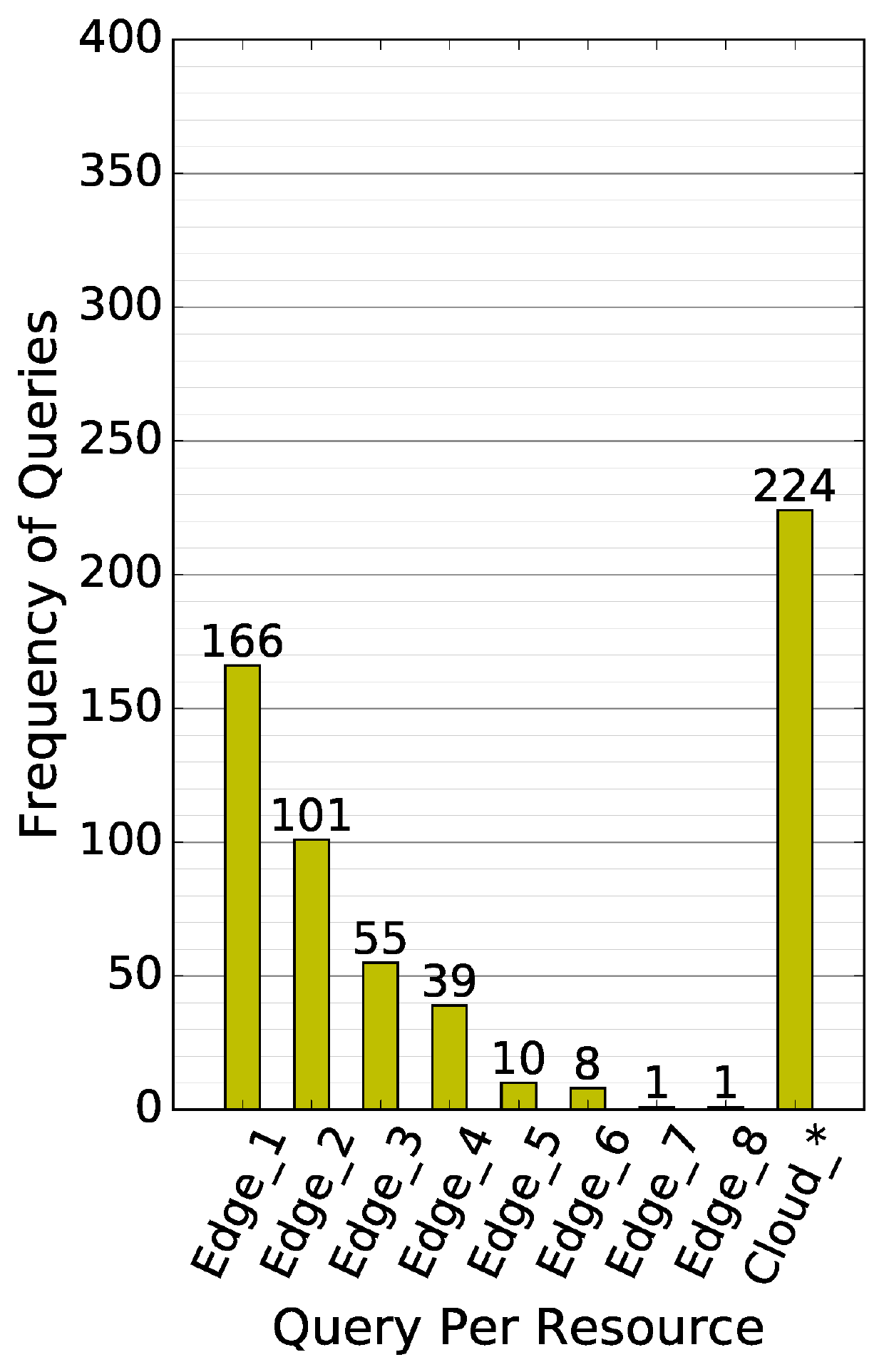}
		\label{fig:queries-edge:100:planetlab:cent}
	}
	\subfloat{%
		\includegraphics[width=0.225\textwidth]{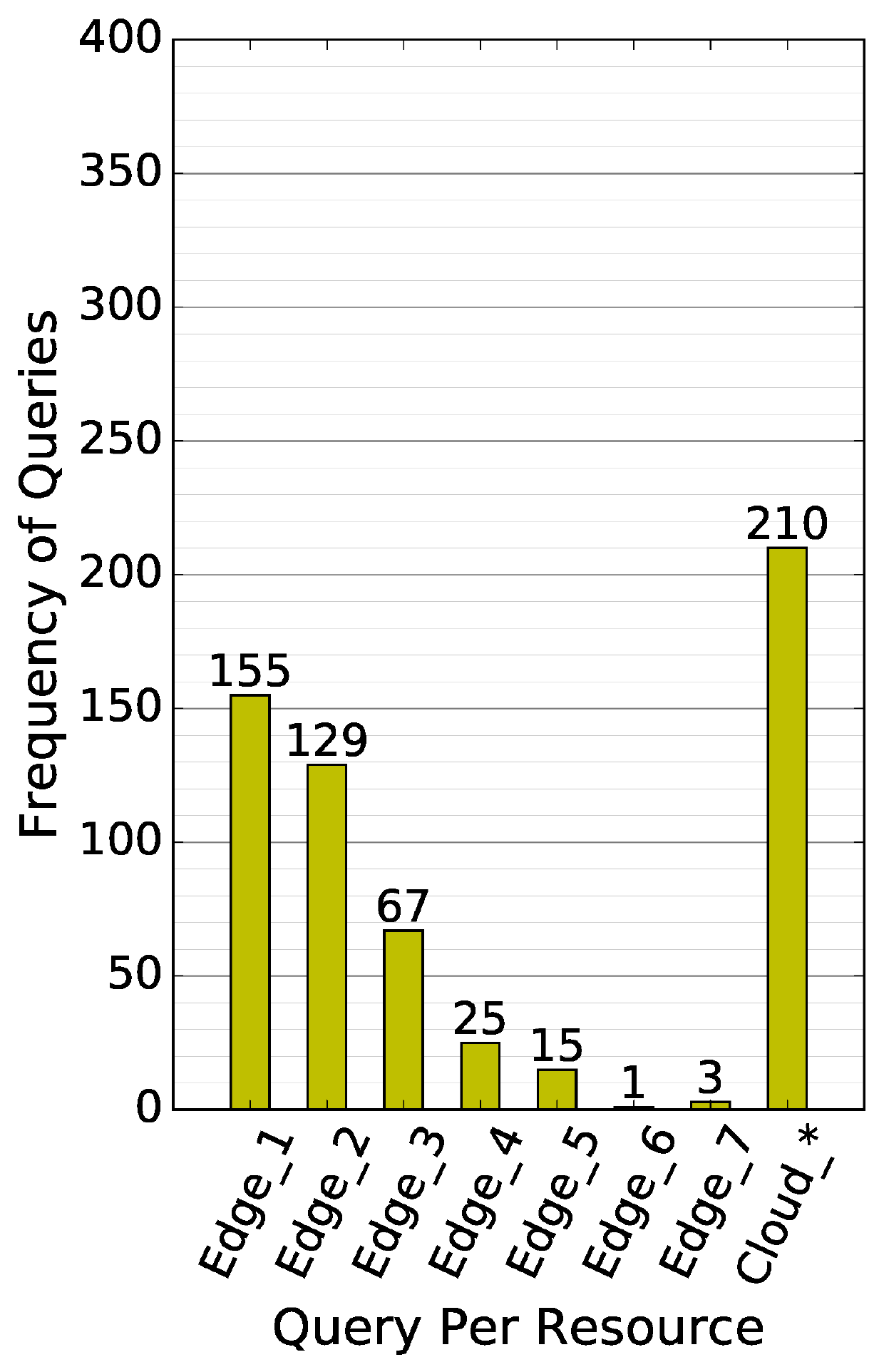}
		\label{fig:queries-edge:1000:planetlab:cent}
	}
	\subfloat{
		\footnotesize
		\begin{rotate}{270}\hspace{-1.4in}(ii)~\emph{Centrist}\end{rotate}
	}\\
	\addtocounter{subfigure}{-10}
	\subfloat[$100~e/sec$]{
		\quad\includegraphics[width=0.225\textwidth]{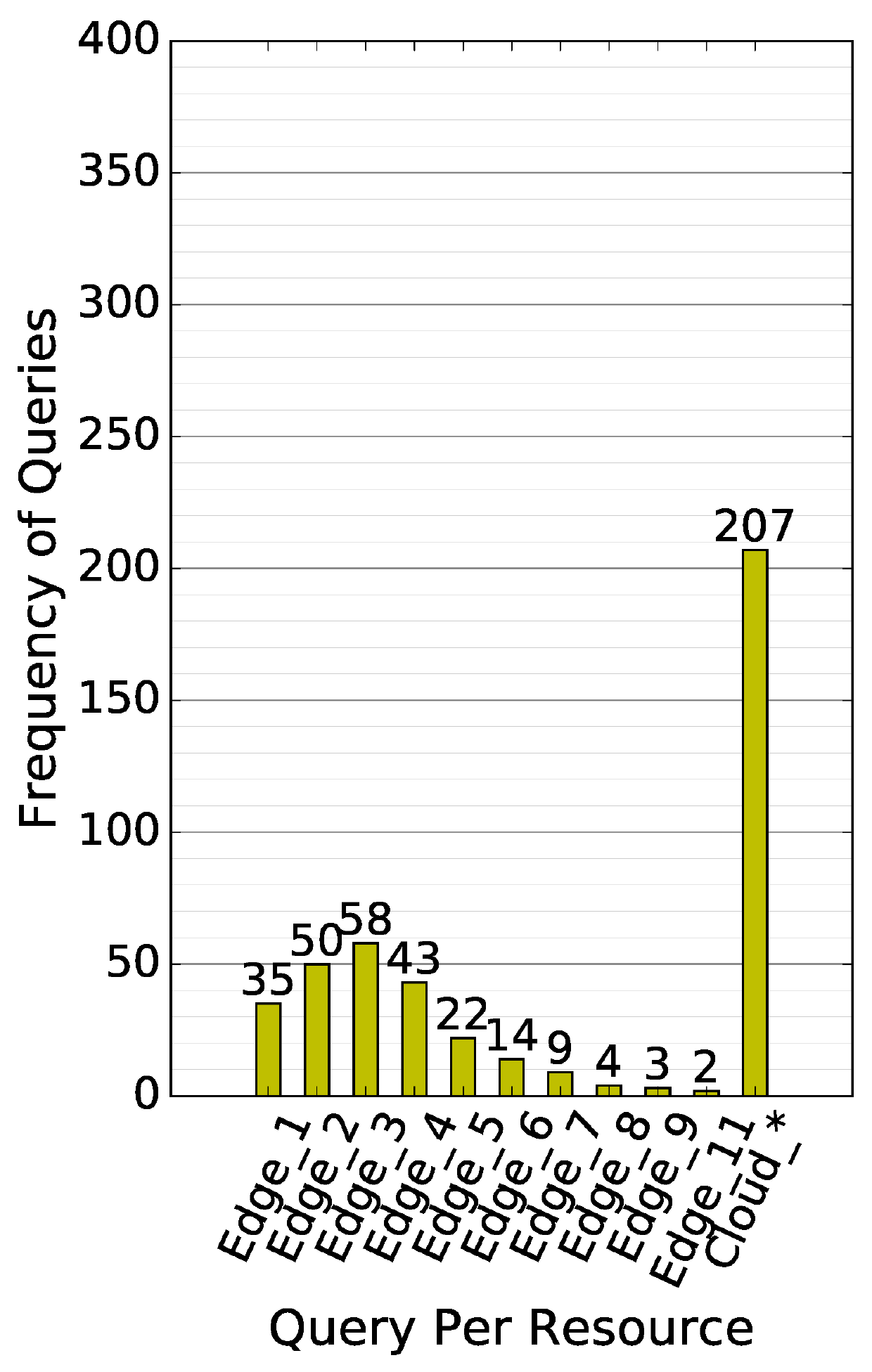}
		\label{fig:queries-edge:100:campus:cons}
	}
	\subfloat[$1000~e/sec$]{
		\includegraphics[width=0.225\textwidth]{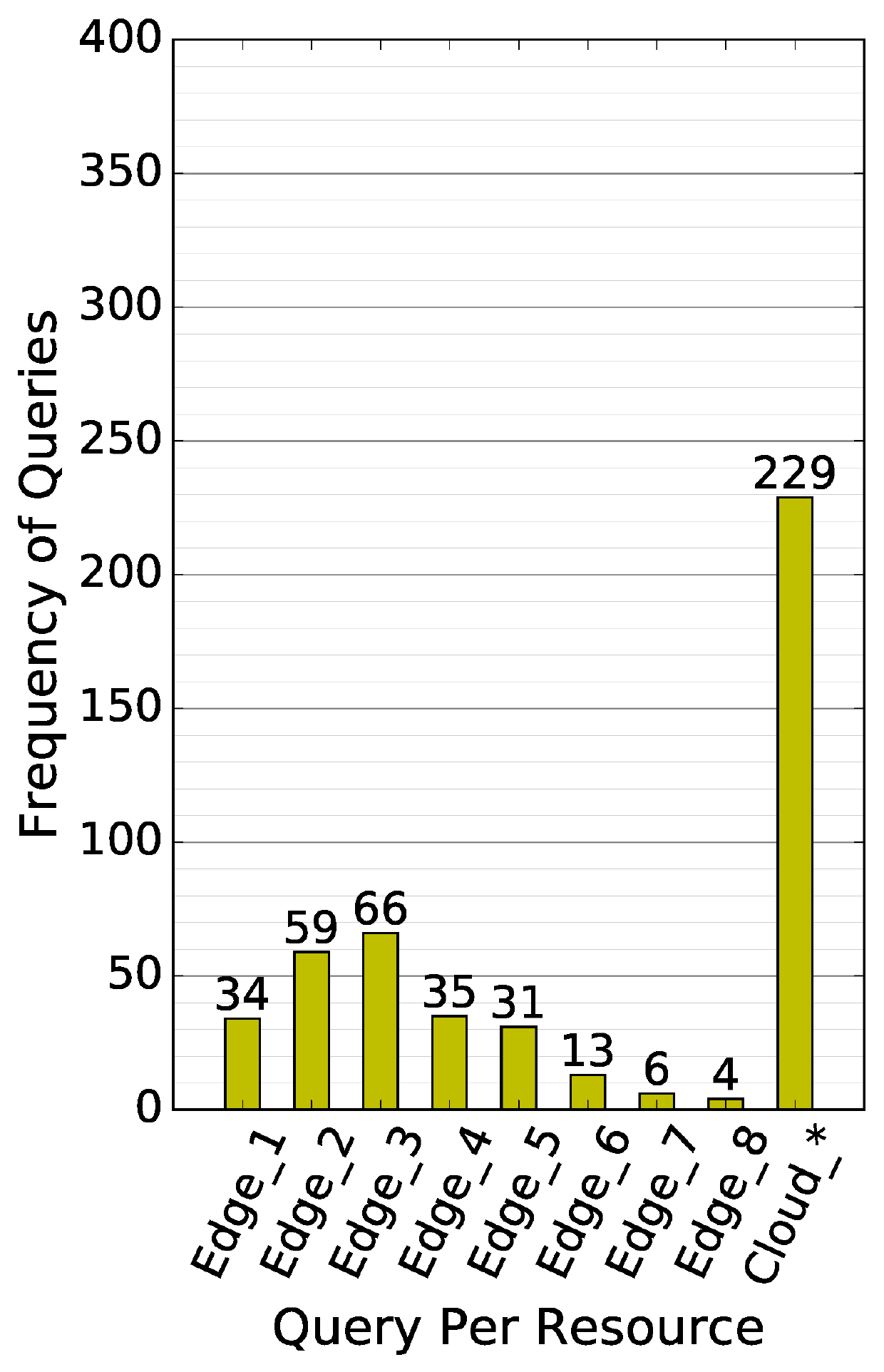}
		\label{fig:queries-edge:1000:campus:cons}
	}~
	\subfloat[$100~e/sec$]{
		\includegraphics[width=0.225\textwidth]{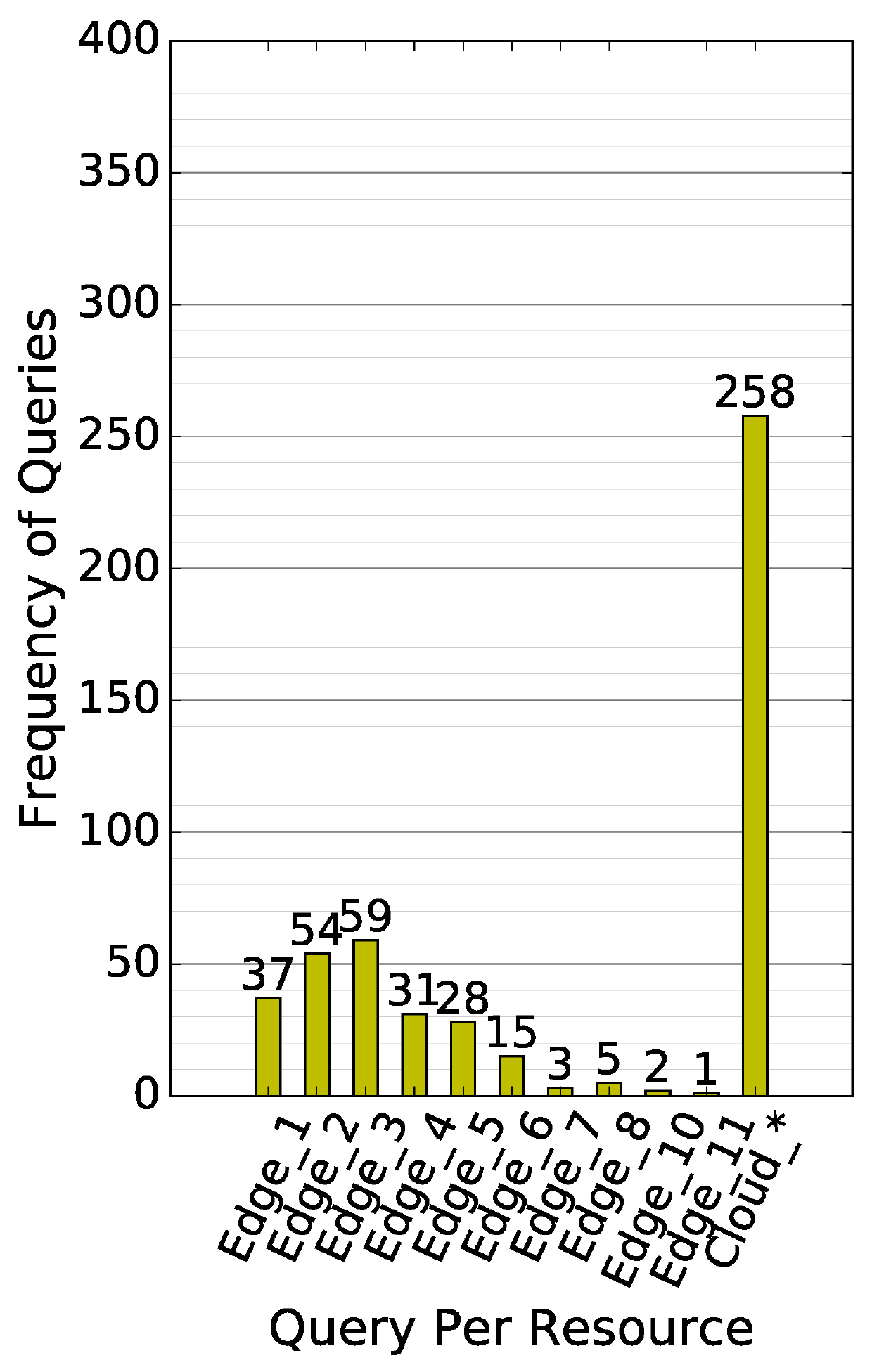}
		\label{fig:queries-edge:100:planetlab:cons}
	}
	\subfloat[$1000~e/sec$]{
		\includegraphics[width=0.225\textwidth]{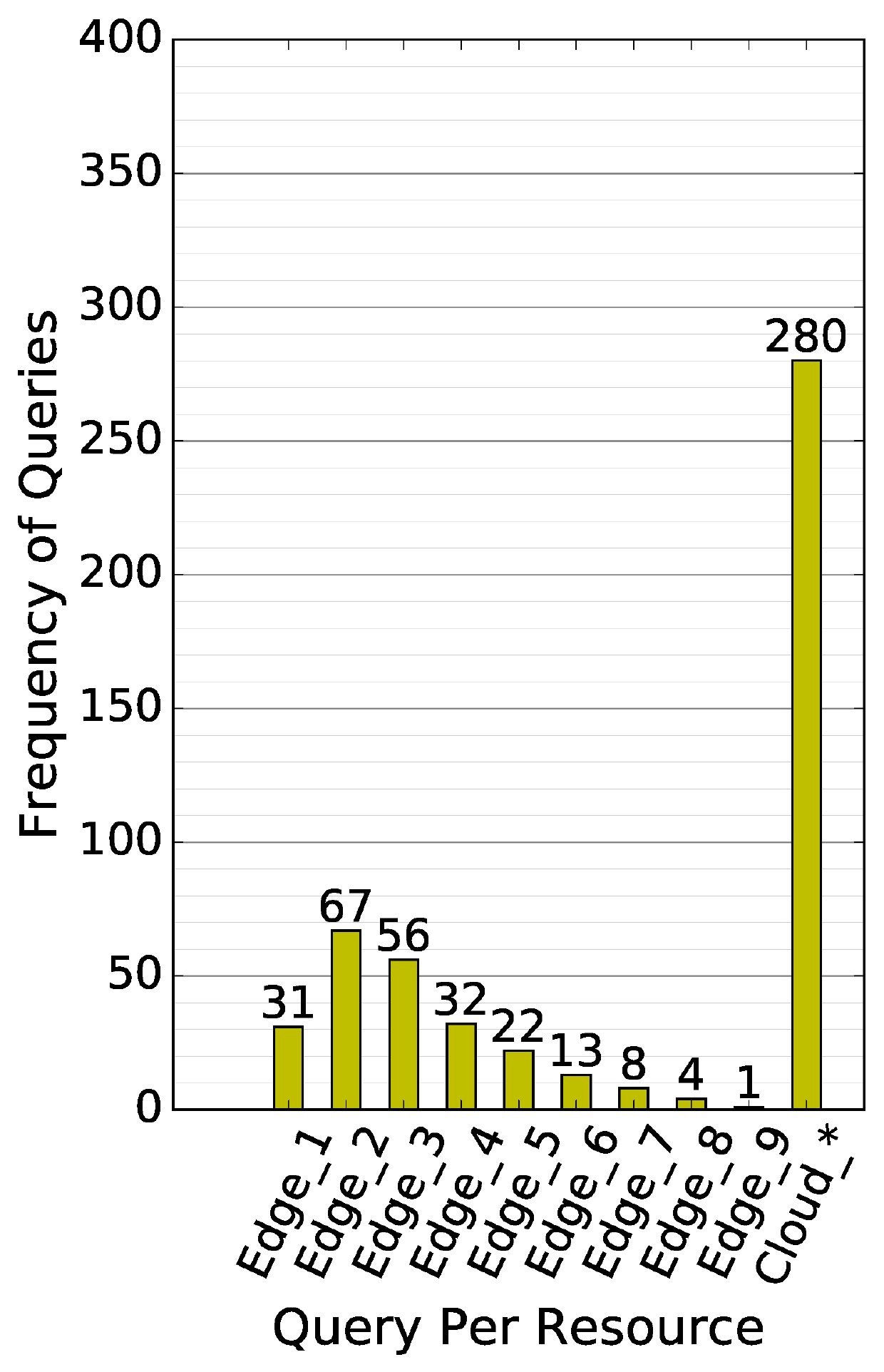}
		\label{fig:queries-edge:1000:planetlab:cons}
	}
	\subfloat{
		\footnotesize
		\begin{rotate}{270}\hspace{-1.5in}(iii)~\emph{Conservative}\end{rotate}
	}\\
	{\footnotesize \begin{tabular}{p{0.45\textwidth}p{0.45\textwidth}}\centering $\leftleftarrows$~\textsc{Campus LAN}~$\rightrightarrows$ & \centering $\leftleftarrows$~\textsc{PlanetLab WAN}~$\rightrightarrows$ \end{tabular}}\\
	
	\caption{Frequency of the number of queries placed on each resource by GA, using different rates, resource allocations, and network setups. [1026 queries per plot]}
	\label{fig:queries-edge}
\end{figure*}

\subsubsection{Query Occupancy on Resources using GA} Fig.~\ref{fig:queries-edge} shows histograms of the number of queries placed on edge and Cloud resources, across all the DAGs using GA. They give the frequency of queries (Y axis) present in edge devices hosting $1, 2, 3,$ \ldots, etc. queries and in the Cloud VM (X Axis), 
totalling to $1026$ queries 
across $45$ DAGs. 
Different input rates \emph{(a,b,c,d)} and resource availability \emph{(i,ii,iii)} for Campus and PlanetLab networks are shown.
We see that for the liberal case, for both network setups and event rates,  
a large fraction of queries ($> 300$) are present in exclusive edge devices, hosting just that query (\texttt{Edge\_1} in X Axis). About half as many queries are paired up on edge devices (\texttt{Edge\_2}), and this decreases sharply for $5-7$ queries present in the same edge. The Cloud VM (\texttt{Cloud\_*}) hosts about as many queries as \texttt{Edge\_2}. 

As we reduce the edge resource availability in the centrist and conservative setups, we see a gradual shift toward more queries placed in the same edge device and in the Cloud VM. With half the number of edges, the centrist setup almost has as many cases of edges having 1 query as 2 queries, and with a conservative setup, the edges with 2 and 3 queries dominate. PlanetLab typically has more queries per resource than the campus LAN since its network latency between edges is higher. So the increase in compute latency due to query collocation on a resource or moving to the VM is smaller than the cost of edge network latency.
We also see that the Cloud VM is assigned more queries, growing to $\approx 20\%$.  
As the number of edge resources decrease, there is a tendency to require more queries to be packed in fewer edge resources. When queries map to the same edge, the maximum rate supported by that edge for each additional query decreases, thus increasing the chance of throughput violations on the edge. Hence, this pushes more of the queries to the Cloud, until the VM violates. 

\subsubsection{Comparison with Baseline Approaches} 
In addition to the optimal BF solution, we consider the merits of GA with respect to a na\"{i}ve \emph{random placement} algorithm (RND) that maps queries randomly on any available edge or Cloud resources in each trial. If a solution is valid (i.e., does not violate constraints) and has a lower latency than a previous best trial, the best solution is updated to the current solution. This is repeated $15,000$ times, similar to the median number of GA generations. Another baseline we report and briefly discuss is a \emph{Cloud-only placement (CO)} approach.

As a cumulative measure of the relative latency quality, we define \emph{percentage latency deviation} of a ``worse'' solution $w$ over a ``better'' one $b$ for a set of $n$ DAGs as:
\begin{equation*}
\mathcal{E}_{b \rightarrow w} = \frac{\sum \limits_{i=1}^{n} (L'_i - L_i)}{n \times \overline{L_i}} \times 100\%
\end{equation*}
where $L'_i$ is the latency for DAG $i$ given by the ``worse'' solution such as GA, $L_i$ is the latency value for the DAG by the ``better'' solution such as BF, and $\overline{L_i}$ is the average latencies for the latter. Here, we consider only DAGs where solutions from both algorithms are valid. Smaller this value, closer the worse solution is to the better one.

A second evaluation measure we use is \emph{percentage invalid}, which reports the fraction of DAGs for which an algorithm was not able to converge to a \emph{valid solution}, that is, one that does not meet all the constraints. Here, the lack of convergence either means that the ``worse'' algorithm under-performs, or, that a feasible solution does not exist.

Lastly, \emph{percentage edge resources used} is the ratio between the number of edge devices on which queries are actually placed by the algorithm compared to the number of edge devices available. Here, a lower fraction means fewer Pi's are used, indicating a better capacity utilization of the \emph{active} devices having at least one query on them.

Table~\ref{tbl:quality} compares these quality metrics for the four approaches, BF optimal, GA, RND and CO. The latency deviation \% values compare BF with GA and RND, and GA with RND and CO, for their relative pairwise performance. In each pair, only DAGs for which valid solutions were available from both algorithms are considered. The invalid \% are evaluated for all $45$  DAGs in GA, RND and CO. The average of the edge used~\% over DAGs with valid solutions is also reported for each approach that uses the edge.


We observe that the latency deviation of both GA and RND for valid solutions is not far from the optimal solution, with under $4\%$ deviation ($\mathcal{E}_{BF\rightarrow GA}$ and $\mathcal{E}_{BF\rightarrow RND}$). 
GA outperforms both RND and CO consistently on latency, improving by $5-15\%$ ($\mathcal{E}_{GA\rightarrow RND}$) and $1.7-6\%$ ($\mathcal{E}_{GA\rightarrow CO}$).  We see that the latency benefits at $1000~e/sec$ is better than at $100~e/sec$ due to the increased difficultly in finding an optimal solution at higher rates.   
While the RND solution has the same fraction of invalid solutions with constraint violations as GA ($11-20\%$), using both edge and Cloud, the CO approach has $15-47\%$ of invalid solutions due to over-allocation of all queries in the DAG on the Cloud VM. 
Overall, GA out-performs RND and CO in terms of the solution quality.

\begin{table}[t]
	\centering
	\caption{Comparison of the quality of the different placement algorithms}
	\resizebox{!}{0.12\textwidth}
	{
		\begin{tabular}{R{0.9cm} R{1.3cm} || R{1cm}R{1cm}R{1cm}R{1.38cm} | R{0.9cm}R{0.9cm}R{0.9cm} | R{0.9cm}R{0.9cm}R{1cm} | R{.9cm}R{.9cm}}
			\hline
			{\bf In Rate} & {\bf Re\-sources} & \multicolumn{4}{c|}{\bf Latency Deviation \%} & \multicolumn{3}{c|}{\bf Invalid\%} & \multicolumn{5}{c}{\bf Avg. Edge Resources Used\%}\\
			\hline
			{\em (e/sec)} & {\bf} & \multicolumn{2}{l}{$\mathcal{E}_{BF\rightarrow GA}$*} & \multicolumn{2}{l|}{$\mathcal{E}_{GA\rightarrow RND}$} & {\em GA} & {\em RND} & {\em CO} & {\em BF*}  & {\em GA*} & {\em RND*} & {\em GA} & {\em RND} \\ 
			&  &  & \multicolumn{2}{l}{$\mathcal{E}_{BF\rightarrow RND}$*} & $\mathcal{E}_{GA\rightarrow CO}$ & &&&&&&& \\ 
			\hline
			\hline
			{$100$} & {liberal} & {$2.92$} & {$3.87$} & {$5.03$} & {$1.92$} & {$0.00$} & {$0.00$} & {$15.56$} & {$41.51$}& {$45.40$} & {$47.87$} & {$55.53$} & {$58.65$}
			\\ \hline
			{$1000$} & {liberal} & {$2.86$} & {$3.20$} & {$15.20$} & {$5.89$} & {$11.11$} & {$11.11$} & {$46.67$} & {$43.10$} & {$45.16$} & {$46.92$} & {$53.83$} & {$54.62$}
			\\ \midrule[0.5pt]
			{$100$} & {centrist} & {$0.69$} & {$0.98$} & {$7.051$} & {$1.75$} & {$0.00$} & {$0.00$} & {$15.56$} & {$61.33$} & {$63.37$} & {$64.73$} & {$72.45$} & {$74.21$}
			\\ \hline
			{$1000$} & {centrist} & {$1.23$} & {$2.40$} & {$13.68$} & {$5.63$} & {$13.33$} & {$13.33$} & {$46.67$} & {$66.61$} & {$69.33$} & {$71.17$} & {$78.80$} & {$80.20$}
			\\ \midrule[0.5pt]
			{$100$} & {conserv.} & {$0.00$} & {$0.16$} & {$7.42$} & {$1.71$} & {$0.00$} & {$0.00$} & {$15.56$} & {$78.57$} & {$78.57$} & {$79.94$} & {$87.59$} & {$88.44$}
			\\ \hline
			{$1000$} & {conserv.} & {$0.00$} & {$0.00$} & {$4.66$} & {$6.01$} & {$20.00$} & {$20.00$} & {$46.67$} & {$82.54$} & {$82.54$} & {$82.54$} & {$90.26$} & {$90.31$}
			\\ \hline
			\multicolumn{12}{l}{*These are calculated for DAGs size upto 12 to compare with available BF solutions.}
		\end{tabular}}
		\label{tbl:quality}
	\end{table}

We also see from Table~\ref{tbl:quality} that the edge resources used~\% for the GA placement solutions are closer to the BF solution for the initial 12 DAGs, and it is also consistently smaller than the RND approach in all cases. A lower value is better here as it indicates a prudent use of available edge resources, allowing idle devices to be turned off. As expected, the edge usage grows as we move from liberal to conservative setups. 

\subsubsection{Alternative Configurations}
While we skip a detailed discussion, we report results from experiments with additional configurations. 
Using $\langle C_k = 8600, \tau_k = 12 \rangle$ and $\langle C_k = 17200, \tau_k = 24\rangle$ as \emph{battery capacities and recharge cycles} both show that GA has latencies that are within $3\%$ of BF for liberal, within $1\%$ for centrist and always converges to BF in conservative case. Further, relative to roulette wheel, both \emph{rank and tournament selections in GA} offer an improvement of $<0.6\%$ in latency for the conservative scenario and a better reduction $1.1\%$ with $1000~e/sec$ for liberal scenario. This is due to reduced overcrowding of chromosomes that can cause pre-mature convergence. There is no change in the fraction of invalid solutions. But for centrist setup, there is $>2\%$ increase in latency with rank and $>1\%$ with tournament method with $100~e/s$. A similar trend also follows for $1000~e/s$ though with $<1\%$ increase. Hence, it is worth examining alternate resource capacities and selection strategies in practice.

\begin{figure*}[t]
	\centering
	\subfloat[Max. \% increase in input rate, without violation (Campus)]{
		~\includegraphics[width=0.5\textwidth]{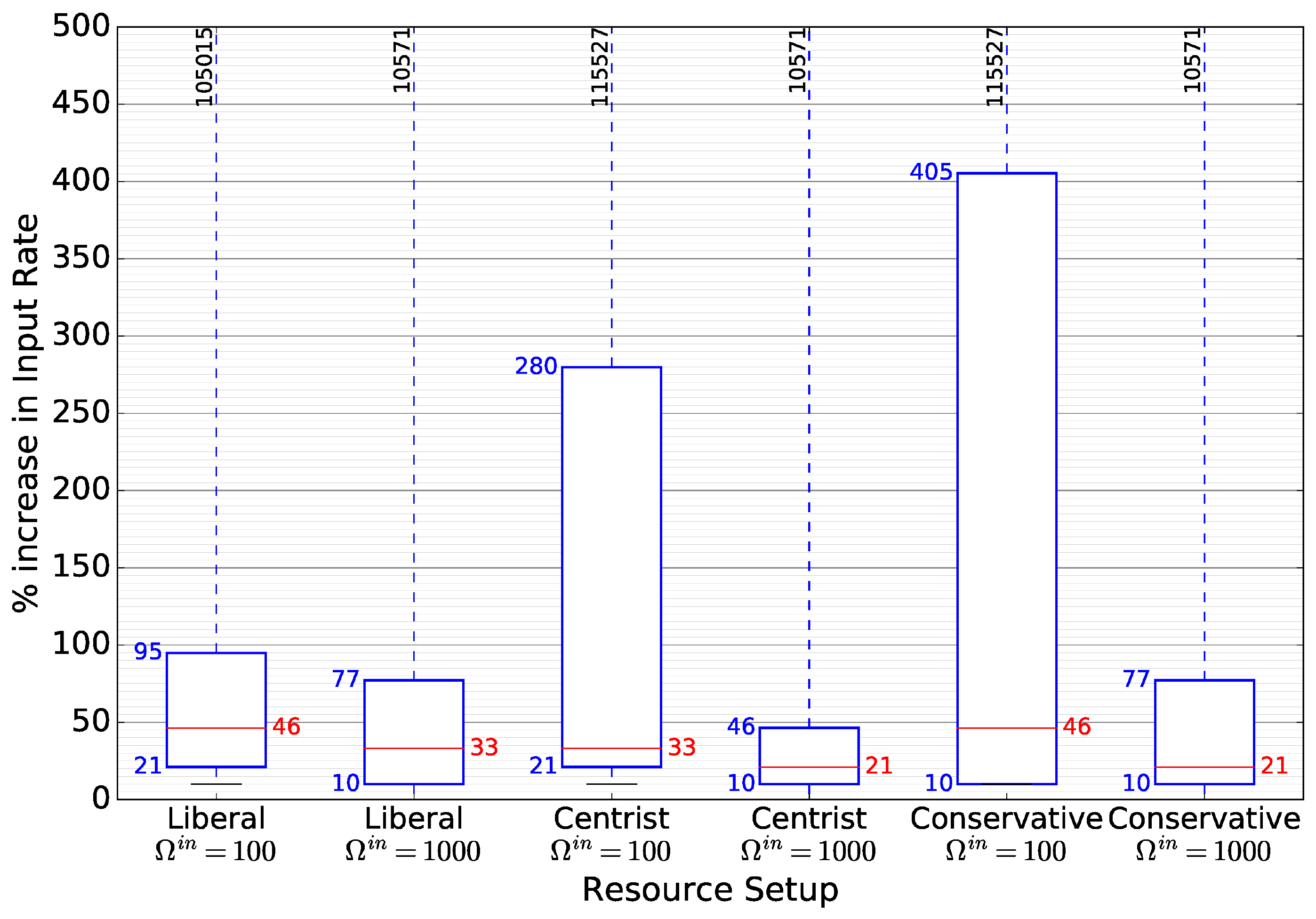}~
		\label{fig:increaseRate:steps}
	}\enskip
	\subfloat[Observed vs. Expected time (Campus)]{
		~\includegraphics[width=0.40\textwidth]{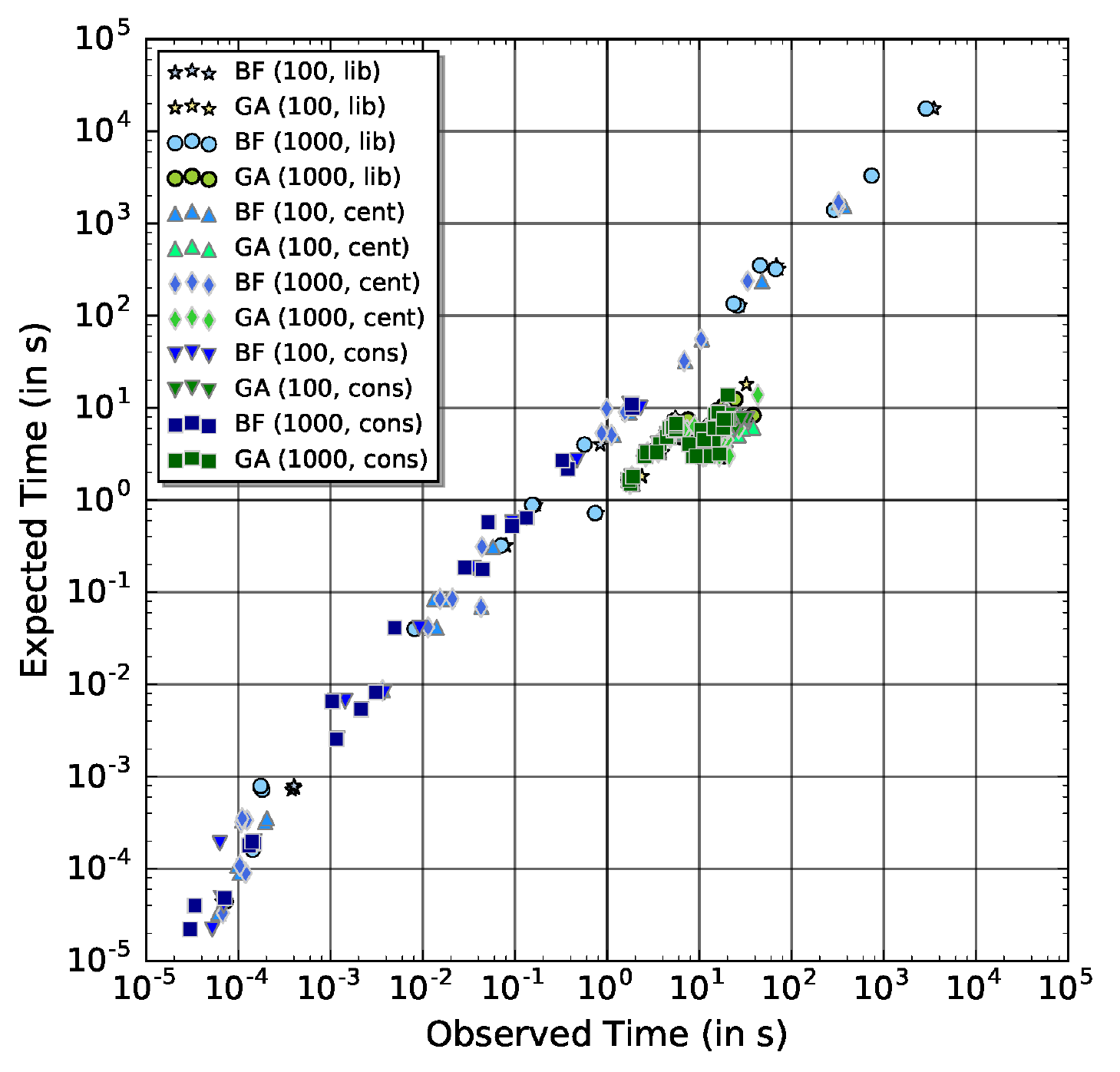}~
		\label{fig:complexity}
	}
	\caption{\emph{Max. \% Increase in rate (a)} for a solution, and \emph{Time complexity (b)} for GA \& BF}
	\label{fig:complexity-ratexp}
\end{figure*}

We also consider the viability of a GA solution under \emph{dynamic input rates}. Fig~\ref{fig:increaseRate:steps} shows the \% increase in input rate to a DAG above a base rate, before the GA placement solved for the base rate violates constraints. For a given solution, the end-to-end latency is not affected by the input rate increase but energy or compute capacity violations may occur. We see that a rate increase of $10\%$ causes violations in $<25\%$ of the DAGs. $75\%$ of the DAGs fail using the original GA solution when the rate increases by $46-405\%$, depending on the setup. This indicates the head-room available for a given GA solution before it has to be updated when the input rate increases.

\subsubsection{Time Complexity of the Solutions}


Fig.~\ref{fig:complexity} shows a scatter plot of the wall clock times to solve various DAGs against the asymptotic time complexity for BF and GA given earlier. 
We multiply the complexity by a constant derived using linear regression. 
BF time till DAG size $12$ that could complete and GA time for all $45$ DAGs are shown.

In this Log-Log plot, we clearly see a strong linear correlation between the expected and observed times for both algorithms. BF's runtime has a wide range, spanning $10\mu s - 10,000$'s of seconds for DAG with $4-12$ queries. 
%
The GA runtimes for all DAGs with $4-50$ queries are tightly clustered at $1-26~sec$. The complexity for GA is proportional to the DAG size and the number of generations. For example, while DAGs \texttt{20\_1\_2} and \texttt{50\_1\_2} take the same number of generations to converge for $100~e/sec$ rate with liberal resources, the wall clock time for the latter is $3\times$ higher than the former. 


We see, as expected, variations in the runtimes for DAGs with the same size but different numbers of source and sink vertices -- these are pre-pinned to specific edge and Cloud resources, respectively, reducing the number of queries to schedule. 
The time taken for centrist and conservative resources is also much lower than the liberal ones since their search space is significantly smaller due to the number of available edge resources halving at each resource scenario.

BF has time complexity that is exponential with the number of resources. For DAGs of size $12$ with one source, its time for a liberal case is $48~mins - 11.8~hours$; it takes minutes for centrist and seconds for conservative. For a DAG size of $14$, BF will take $\approx 107$ days using liberal resources (which we could not complete), while it takes $13~mins$ for conservative and $4~mins$ for centrist.
GA takes $< 26~secs$ for all cases we examine -- DAG sizes till 50 and up to 50 resources, and with runtime matching the expected time complexity. This makes it generally usable for diverse configurations. However, since we run GA for at least $15,000$ iterations, BF can be faster than GA when fewer resources are available and for smaller DAGs. In such cases, BF is a feasible option.

\section{Discussion}
\label{sec:discuss}
In any emerging area that sees the confluence on multiple technologies such as Internet of Things, Cloud computing, Big Data, and mobile platforms, addressing specific research problems opens the door for more such problems and opportunities that exist. Here, we summarize our key findings, and highlight a host of future experiments and research that arise from this study.

\subsection{Key Outcomes}
There are several key takeaways from this article, beyond what was discussed in the analysis sections, that have a broader impact beyond just the immediate problem we address.

\textbf{Edge and Cloud.} We see that edge devices like the Pi perform $\frac{1}{3}^{rd}$ as well as similar Cloud VMs for CEP event analytics, and both have similar performance trends for different query categories. They are also cheaper than the Cloud in the long-term, if they are already deployed and available as part of IoT deployments. 

However, Clouds are still useful when we consider aggregation across many streams, or from edge devices that span private networks, and when the throughput limits required are very high. The ability to have many cores in a single VM helps as well, but allowing in-memory communication between many queries present in the same VM. That said, we also see that the punitive cost in the end-to-end latency is the network latency between edge and Cloud, and its variability as well. So for highly time-sensitive applications, a much-closer data center or a private Cloud on the same network will be necessary. The bandwidth appears to be less of a concern, given the small event sizes which even cumulatively or at a high rate are tolerable.

We have used real-world, high-precision measurements of energy usage in our edge resources for a variety of event queries. This goes beyond current literature that limits itself to examining CPU, memory and network usage, which are poorer approximations of energy use. For e.g., in all cases, while the CPU and memory utilization by each query is stable at about $98\%$ (single core) 
and $3\%$, 
respectively, we see two discrete energy levels for the sequence-like and aggregate queries. This shows the importance of practical validation. 

\textbf{Scheduling Approaches.} Our benchmarks also show that the Pi consumes discrete levels of power for the different queries, that allows for predictable energy modeling. In fact, given the tight bounds of these consumption levels, we can approximate the power levels to just three categories: base load, filter-like queries, and aggregate queries. 

The GA meta-heuristic has shown to be robust and scalable in solving the non-linear optimization problem. It gives optimal or near-optimal solutions, where it is possible to compare against the optimal BF; gives results with low end-to-end latency values for larger problems; has a limited number of cases where it was unable to provide feasible solution (in one was indeed possible); and can be consistently solved within seconds. With the complexity plots showing a high correlation between expected and observed runtime, the GA holds promise for providing good placement solutions for much larger IoT deployments, on the order of thousands of resources and DAGs. 

That being said, for small scale IoT deployments with under ten resources and small DAGs, BF offers optimal solutions within a reasonable time and should be chosen. BF can also be complemented with techniques like Dynamic Programming to possible speed up the time, though it would not have a tangible impact on exponential time complexity. GA however offers the flexibility of trading-off runtime and solution quality. It can be limited to finding a solution within a fixed time budget, and the GA evolves for that duration and results the best solution seen that far.

\textbf{Supplementary benefits.} A variation of this simulation study is to estimate the least number of edge and Cloud resources required to support a certain number of streams and query workloads while meeting specific QoS required for the application. Such ``What if'' studies are crucial for emerging domains in IoT to better plan deployments that may take months and millions of dollars, and conserve the resources required for future workloads. 

\subsection{Future Experiments}
There are additional experiments and studies that can be considered to validate the proposed solution in a more diverse environment.

\begin{enumerate}
\item We do not consider multi-threaded or multi-core execution in our study. Siddhi supports a limited for of multi-threading, which we did not leverage in our study to simplify the experiments, and one can always run multiple copies of Siddhi on independent cores. Given the prevalence multi-core CPUs in even edge devices, this needs to be considered. This will make edge devices like the Pi even more favorable. 

\item While our benchmarks considered the most common CEP patterns, these could be complemented with a wider variety of sample queries from among these types. Using queries from real-world deployments will also make the workloads more representative. Also, while our benchmarks considered pattern queries with low selectivity, these were not used in our synthetic DAGs since they produced DAGs with very low output rates and selectivities. A special class of DAGs including such low-rate queries can be considered. In a similar vein, we should also consider a wider variety event types, with different payloads. Our prior work on benchmarking for distributed stream processing systems offers some possibilities~\cite{shukla-tpctc}.  

\item Our benchmarks do not consider the energy cost for the network (LAN/WLAN) transmission. In this work current drawn by Pi is measured only for running Siddhi queries on the Pi, with events being generated locally. For a more realistic scenario, the energy cost for both the wireless and the LAN interface need to be measured, and included in the simulation study.  

\item We have seen scenarios with just two types of networks, private campus and public Cloud. However, even within these networks, there is bound to be variability. Edge devices (or VMs) could be at different parts of the topology in the private network (or the data center), and the latency costs may be different due to multiple switches coming in the way. While we used a high speed uplink from campus to the public Internet, the network behavior from edge to Cloud may be different when using a home broadband or cellphone carriers. There is also a growth in the number of Cloud data centers with, for e.g., three new Azure data centers coming online in India as we are writing this article. Choosing the best/nearest data center when having a multi-city deployment, and the network costs between the data centers need attention too. 

\item Even within edge devices, we have considered the Raspberry Pi 2 Model B, and there exist newer/faster models like the Pi 3, embedded versions like the Pi Zero, and other DIY platforms like the Arduino, Intel Edison, etc. While our current work targets platforms that run Linux and Java for the analytics platforms, these concepts can also be extended to more constrained devices and recent platforms. Clouds offer different VM favors as well that could be considered.

\item It would be useful to understand the appropriate mix of the different numbers of edge and Cloud resources, e.g. more Cloud VMs, fewer edges, fixed number of edge devices for different DAGs, etc. These would offer better insight on the resource usage by solutions from the optimization solver.

\item While we have considered two input rates for our simulation study, it would be useful to observe the impact on latency, infeasible solutions generated and resource usage as we increase the rate to higher levels as well.

\item We have used real-world benchmarks to drive the simulation study. However, to offer even higher guarantees of the practical viability and relevance of our work, the placement solutions obtained from the optimization solvers should be tested with real life deployments having  multiple edge devices and the Cloud. This will be yet another stepping stone toward translating research into practice for  analytics across edge and Cloud for IoT.

\end{enumerate}

\subsection{Future Research}
There are several promising research avenues to explore further in this emerging area of event analytics across edge and Cloud.
\begin{enumerate}
\item Resource usage was not a primary consideration for our problem definition, even though we reported the resource usage on the edge for different solutions. While the constraints ensured that solutions were limited to the full compute and energy capacity of a resource, our optimization goals did not consider resource usage across devices -- whether to ensure the utilization of the edge was balanced (e.g. to ensure no single Pi is overloaded, and they all drain their battery at the same rate), or to ensure that the utilization on edge and/or Cloud was high for active devices/VMs (e.g. to ensure we get full value for VMs that are paid for, or to allow some inactive edge devices to be turned-off or duty cycled if others with high utilization can take the workload).

\item This article considers the problem of scheduling a single DAG on to the edge and Cloud that are fully available. However, practical situations have DAGs that may arrive periodically, or exit after a few days or weeks. There may also be analytics from multiple domains that share the same IoT fabric and devices.  
In such cases it is required to place multiple DAGs on to the same set of resources, or place a DAG on edge and Cloud resources that only have partial capacities available. Knowing the entry and exit schedules of the DAGs will also better inform us as to plan for future submissions or capacity availability. 

\item Model VM cost into the equation. In our work VM cost has not been included, but this cost may become significant as we increase the number of VMs. This calls for conservative use of VMs and keep queries on edge as much as possible.

\item Our problem dealt with input rates that arrive at a constant rate, and this is reasonable since many sensors are deployed to generate events at a constant sampling interval. The impact of input rate variability on the optimization solution was not considered. While we observe that there is not a lot of variability in the energy usage for different event rates, it may be that changing the input rate to the DAG will cause different solutions to be generated. Given the long-running nature of the event dataflows, the impact of variable input rates or periodic changes to the input rates on the solutions that are generated should be considered. 

More generally, we assume a fixed set of edge and Cloud resources in our problem, and a single solution that is deployed when the DAG is submitted. Since such event analytics run for days or weeks at a time, many factors may change in this period: event rates may change significantly, edge devices may fail or be taken down for planned maintenance, solar energy generation may be lower due to a cloudy day causing edges with longer recharge cycle, network behavior may vary, and connectivity between edge-edge or edge-Cloud may go down all together, and so on. So we should consider our ability to change the solution on the fly as the environmental conditions change, and also to provide robustness to guarantee the latency QoS. Additional strategies to consider may include dynamically moving tasks between edge and Cloud or vice versa, replicating the queries across multiple devices, etc.

\item Lastly, one aspect that we had introduced briefly in an earlier work and remains relevant still is that of planning placement of queries to preserve the privacy of data~\cite{Nithya14}. IoT deployments offer an  unprecedented ability to observe the environment around us. As a result, some of the sensor streams on which we perform analytics may contain sensitive information that would be embarrassing or illegal if compromised~\cite{uber:2012,weber:2010}. Incorporating privacy constraints as a first-class entity in the placement problem across edge and Cloud is important. Here, we may wish to limit the event streams that go out of the private network, have variable trust in different edge resources, or introduce ``anonymizing'' queries at the trust boundaries. This is a vast and important area that requires exploration.

\end{enumerate}

\section{Conclusion}
\label{sec:conclusion}
Existing literature on using edge and Cloud resources focus on a few application quality parameters, such as latency and throughput; limited system characteristics like CPU, network and power; or specific architectures such as Cloud-only, Mobile-Cloud and Fog Computing. 
In this paper, we have identified a unique combination of these dimensions essential for IoT: reducing latency for streaming dataflows across edge and Cloud, while conserving energy and bound by the compute capabilities of the devices.  

Our micro-benchmark results offer a novel glimpse on the compute, network and energy performance of edge devices and Cloud VMs for individual CEP queries. The diverse experiments with different query types and event rates offer a broad set of performance distributions that are valuable to evaluate other resource platforms for event analytics, and for realistic simulation studies, as we have described.

We formulate the query placement problem for a CEP dataflow on to edge and Cloud resources as an optimization problem, with constraints on the compute and energy capabilities of the resource based on realistic IoT deployments. We propose a Brute Force (BF) approach to solving it optimally, and also map this problem to a Genetic Algorithm (GA) pattern that is solved while considering the constraints. 

We validate and evaluate the problem and solution approaches using a comprehensive simulation study that includes a diverse set of synthetic DAGs that are embedded with static and runtime properties that are sourced from the real-world distributions. We have obtained results for 45 DAGs having 17 query types, with 3 resource configurations, 2 input rates, 2 network setups, using the BF and GA approaches and a random and a Cloud-only baselines. Sampling the parameter values from the benchmark distributions during these simulations mimics variability of real IoT deployments.

Our analysis shows that GA gives optimal or near-optimal solutions comparable to BF, and provides a better trade-off between lower latency and more frequent feasible solutions than the random or Cloud-only placement baselines. It also offers solutions within seconds for even DAGs as large as $50$ queries on $50$ edge and Cloud resources, while the BF takes $12~hours$ for a $12$-query DAG using liberal resources, and weeks for larger DAGs. GA also tells users of the resource constraint that caused a solution to be infeasible, helping with capacity planning. $75\%$ of the DAGs can also withstand a $10\%$ increase in input rate without a constraint violation. These are promising results that can inform practical IoT deployments using sound theory and experimental results.

In future, these experiments can be enhanced by considering network energy costs, more diverse edge/Cloud resources and networks, additional input rates, and variable ratios of edge devices and Cloud VMs. It will also help to actually deploy and validate the solution. There is also a swathe of new research ideas to pursue in this nascent area. Defining placement problems to collocate multiple DAGs on the same set of resources will allow multiple users and domains to share the same set of IoT resources. Finding the minimum number of edge resources and VMs required for a given workload, with associated costs, becomes a related problem to plan IoT installations. Temporal dynamism in all respects -- input rates event, availability of edges, performance of the network, period between battery recharge -- requires sustained attention and improved heuristics. Network links and (edge) resources may suffer from transient failures, requiring fault-resilience. Variable-rate events generated from probability distributions such as Poisson or Zipfian, or realistic IoT data streams can be used for empirical validation~\cite{Akdere:2008,shukla-tpctc}. Lastly, a key research problem is to design approximation algorithms with stronger quality guarantees and formal bounds than the GA heuristic, while still offering a practically usable computational complexity that is validated with realistic DAGs.


\section{Acknowledgments}
\label{sec:ack}
We wish to thank several students, interns and staff members of the DREAM:Lab, CDS, IISc, who helped with various aspects of the problem formulation, experiments and discussions, including Nithyashri Govindarajan, Shashank Shekhar, Pranav Konanur and Siva Prakash Reddy Komma. We also thank Dr. T.V. Prabhakar from the Department of Electronic Systems Engineering (DESE) at IISc, and his staff, Madhusudan Koppal and Abhirami Sampath, for enabling the energy measurements for the Raspberry Pi devices. We also thank our research sponsors, Robert Bosch Center for Cyber Physical Systems (RBCCPS) at IISc and the Ministry of Electronics and Information Technology (MeitY). We acknowledge Microsoft for their Azure for Research grant that provided us with the Cloud resources for our experiments.

\bibliographystyle{IEEEtran}
\bibliography{main}

\begin{thebibliography}{10}
\providecommand{\url}[1]{#1}
\csname url@samestyle\endcsname
\providecommand{\newblock}{\relax}
\providecommand{\bibinfo}[2]{#2}
\providecommand{\BIBentrySTDinterwordspacing}{\spaceskip=0pt\relax}
\providecommand{\BIBentryALTinterwordstretchfactor}{4}
\providecommand{\BIBentryALTinterwordspacing}{\spaceskip=\fontdimen2\font plus
\BIBentryALTinterwordstretchfactor\fontdimen3\font minus
  \fontdimen4\font\relax}
\providecommand{\BIBforeignlanguage}[2]{{%
\expandafter\ifx\csname l@#1\endcsname\relax
\typeout{** WARNING: IEEEtran.bst: No hyphenation pattern has been}%
\typeout{** loaded for the language `#1'. Using the pattern for}%
\typeout{** the default language instead.}%
\else
\language=\csname l@#1\endcsname
\fi
#2}}
\providecommand{\BIBdecl}{\relax}
\BIBdecl

\bibitem{perera2014sensing}
C.~Perera, A.~Zaslavsky, P.~Christen, and D.~Georgakopoulos, ``Sensing as a
  service model for smart cities supported by internet of things,''
  \emph{Trans. Emerg. Telecommun. Technol.}, vol.~25, no.~1, pp. 81--93, Jan.
  2014.

\bibitem{smart-health}
J.~Wei, ``How wearables intersect with the cloud and the internet of things :
  Considerations for the developers of wearables.'' \emph{IEEE Consumer
  Electronics Magazine}, vol.~3, no.~3, pp. 53--56, July 2014.

\bibitem{simmhan:cise:2012}
Y.~Simmhan, V.~Prasanna, S.~Aman, A.~Kumbhare, R.~Liu, S.~Stevens, and Q.~Zhao,
  ``Cloud-based software platform for big data analytics in smart grids,''
  \emph{Computing in Science and Engg.}, vol.~15, no.~4, pp. 38--47, Jul. 2013.

\bibitem{siddhi11}
S.~Suhothayan, K.~Gajasinghe, I.~Loku~Narangoda, S.~Chaturanga, S.~Perera, and
  V.~Nanayakkara, ``Siddhi: A second look at complex event processing
  architectures,'' in \emph{Workshop on Gateway Computing Environments}.\hskip
  1em plus 0.5em minus 0.4em\relax NY, USA: ACM, 2011, pp. 43--50.

\bibitem{Ahmad:2004}
Y.~Ahmad and U.~\c{C}etintemel, ``Network-aware query processing for
  stream-based applications,'' in \emph{International Conference on Very Large
  Data Bases}, ser. VLDB.\hskip 1em plus 0.5em minus 0.4em\relax Toronto,
  Canada: VLDB Endowment, 2004, pp. 456--467.

\bibitem{storm-twitter}
A.~Toshniwal, S.~Taneja, A.~Shukla, K.~Ramasamy, J.~M. Patel, S.~Kulkarni,
  J.~Jackson, K.~Gade, M.~Fu, J.~Donham, N.~Bhagat, S.~Mittal, and D.~Ryaboy,
  ``Storm@twitter,'' in \emph{ACM International Conference on Management of
  Data}, ser. SIGMOD.\hskip 1em plus 0.5em minus 0.4em\relax NY, USA: ACM,
  2014, pp. 147--156.

\bibitem{cep-survey}
G.~Cugola and A.~Margara, ``Processing flows of information: From data stream
  to complex event processing,'' \emph{ACM Comput. Surv.}, vol.~44, no.~3, pp.
  15:1--15:62, Jun. 2012.

\bibitem{debs-challenge-soccer}
C.~Mutschler, H.~Ziekow, and Z.~Jerzak, ``The debs grand challenge,'' in
  \emph{International Conference on Distributed Event-based Systems}, ser.
  DEBS.\hskip 1em plus 0.5em minus 0.4em\relax NY: ACM, 2013, pp. 289--294.

\bibitem{Hirzel12}
M.~Hirzel, ``Partition and compose: Parallel complex event processing,'' in
  \emph{ACM International Conference on Distributed Event-Based Systems}, ser.
  DEBS.\hskip 1em plus 0.5em minus 0.4em\relax NY, USA: ACM, 2012, pp.
  191--200.

\bibitem{Woods10}
L.~Woods, J.~Teubner, and G.~Alonso, ``Complex event detection at wire speed
  with fpgas,'' \emph{Proc. VLDB Endow.}, vol.~3, no. 1-2, pp. 660--669, Sep.
  2010.

\bibitem{clonecloud11}
B.-G. Chun, S.~Ihm, P.~Maniatis, M.~Naik, and A.~Patti, ``Clonecloud: Elastic
  execution between mobile device and cloud,'' in \emph{Conference on Computer
  Systems}, ser. EuroSys.\hskip 1em plus 0.5em minus 0.4em\relax NY, USA: ACM,
  2011, pp. 301--314.

\bibitem{mobile:fernando:2013}
N.~Fernando, S.~W. Loke, and W.~Rahayu, ``Mobile cloud computing,''
  \emph{Future Gener. Comput. Syst.}, vol.~29, no.~1, pp. 84--106, Jan. 2013.

\bibitem{Shi12}
C.~Shi, V.~Lakafosis, M.~H. Ammar, and E.~W. Zegura, ``Serendipity: Enabling
  remote computing among intermittently connected mobile devices,'' in
  \emph{ACM International Symposium on Mobile Ad Hoc Networking and Computing},
  ser. MobiHoc.\hskip 1em plus 0.5em minus 0.4em\relax NY, USA: ACM, 2012, pp.
  145--154.

\bibitem{varshney:icfec:2017}
P.~Varshney and Y.~Simmhan, ``Demystifying fog computing: Characterizing
  architectures, applications and abstractions,'' in \emph{IEEE International
  Conference on Fog and Edge Computing}.\hskip 1em plus 0.5em minus 0.4em\relax
  New York, NY: {IEEE} Computer Society, 2017, pp. 115--124.

\bibitem{p2p:record:2003}
K.~Aberer, P.~Cudr{\'e}-Mauroux, A.~Datta, Z.~Despotovic, M.~Hauswirth,
  M.~Punceva, and R.~Schmidt, ``P-grid: A self-organizing structured p2p
  system,'' \emph{SIGMOD Rec.}, vol.~32, no.~3, pp. 29--33, Sep. 2003.

\bibitem{Srivastava05}
U.~Srivastava, K.~Munagala, and J.~Widom, ``Operator placement for in-network
  stream query processing,'' in \emph{ACM Symposium on Principles of Database
  Systems}, ser. PODS.\hskip 1em plus 0.5em minus 0.4em\relax NY, USA: ACM,
  2005, pp. 250--258.

\bibitem{Nithya14}
N.~Govindarajan, Y.~Simmhan, N.~Jamadagni, and P.~Misra, ``Event processing
  across edge and the cloud for internet of things applications,'' in
  \emph{International Conference on Management of Data}, ser. COMAD.\hskip 1em
  plus 0.5em minus 0.4em\relax Mumbai, India: Computer Society of India, 2014,
  pp. 101--104.

\bibitem{smartx}
\BIBentryALTinterwordspacing
{SmartX}, ``Iisc smart campus: Closing the loop from network to knowledge,''
  2016. [Online]. Available: \url{http://smartx.cds.iisc.ac.in}
\BIBentrySTDinterwordspacing

\bibitem{usc-smart-grid}
Q.~Zhou, Y.~Simmhan, and V.~Prasanna, ``Incorporating semantic knowledge into
  dynamic data processing for smart power grids,'' in \emph{International
  Semantic Web Conference}, ser. ISWC.\hskip 1em plus 0.5em minus 0.4em\relax
  Berlin, Heidelberg: Springer, 2012, pp. 257--273.

\bibitem{cloudlet:satyanarayanan:2009}
M.~Satyanarayanan, P.~Bahl, R.~Caceres, and N.~Davies, ``The case for vm-based
  cloudlets in mobile computing,'' \emph{IEEE Pervasive Computing}, vol.~8,
  no.~4, pp. 14--23, Oct. 2009.

\bibitem{Yang14}
S.~Yang, D.~Kwon, H.~Yi, Y.~Cho, Y.~Kwon, and Y.~Paek, ``Techniques to minimize
  state transfer costs for dynamic execution offloading in mobile cloud
  computing,'' \emph{IEEE Transactions on Mobile Computing}, vol.~13, no.~11,
  pp. 2648--2660, Nov 2014.

\bibitem{yang12}
L.~Yang, J.~Cao, S.~Tang, T.~Li, and A.~T.~S. Chan, ``A framework for
  partitioning and execution of data stream applications in mobile cloud
  computing,'' in \emph{IEEE International Conference on Cloud Computing}, ser.
  CLOUD.\hskip 1em plus 0.5em minus 0.4em\relax Honolulu, HI, USA: IEEE, 2012,
  pp. 794--802.

\bibitem{Chandramouli13}
B.~Chandramouli, S.~Nath, and W.~Zhou, ``Supporting distributed feed-following
  apps over edge devices,'' \emph{Proc. VLDB Endowment}, vol.~6, no.~13, pp.
  1570--1581, Aug. 2013.

\bibitem{fog:bonomi:2014}
F.~Bonomi, R.~Milito, P.~Natarajan, and J.~Zhu, \emph{Fog Computing: A Platform
  for Internet of Things and Analytics}.\hskip 1em plus 0.5em minus 0.4em\relax
  Cham: Springer International Publishing, 2014, pp. 169--186.

\bibitem{fog:vaquero:2014}
L.~M. Vaquero and L.~Rodero-Merino, ``Finding your way in the fog: Towards a
  comprehensive definition of fog computing,'' \emph{SIGCOMM Comput. Commun.
  Rev.}, 2014.

\bibitem{Lesser04}
S.~Abdallah and V.~Lesser, ``Organization-based cooperative coalition
  formation,'' in \emph{IAT}, 2004.

\bibitem{Karrels13}
D.~R. Karrels, G.~L. Peterson, and B.~E. Mullins, ``Large-scale cooperative
  task distribution on peer-to-peer networks,'' \emph{Web Intelli. and Agent
  Sys.}, vol.~11, no.~1, pp. 67--79, Jan. 2013.

\bibitem{Gehrke03}
Y.~Yao and J.~Gehrke, ``Query processing in sensor networks,'' in
  \emph{Proceeding of the 1st Biennial Conference on Innovative Data Systems},
  ser. CIDR.\hskip 1em plus 0.5em minus 0.4em\relax Asilomar, CA, USA: CIDR,
  2003, pp. 233--244.

\bibitem{telegraphcq}
S.~Chandrasekaran, O.~Cooper, A.~Deshpande, M.~J. Franklin, J.~M. Hellerstein,
  W.~Hong, S.~Krishnamurthy, S.~R. Madden, F.~Reiss, and M.~A. Shah,
  ``Telegraphcq: Continuous dataflow processing,'' in \emph{SIGMOD}, 2003.

\bibitem{Nocedal06}
J.~Nocedal and S.~J. Wright, \emph{Numerical Optimization}, 2nd~ed.\hskip 1em
  plus 0.5em minus 0.4em\relax NY, USA: Springer, 2006.

\bibitem{Kwok99}
Y.-K. Kwok and I.~Ahmad, ``Static scheduling algorithms for allocating directed
  task graphs to multiprocessors,'' \emph{ACM Comput. Surv.}, vol.~31, no.~4,
  pp. 406--471, Dec. 1999.

\bibitem{Garey75}
M.~R. Garey and D.~S. Johnson, ``Complexity results for multiprocessor
  scheduling under resource constraints,'' \emph{SIAM Journal on Computing},
  vol.~4, no.~4, pp. 397--411, 1975.

\bibitem{Michalewicz96}
Z.~Michalewicz, \emph{Genetic Algorithms + Data Structures = Evolution
  Programs}, 3rd~ed.\hskip 1em plus 0.5em minus 0.4em\relax London, UK:
  Springer-Verlag, 1996.

\bibitem{Hogenboom:2009}
A.~Hogenboom, V.~Milea, F.~Frasincar, and U.~Kaymak, ``Rcq-ga: Rdf chain query
  optimization using genetic algorithms,'' in \emph{10th International
  Conference on Electronic Commerce and Web Technologies}.\hskip 1em plus 0.5em
  minus 0.4em\relax Berlin, Heidelberg: Springer, 2009, pp. 181--192.

\bibitem{Shukla:2015}
A.~Shukla, H.~M. Pandey, and D.~Mehrotra, ``Comparative review of selection
  techniques in genetic algorithm,'' in \emph{International Conference on
  Futuristic Trends on Computational Analysis and Knowledge Management}, ser.
  ABLAZE.\hskip 1em plus 0.5em minus 0.4em\relax Noida, India: IEEE, Feb 2015,
  pp. 515--519.

\bibitem{Steinbrunn1997}
M.~Steinbrunn, G.~Moerkotte, and A.~Kemper, ``Heuristic and randomized
  optimization for the join ordering problem,'' \emph{The VLDB Journal},
  vol.~6, no.~3, pp. 191--208, Aug. 1997.

\bibitem{ga-quality1}
P.~A. Diaz-Gomez and D.~F. Hougen, ``Three interconnected parameters for
  genetic algorithms,'' in \emph{Conference on Genetic and Evolutionary
  Computation}.\hskip 1em plus 0.5em minus 0.4em\relax NY, USA: ACM, 2009, pp.
  763--770.

\bibitem{ga-quality2}
R.~Hassan, B.~Cohanim, O.~De~Weck, and G.~Venter, ``A comparison of particle
  swarm optimization and the genetic algorithm,'' in
  \emph{AIAA/ASME/ASCE/AHS/ASC Structures, Structural Dynamics and Materials
  Conference}.\hskip 1em plus 0.5em minus 0.4em\relax USA: American Institute
  of Aeronautics and Astronautics, 2005, p. 1897.

\bibitem{RTRG}
R.~A. Shafik, B.~M. Al-Hashimi, and K.~Chakrabarty, ``Soft error-aware design
  optimization of low power and time-constrained embedded systems,'' in
  \emph{Conference on Design, Automation and Test in Europe}, ser. DATE.\hskip
  1em plus 0.5em minus 0.4em\relax Leuven, Belgium: EDAA, 2010, pp. 1462--1467.

\bibitem{PlanetLab}
B.~Chun, D.~Culler, T.~Roscoe, A.~Bavier, L.~Peterson, M.~Wawrzoniak, and
  M.~Bowman, ``Planetlab: An overlay testbed for broad-coverage services,''
  \emph{SIGCOMM Comput. Commun. Rev.}, vol.~33, no.~3, pp. 3--12, Jul. 2003.

\bibitem{e2e1}
R.~Zhu, B.~Liu, D.~Niu, Z.~Li, and H.~V. Zhao, ``Network latency estimation for
  personal devices: A matrix completion approach,'' \emph{IEEE/ACM Transactions
  on Networking}, vol.~25, no.~2, pp. 724--737, April 2017.

\bibitem{e2e2}
L.~Eyraud-Dubois and P.~Uznanski, ``{Bedibe: Datasets and Software Tools for
  Distributed Bandwidth Prediction},'' in \emph{{Rencontres Francophones sur
  les Aspects Algorithmiques des T{\'e}l{\'e}communications (AlgoTel)}}.\hskip
  1em plus 0.5em minus 0.4em\relax La Grande Motte, France: HAL Archives, May
  2012, pp. 1--5.

\bibitem{wisc}
K.~He, A.~Fisher, L.~Wang, A.~Gember, A.~Akella, and T.~Ristenpart, ``Next
  stop, the cloud: Understanding modern web service deployment in ec2 and
  azure,'' in \emph{Conference on Internet Measurement Conference}, ser.
  IMC.\hskip 1em plus 0.5em minus 0.4em\relax NY, USA: ACM, 2013, pp. 177--190.

\bibitem{gaming2}
R.~Shea, J.~Liu, E.~C.~H. Ngai, and Y.~Cui, ``Cloud gaming: architecture and
  performance,'' \emph{IEEE Network}, vol.~27, no.~4, pp. 16--21, July 2013.

\bibitem{shukla-tpctc}
A.~Shukla and Y.~Simmhan, ``Benchmarking distributed stream processing
  platforms for iot applications,'' in \emph{8th TPC Technology Conference},
  ser. TCPC.\hskip 1em plus 0.5em minus 0.4em\relax Cham: Springer
  International, 2016, pp. 90--106.

\bibitem{uber:2012}
U.~I. Voytek, ``Rides of glory,'' 2012.

\bibitem{weber:2010}
R.~H. Weber, ``Internet of things--new security and privacy challenges,''
  \emph{Computer Law \& Security Review}, 2010.

\bibitem{Akdere:2008}
M.~Akdere, U.~\c{C}etintemel, and N.~Tatbul, ``Plan-based complex event
  detection across distributed sources,'' \emph{Proc. VLDB Endowment}, vol.~1,
  no.~1, pp. 66--77, Aug. 2008.

\end{thebibliography}

\end{document}